\documentclass{article}
\usepackage[margin=1in]{geometry}
\usepackage{graphicx}
\usepackage[colorlinks=true,linkcolor=blue, urlcolor=blue, citecolor=blue]{hyperref}
\usepackage{xcolor}
\usepackage{pdfpages}
\usepackage{amsmath}       
\usepackage{amssymb}       
\usepackage{bm}            
\usepackage{longtable}     
\usepackage{array}         
\usepackage{calc}          
\usepackage{booktabs}      
\usepackage{tabularx}      
\long\def\todo#1{{\bfseries\color{red}[TODO: #1]}}

\usepackage[backend=biber,sorting=none]{biblatex}
\usepackage{xspace}   
\usepackage{textcomp} 
\usepackage{needspace}   
\usepackage{placeins}    
\usepackage{float}       
\usepackage{etoolbox}    
\usepackage{fancyvrb}    
\usepackage{color}       
\usepackage{subfig}      
\usepackage{tikz}        
\usetikzlibrary{decorations.pathmorphing,decorations.markings,arrows.meta}
\usepackage{colortbl}    
\newcommand{\outdated}{\rowcolor{black!8}}  

\newif\ifdelphi      \delphifalse
\newif\ifdelphidiff  \delphidifffalse
\newif\ifdelphiprompt \delphiprompttrue
\newif\ifreviewfindings \reviewfindingstrue
\newcommand{\dminus}[1]{\ifdelphidiff\textcolor{purple}{#1}\else\ifdelphi#1\fi\fi}
\newcommand{\dplus}[1]{\ifdelphidiff\textcolor{green!60!black}{#1}\else\ifdelphi\else#1\fi\fi}
\newcommand{\delphirow}{\ifdelphidiff\rowcolor{purple!12}\fi}

\DefineVerbatimEnvironment{Highlighting}{Verbatim}{commandchars=\\\{\}}

\makeatletter
\newsavebox\pandoc@box
\newcommand*\pandocbounded[1]{
  \sbox\pandoc@box{#1}%
  \Gscale@div\@tempa{\textheight}{\dimexpr\ht\pandoc@box+\dp\pandoc@box\relax}%
  \Gscale@div\@tempb{\linewidth}{\wd\pandoc@box}%
  \ifdim\@tempb\p@<\@tempa\p@\let\@tempa\@tempb\fi%
  \ifdim\@tempa\p@<\p@\scalebox{\@tempa}{\usebox\pandoc@box}%
  \else\usebox{\pandoc@box}%
  \fi%
}
\makeatother

\AtBeginEnvironment{longtable}{\small}
\newcommand{\slop} {{\textsc{JFC}}\xspace}

\usepackage{titlesec}
\titleclass{\subsubsubsection}{straight}[\subsubsection]
\newcounter{subsubsubsection}[subsubsection]
\renewcommand{\thesubsubsubsection}{\thesubsubsection.\arabic{subsubsubsection}}
\titleformat{\subsubsubsection}{\normalfont\normalsize\bfseries}{\thesubsubsubsection}{1em}{}
\titlespacing*{\subsubsubsection}{0pt}{3.25ex plus 1ex minus .2ex}{1.5ex plus .2ex}
\titleclass{\subsubsubsubsection}{straight}[\subsubsubsection]
\newcounter{subsubsubsubsection}[subsubsubsection]
\renewcommand{\thesubsubsubsubsection}{\thesubsubsubsection.\arabic{subsubsubsubsection}}
\titleformat{\subsubsubsubsection}{\normalfont\normalsize\itshape}{\thesubsubsubsubsection}{1em}{}
\titlespacing*{\subsubsubsubsection}{0pt}{3.25ex plus 1ex minus .2ex}{1.5ex plus .2ex}

\usepackage{authblk}
\usepackage{orcidlink}  

\usepackage{calc} 

\newlength{\thanksindent}
\title{AI Agents Can Already Autonomously Perform \\ Experimental High Energy Physics}

\author[1,2]{Eric A. Moreno\,\orcidlink{0000-0001-5666-3637}\thanks{%
These authors contributed equally.\par
\hangindent=\thanksindent
\noindent\hspace*{\thanksindent}%
$^\dagger$\href{mailto:emoreno@mit.edu}{emoreno@mit.edu} \,
$^\ddagger$\href{mailto:sambt@mit.edu}{sambt@mit.edu} \,
$^\S$\href{mailto:novaka@mit.edu}{novaka@mit.edu} \,
$^\sharp$\href{mailto:dolores.garcia@cern.ch}{dolores.garcia@cern.ch} \,
$^\flat$\href{mailto:pcharris@mit.edu}{pcharris@mit.edu} \,
$^\natural$\href{mailto:yiyang26@mit.edu}{yiyang26@mit.edu}%
}\textsuperscript{$\dagger$}}
\author[1,2]{Samuel Bright-Thonney\,\orcidlink{0000-0003-1889-7824}\textsuperscript{*$\ddagger$}}
\author[1,2]{Andrzej Novak\,\orcidlink{0000-0002-0389-5896}\textsuperscript{*$\S$}}
\author[3]{Dolores Garcia\,\orcidlink{0000-0002-0120-8757}\textsuperscript{$\sharp$}}
\author[1,2]{Yiyang Zhao\,
\orcidlink{0009-0000-2290-1828}\textsuperscript{$\natural$}}
\author[1,2]{Philip Harris\,\orcidlink{0000-0001-8189-3741}\textsuperscript{$\flat$}}
\date{\today}

\affil[1]{Department of Physics, Massachusetts Institute of Technology}
\affil[2]{NSF AI Institute for Artificial Intelligence and Fundamental Interactions}
\affil[3]{CERN}

\begin{document}

\maketitle

\begin{abstract}
Large language model-based AI agents are now able to autonomously execute substantial portions of a high energy physics (HEP) analysis pipeline with minimal expert-curated input. Given access to a HEP dataset, an execution framework, and a corpus of prior experimental literature, we find that Claude Code succeeds in automating all stages of a typical analysis: event selection, background estimation, uncertainty quantification, statistical inference, and paper drafting.
We argue that the experimental HEP community is underestimating the current capabilities of these systems, and that most proposed agentic workflows are too narrowly scoped or scaffolded to specific analysis structures.
We present a proof-of-concept framework, \emph{Just Furnish Context} (\slop), that integrates autonomous analysis agents with literature-based knowledge retrieval and multi-agent review, and show that this is sufficient to plan, execute, and document a credible high energy physics analysis. We demonstrate this by conducting analyses on open data from ALEPH, DELPHI, and CMS to perform electroweak, QCD, and Higgs boson measurements. We present two of those results in a condensed short paper form --- a CMS Run1 Open Data $H\to \tau^+\tau^-$ to demonstrate performance on a well-established result, and the first Lund plane measurement on LEP data --- a genuinely novel result and, to our knowledge, the first produced autonomously by an AI agent.
Rather than replacing physicists, these tools promise to offload the repetitive technical burden of analysis code development, freeing researchers to focus on physics insight, truly novel method development, and rigorous validation.
Given these developments, we advocate for new strategies for how the community trains students, organizes analysis efforts, and allocates human expertise.
\end{abstract}

\vspace{0.5em}

\section{Introduction}
\label{sec:intro}

A typical experimental high energy physics analysis is a years-long endeavor, often spanning the majority of a graduate student or postdoc's time at an academic institution. For large collider experiments the process is algorithmic: a physicist (a) identifies an interesting measurement channel or new physics signature%
, (b) studies existing literature to understand how similar measurements have been done, (c) designs and validates an event selection procedure using Monte Carlo (MC) simulation, quantifies all relevant sources of systematic uncertainty, and finally (d) ``unblinds" on real experimental data and performs statistical hypothesis tests to extract measurements or limits.

This process invariably involves writing thousands of lines of code to process data and apply standard, though not off-the-shelf, analysis techniques, most of which is structurally similar to code written in dozens of other analyses within the same collaboration. While this can be a useful exercise for developing programming skills, it often consumes a substantial fraction of a physicist's working time and demands little physics reasoning or insight. This is frequently exacerbated by the absence of high-quality, centralized, or up-to-date documentation for much of the core software infrastructure within experimental collaborations. The result is a slow, error-prone, and tedious process that, beyond a brief initial learning phase, adds little to a student's education and robs all practitioners of valuable time that could be better spent on higher-level questions.

In this paper, we argue that the majority of this process can be delegated to AI agents. Modern coding assistants can autonomously write and execute code, access documentation (both local and online), iteratively critique and debug their own output, and document their progress for human overseers \cite{jimenez2023swe, achiam2023gpt}. As large language models (LLMs) and commercial agentic frameworks continue to improve, these tools will only become more reliable.

This is not a speculative claim about future capabilities. In this paper we show that, given a general framework distilling the processes used in practice at large HEP collaborations and an initial high-level physics prompt, Claude Code\footnote{As well as other commercial frontier models.} is able to produce a complete analysis: event selection, background estimation, systematic uncertainty evaluation, statistical inference, and a written report with publication-grade figures. The report quality is typically, at a cursory expert review, indistinguishable from a report produced by a junior graduate student (or an expert under time constraints). Recent work by Badea et al.\ \cite{badea2026agentic} demonstrated that an AI agent, with iterative physicist supervision at each step, can contribute to a measurement using LEP open data. We build on this direction but take a fundamentally different approach: rather than using a supervised \emph{human-in-the-loop} approach, \slop delegates review to specialized AI agents, concentrating human oversight at a single unblinding gate, yielding a nearly fully autonomous framework.

The framework specification distills the two things a research group supplies to a new student: a methodology for how an analysis proceeds and the domain-specific conventions of the field. It then supplements those with a behavioral contract governing how the agents communicate and are reviewed, shown in detail in Section~\ref{sec:framework}.

\begin{figure}[ht]
  \centering
  \includegraphics[width=0.95\linewidth]{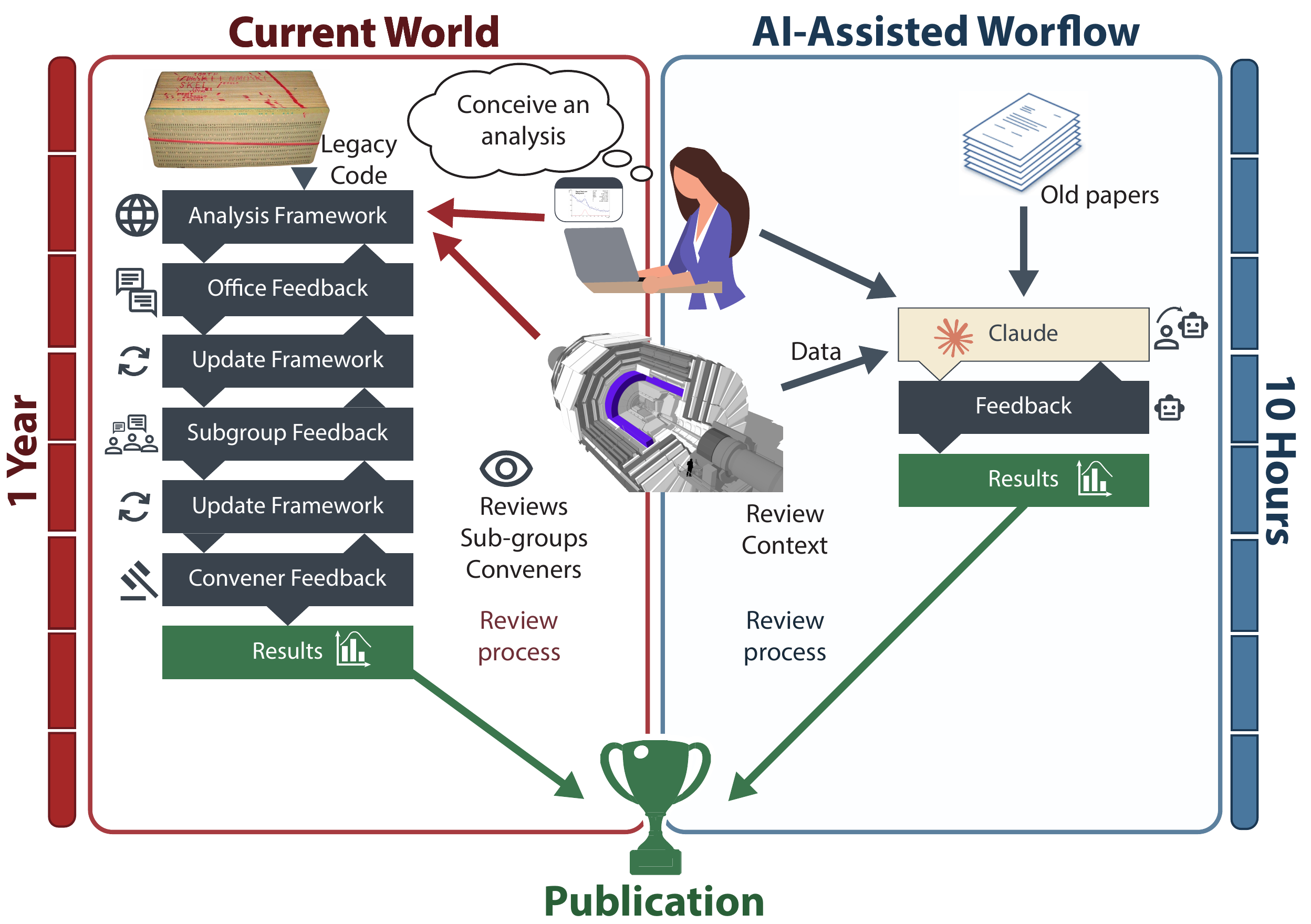}
  \caption{ Diagram of how an AI-agent workflow can mirror the typical high-energy physics analysis workflow. On the left, we show the conventional analysis pipeline, which usually starts with legacy code that is then modified to perform the analysis. Analyses normally involve 3 or more levels of review, starting with feedback from other postdocs, students, and faculty collaborating on the analysis (office feedback). The next tier sits within an analysis subgroup (colloquially referred to as a level 3 group). The second tier of review involves a pre-approval phase, in which physics group conveners review an analysis, followed by a formal collaboration review leading to a result and a submission for publication. On the right side, an equivalent interactive workflow can be entirely handled by AI agents, from the conception of an idea through a result that would then undergo a similar collaboration review, followed by publication. } 
  \label{fig:vision}
\end{figure}

The analyses presented in this paper were produced entirely by AI agents operating on archived open ALEPH and DELPHI\footnote{We also ran a set of DELPHI analyses as earlier exploratory runs on an earlier model (Claude Opus~4.6); we include them to illustrate the framework's reach (e.g.\ Fig.~\ref{fig:timings}) but do not present their numerical results as validated measurements in this paper.} data and Monte Carlo simulation from the Large Electron-Positron (LEP) collider and CMS Open Data from the Large Hadron Collider (LHC)\cite{CERNOpenData}, with the agents optionally accessing domain knowledge from published LEP papers through a structured retrieval system or internet access.

Importantly, \textbf{we present these analyses primarily to showcase the level of HEP analysis that current agentic systems can produce entirely autonomously, rather than as legitimate scientific results}, based only on a short physics prompt and the \slop framework. In addition, we present two results that have been evaluated more closely for correctness
as distilled short-form papers (Appendices~\ref{app:tautau_pub} and~\ref{app:lund_pub}). \textbf{We believe this warrants rethinking how mainstream analysis work is done in HEP collaborations and how student labor is utilized.} Beyond routine analysis, these tools also open the door to systematic reproducibility studies, large-scale reanalysis of archived datasets, and rigorous documentation of analysis workflows --- all longstanding challenges that the required human effort and expertise have historically kept impractical to address at scale.

We emphasize that we are not calling for AI agents to supplant humans in producing scientific results. The final output of any AI-assisted analysis must be thoroughly checked, understood, and validated by domain experts before it can be considered a scientific result, and must undergo the same internal scrutiny as any other collaboration paper. We also do not claim that current agents can handle every aspect of every analysis; complex analyses involving novel techniques, bespoke reconstruction algorithms, or subtle interplay between multiple systematic uncertainties will continue to require substantial hands-on human involvement. Agents make mistakes, sometimes subtle ones, and humans must remain responsible for judging their outputs and be held publicly accountable for their mistakes. Most importantly, it is our creativity, expertise, and judgment that defines the research agenda these tools will help us pursue.

The remainder of this paper describes our proof-of-concept framework, the lessons learned from applying it to ALEPH, DELPHI, and CMS open data, and its current limitations.
Section~\ref{sec:related} reviews the rapidly growing landscape of agentic AI for science and for HEP specifically.
Section~\ref{sec:framework} describes the \slop framework, including its agent architecture, knowledge retrieval system, and multi-agent review process.
Section~\ref{sec:results} surveys the analyses ``we'' have reproduced using \slop and discusses their quality, throughput and cost, and limitations. Section~\ref{sec:discussion} then examines the broader implications for analysis workflows, legacy data reanalysis, and graduate training.
Section~\ref{sec:conclusion} concludes with a call to action. Links to the code and documentation for all agent-generated analyses can be found at \url{https://jfc-mit.github.io/}.

\section{Related Work}
\label{sec:related}

The application of AI agents to scientific research is a rapidly growing field spanning many domains, from chemistry and drug discovery to materials science and mathematics.
In chemistry, ChemCrow~\cite{bran2023chemcrow} augments large language models with expert-designed chemistry tools to autonomously plan and execute synthesis tasks, while Coscientist~\cite{boiko2023autonomous} integrates LLMs with internet search, code execution, and robotic lab equipment to design and carry out chemical experiments end-to-end.
The AI Scientist~\cite{lu2024ai} demonstrated a fully autonomous pipeline, from idea generation through experiment execution to paper writing.
Multi-agent architectures have also been explored for scientific discovery: SciAgents~\cite{ghafarollahi2025sciagents} combines LLMs with knowledge graphs in a multi-agent framework for materials science, while PaperQA2~\cite{skarlinski2024language} achieves superhuman performance on scientific literature synthesis through an agentic retrieval-augmented generation pipeline.
These efforts illustrate a broad trend toward autonomous AI-driven research, but each domain brings its own challenges in terms of data complexity, validation requirements, and the need for domain-specific reasoning.

Within high energy physics (HEP), interest in applying LLM-based agents to analysis workflows has grown significantly in the past year, with several concurrent and complementary efforts emerging.
A community vision paper~\cite{Aarrestad:2026xrs} recently outlined grand challenges for building an AI-native research ecosystem for experimental particle physics, providing a high-level roadmap for how current and future facilities can benefit from AI integration.
We review the existing landscape of HEP-focused agentic AI efforts below, organized by the type of task they address.

\subsection{Agents for data analysis}

The most directly relevant prior work is that of Gendreau-Distler et al.~\cite{Gendreau-Distler:2025fsj}, who present an LLM-agent-driven data analysis framework.
Their system pairs an LLM-based supervisor-coder agent with the Snakemake workflow manager to automate a Higgs boson diphoton cross-section measurement using ATLAS Open Data.
The workflow manager enforces reproducibility and determinism, while the agent generates, executes, and iteratively corrects analysis code.
The authors benchmark several state-of-the-art LLMs spanning the Gemini, GPT, and Claude families, as well as leading open-weight models.
Notably, however, the authors themselves state that ``multi-step task planning is beyond the current scope'' of their system~\cite{Gendreau-Distler:2025fsj}, highlighting that the agent operates within a fairly rigid, pre-defined analysis structure rather than autonomously planning and executing a full analysis from a high-level specification.

Diefenbacher et al.~\cite{Diefenbacher:2025zzn} investigate LLM-based agents for anomaly detection using the LHC Olympics dataset~\cite{Kasieczka:2021xcg}, demonstrating that agentic setups can develop and test analysis methods comparable to state-of-the-art human-designed approaches.
Their study provides a systematic comparison of prompting strategies and LLM models, benchmarking stability, cost, and reproducibility across multiple runs.
This work is notable for showing that agents can independently arrive at sensible analysis strategies---such as combining bump hunts with weakly supervised approaches like CWoLa---without being explicitly instructed to do so.

Menzo et al.\ present HEPTAPOD~\cite{Menzo:2025cim}, an orchestration framework that enables large language models to interface with domain-specific HEP tools, construct simulation workflows, and manage multi-step research pipelines.
The system uses schema-validated operations and run-card-driven configuration to ensure reproducibility, and is demonstrated on a representative BSM Monte Carlo validation pipeline spanning model generation, event simulation, and downstream analysis within a unified workflow.
HEPTAPOD provides a structured, auditable layer between researchers, LLMs, and computational infrastructure, focusing on upstream simulation and workflow management rather than end-to-end data analysis, with human-in-the-loop oversight throughout.

Most recently, Badea et al.~\cite{badea2026agentic} present a proof-of-concept measurement using LEP open data with an agentic AI--physicist collaboration.
Their approach relies on an iterative human-in-the-loop cycle in which the physicist provides detailed feedback after each agent attempt, guiding the analysis toward a correct result.
While that work shows that agents can play a substantive role in a real measurement, its supervised workflow differs fundamentally from the autonomous approach we propose: in \slop, multi-agent review replaces the human feedback loop during analysis development, and human oversight is concentrated at a single formal unblinding gate rather than distributed throughout the process.

\subsection{Agents for simulation and experiment design}

GRACE~\cite{Hill:2026naa} takes a different approach, targeting the upstream problem of experimental design rather than data analysis.
Given a natural-language prompt or a published experimental paper, GRACE extracts a structured representation of the experiment, constructs a runnable simulation, and autonomously explores design modifications using Monte Carlo methods.
The system demonstrates that an agentic approach to detector geometry optimization can identify directions consistent with known upgrade priorities, using only baseline simulation inputs.
ColliderAgent~\cite{Qiu:2026iby} presents a language-driven multi-agent system for end-to-end collider phenomenology, capable of executing workflows from a theoretical Lagrangian to final phenomenological outputs---including parton-level and detector-level analyses, and large-scale parameter scans---without requiring package-specific code.
Their hierarchical agent architecture is coupled to a unified execution backend for phenomenological calculation and simulation toolchains, and is validated on representative BSM scenarios including leptoquark and axion-like-particle models.

\subsection{LLM-assisted toolkits}

CoLLM~\cite{Esmail:2026jpb} provides an end-to-end pipeline from plain-language analysis specifications to trained deep learning classifiers for collider analyses.
While it uses LLMs to generate analysis code under physics constraints, it functions more as an AI-assisted toolkit with a graphical user interface than as a fully autonomous agent, and focuses specifically on the deep learning classification step of an analysis.
Xiwu~\cite{Zhang:2024kws} is an LLM assistant for HEP that can switch between open-weight models and be fine-tuned on domain knowledge.

\subsection{Benchmarks and community efforts}

The CelloAI benchmarks~\cite{Atif:2026eju} provide a framework for evaluating AI assistants in HEP contexts, focusing on code documentation and generation tasks.
Most recently, Collider-Bench~\cite{colliderbench} introduced a benchmark in which LLM agents must reproduce published LHC searches as executable simulation-and-selection pipelines, scored on predicted signal-region yields, and reports that no agent reliably beats a physicist-in-the-loop baseline.
Both CelloAI and Collider-Bench are deliberately scoped to controlled, automatically scorable tasks --- code documentation and generation in the former, and in the latter the reconstruction of a published search as a Monte-Carlo simulation-and-selection pipeline graded on predicted signal-region yields. The framework we present targets a substantially broader portion of the analysis workflow: operating autonomously on real experimental open data, it designs the event selection, estimates backgrounds, evaluates systematic uncertainties, performs the blinded statistical inference, and drafts the documentation --- in one case delivering a measurement with no prior $e^+e^-$ counterpart. The two directions are complementary: Collider-Bench supplies the controlled, quantitative evaluation of agent capability that our single-run demonstrations do not, while \slop probes how far an autonomous pipeline can carry a complete measurement on real data.

\subsection{Knowledge retrieval for physics analysis}
\label{sec:related:rag}

A key challenge for autonomous analysis agents is accessing and applying domain knowledge from the existing literature.
Standard retrieval-augmented generation (RAG) approaches struggle with scientific text because relevant information often spans multiple sections, figures, and even multiple papers.
McGreivy et al.~\cite{McGreivy:2025rrz} address this with SciTreeRAG, which exploits the hierarchical structure of experimental physics papers to build a tree representation of a document corpus, enabling more contextually coherent retrieval than flat chunking approaches.
They additionally introduce SciGraphRAG, which transforms unstructured literature into structured knowledge graphs to capture cross-document relationships.
Both systems are demonstrated on the LHCb experiment corpus.
PaperQA2~\cite{skarlinski2024language} takes a complementary approach, treating retrieval and response generation as a multi-step agent task that can revise its search parameters and examine candidate answers before producing a final response, achieving superhuman accuracy on scientific literature synthesis benchmarks.

\subsection{Summary and gaps}

The existing landscape reveals several important patterns.
First, most HEP-focused agentic systems operate within highly scaffolded workflows where the analysis structure is pre-defined and the agent's role is limited to code generation within fixed steps.
Second, none of the existing systems combine autonomous multi-step planning, domain knowledge retrieval from the literature, and multi-agent review into an integrated framework.
Third, while a benchmark for agent-based reproduction of LHC searches at the simulation level has very recently appeared~\cite{colliderbench}, the community still lacks benchmarks for evaluating agents on the realistic, end-to-end analysis of real experimental data---the kind that require event selection, background estimation, systematic uncertainty quantification, and statistical inference.
These gaps motivate the framework we present in the following sections.

\section{The \slop Framework}

\slop is an agentic framework for autonomous execution of HEP analysis pipelines.
An orchestrator agent delegates execution and review work to subagents across sequential phases.
Each phase must produce a written artifact and pass an independent review before the next phase can begin, providing structured task decomposition and fresh context management.
In this section, we describe its domain-specific capabilities: physics-aware analysis planning, literature-based knowledge retrieval, and multi-agent review with role-specialized critics.

\label{sec:framework}
\begin{figure}[ht!]
  \centering
  \includegraphics[width=0.8\linewidth]{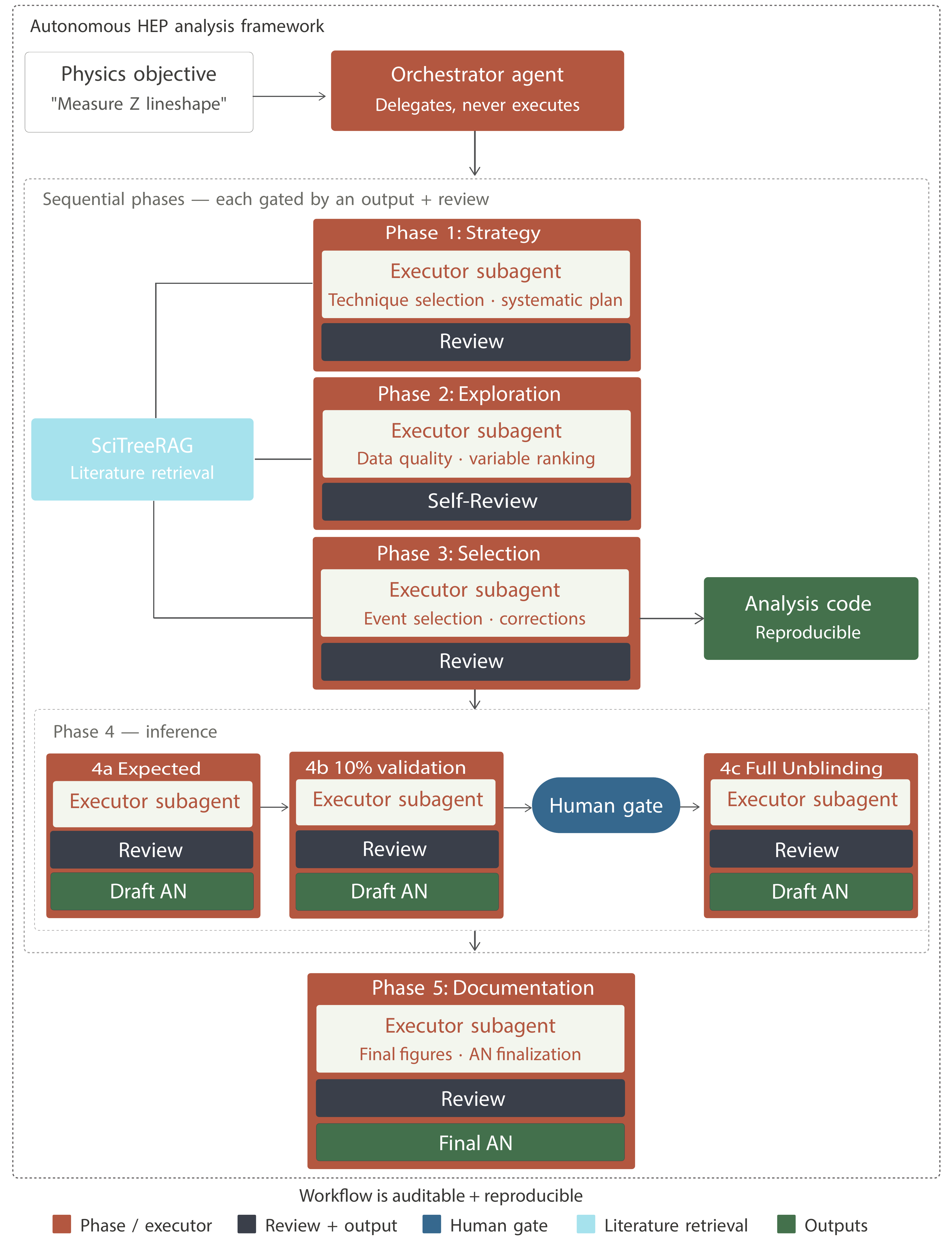}
  \caption{ The \slop framework. A high-level physics objective is
           passed to an autonomous analysis agent, which plans and executes the full
           pipeline while querying a literature retrieval system (SciTreeRAG) for
           domain knowledge. The resulting analysis undergoes multi-agent review;
           if any reviewer flags an issue the agent revises and resubmits until all
           reviewers approve. Eventually the final document (a rich analysis note
           \protect\footnote{An analysis note is the standard HEP analysis record used for
           internal review akin to a thesis draft.}) is 
           passed to human physicists for evaluation. }
  \label{fig:gad-framework}
\end{figure}

\subsection{Architecture overview}

The \slop framework is developed and deployed with \textbf{Claude Code}, Anthropic's agentic coding CLI, which serves as both the LLM backend and the agent runtime with the Opus~4.8 model with a one-million-token context window, used for all executor, reviewer, and arbiter subagents.

The key architectural principle is that the agent operates with genuine autonomy over the analysis workflow.
Unlike prior systems that constrain the agent to code generation within pre-defined analysis steps~\cite{Gendreau-Distler:2025fsj}, the \slop analysis agent receives a high-level physics objective---such as ``measure the hadronic cross-section of the Z boson using ALEPH data''---and must autonomously determine the full analysis strategy: what event selection to apply, how to estimate backgrounds, which systematic uncertainties to evaluate, what statistical framework to use, and how to present the results.
The agent is not given a template or skeleton analysis to fill in; it must plan and execute the entire pipeline from scratch, consulting the literature as needed.

The framework specification splits roughly into three independent components:
\begin{itemize}
    \item \textbf{Methodology.} We encode what a typical particle physics analysis workflow looks like, from planning and data exploration to statistical analysis. This structure, while not usually explicitly spelled out, generally lives as institutional knowledge in scientific collaborations and research groups; roughly, it corresponds to the guidance a graduate student would receive as they progress through their thesis research. A separate aspect of this methodology is how a collaboration ensures its published results are accurate: an official review process that progresses in tiers from junior postdocs, to senior postdocs leading analysis groups, to senior academics who function as internal journal reviewers. We simulate this by encoding tiers of review subagents at every step of the analysis.
    \item \textbf{Domain-specific conventions.} We specify a well-isolated set of conventions --- tool use, visualization standards, and more obscure or recent analysis techniques. Encoding this baseline seems unavoidable, as general models cannot reliably guess the specific conventions that vary greatly by field.
    \item \textbf{Agent behavior.} To manage context windows, the analysis steps outlined in the methodology are dispatched to separate subagents that receive a pared-down overall analysis context and plan together with their dedicated instruction sets. We therefore fix a strict specification of what each agent receives and outputs (operationally, no physics information), together with an append-only log that lets the process be explored and verified by humans and that facilitates agent-level debugging of complex failures.
\end{itemize}
Where the methodology tells the agent that it must, say, evaluate systematic uncertainties, the conventions tell it which sources are standard and why; together they constitute an operational context analogous to a graduate course curriculum establishing the standards and domain knowledge against which the agent's decisions are made and reviewed.

\subsection{Task decomposition into phases}

The orchestrator decomposes the analysis into sequential phases, each gated by the production of a written artifact on disk and passage of an independent review.
Within each inference phase (4a--4c) the \texttt{note-writer} and \texttt{typesetter} agents produce or update the analysis note alongside the inference artifact (the documentation steps, commit-tagged \texttt{doc4a}--\texttt{doc4c}); a final documentation phase (Phase~5) then produces the publication figures, the final note, and the typeset PDF:

\begin{enumerate}
    \item \textbf{Phase~1 --- Strategy} (4-bot review): The \texttt{executor} agent queries the literature corpus, identifies signal and background processes, proposes event selection, defines the blinding variable, and outlines the systematic uncertainty program.
    A mandatory reference analysis survey (2--3 published analyses) and a conventions compliance table are required.
    The output artifact is \texttt{STRATEGY.md}.

    \item \textbf{Phase~2 --- Exploration} (self + plot-validator review): The \texttt{executor} agent performs data reconnaissance (sample inventory, data quality, variable ranking) and literature search (reference analyses, modern methods, published inputs).
    Output artifact: \texttt{EXPLORATION.md}.

    \item \textbf{Phase~3 --- Processing} (1-bot + plot-validator review): The \texttt{executor} agent implements the Phase~1 strategy: event selection, control and validation regions, background estimation or correction chain, closure tests, and stress tests.
    Output: \texttt{SELECTION.md}.

    \item \textbf{Phase~4a --- Expected Results} (review by physics, critical, constructive, plot-validator, and bibtex reviewers, adjudicated by the arbiter): The \texttt{executor} agent evaluates systematic uncertainties, constructs the statistical model, performs Asimov fits, signal injection tests, and mandatory fit diagnostics; the \texttt{note-writer} and \texttt{typesetter} then produce the first complete analysis note, compiled to PDF via \texttt{tectonic}.
    Outputs: \texttt{INFERENCE\_EXPECTED.md} and \texttt{ANALYSIS\_NOTE\_4a\_v1}.

    \item \textbf{Phase~4b --- Partial Unblinding} (same review panel $\to$ human gate): The \texttt{executor} runs the full analysis chain on a 10\% signal-region data subsample (fixed random seed) and validates consistency with expected results; the \texttt{note-writer} updates the analysis note with the 10\% results and an unblinding checklist.
    Outputs: \texttt{INFERENCE\_PARTIAL.md} and \texttt{ANALYSIS\_NOTE\_4b\_v1}.
    After the review passes, the pipeline \emph{pauses} and presents a structured summary to the human analyst, who must explicitly approve, request changes, or halt. This step strictly mirrors established practices and is where the collaboration (or in this case human supervisor) can raise concerns with the methodology without biasing the result.

    \item \textbf{Phase~4c --- Full Unblinding} (1-bot review: critical reviewer + plot-validator): Executes only after human approval.
    The \texttt{executor} runs on the complete dataset and validates consistency with partial and expected results; the \texttt{note-writer} updates the analysis note with the full-data results.
    Outputs: \texttt{INFERENCE\_OBSERVED.md} and \texttt{ANALYSIS\_NOTE\_4c\_v1}.

    \item \textbf{Phase~5 --- Documentation} (full panel: physics, critical, constructive, plot-validator, rendering, and bibtex reviewers, adjudicated by the arbiter): The \texttt{note-writer} and \texttt{typesetter} produce the final publication figures, the final analysis note, and the typeset PDF.
    Output: \texttt{ANALYSIS\_NOTE\_5\_v1}, the compiled PDF, and a \texttt{results/} directory.
\end{enumerate}

The orchestrator agent itself never writes analysis code or reads full data files; it delegates all computational work to subagents.
Context is passed between phases exclusively through written artifacts and an append-only experiment log (\texttt{experiment\_log.md}) that records decisions, failures, and their mechanisms.

\subsection{Tools and software stack}

The framework enforces a specific, pure-Python HEP software stack, documented in Table~\ref{tab:toolstack}.

\begin{table}[!ht]
\centering
\small
\begin{tabularx}{\textwidth}{@{}l >{\raggedright\arraybackslash}X >{\raggedright\arraybackslash}X@{}}
\hline
\textbf{Task} & \textbf{Required tool} & \textbf{Explicitly forbidden} \\
\hline
ROOT file I/O & \texttt{uproot}~\cite{Pivarski_Uproot_2017} ($\geq$5.0) & PyROOT, ROOT C++ macros \\
Array operations & \texttt{awkward-array}~\cite{Pivarski_Awkward_Array_2018}, \texttt{numpy}~\cite{harris2020array} & pandas (for event data) \\
Histogramming & \texttt{hist}~\cite{Schreiner_hist}, \texttt{boost-histogram}~\cite{boost_histogram} & ROOT \texttt{TH1}, \texttt{numpy.histogram} \\
Plotting & \texttt{matplotlib}~\cite{Hunter:2007} + \texttt{mplhep}~\cite{Novak_mplhep_2020} & ROOT \texttt{TCanvas}, plotly \\
Statistical modeling & \texttt{pyhf}~\cite{Heinrich_pyhf_v0_7_6} (binned), \texttt{zfit}~\cite{zfit} (unbinned) & RooFit, RooStats \\
4-vectors & \texttt{vector}~\cite{Chopra_Vector_JIT-compilable_mathematical_2025}, \texttt{particle}~\cite{Rodrigues_Particle} & Manual calculations \\
Document preparation & LaTeX + \texttt{tectonic} & LLM-based conversion \\
Environment management & \texttt{pixi}~\cite{fischer2025pixi} & manual \texttt{pip install} \\
\hline
\end{tabularx}
\caption{Mandated software stack. Optional packages include \texttt{coffea} for columnar event processing, \texttt{xgboost}/\texttt{scikit-learn} for MVA, \texttt{iminuit}/\texttt{cabinetry} for fitting, and \texttt{fastjet} for jet clustering, activated as needed per analysis.}
\label{tab:toolstack}
\end{table}

All analysis code is written in a columnar style: selections are boolean masks over arrays, not event loops.
Agents prototype on $\sim$1000-event slices before scaling to the full dataset, with automatic scale-out rules: single-core for jobs under 2 minutes, \texttt{ProcessPoolExecutor} for 2--15 minutes, and SLURM batch submission for longer runs.

\subsection{Literature-based knowledge retrieval}
\label{sec:knowledge}

A critical component of the \slop framework is its integration with a literature-based knowledge retrieval system built on SciTreeRAG~\cite{McGreivy:2025rrz}.
An analysis agent operating without access to the existing literature would need to rely entirely on knowledge absorbed during pre-training---a brittle foundation, since the specific details of how a particular experiment's event selection was designed, what detector effects must be accounted for, or how a particular systematic uncertainty was historically evaluated are rarely represented in an LLM's training data with sufficient fidelity to be directly useful.

The primary retrieval system indexes a corpus of published LEP (ALEPH, DELPHI) papers using SciTreeRAG, exploiting the hierarchical document structure (sections, subsections, figures, tables) to build contextually coherent retrievals rather than the flat text chunks used by standard RAG systems.
This is complemented by an arXiv MCP server for modern methodology search and INSPIRE-HEP/HEPData queries for recent measurements and published numerical inputs.

The ALEPH catalog contains 1,503 total entries (399 papers, 736 proceedings, 368 theses). Of those, 721 have source material available, and 575 were successfully converted to markdown. The DELPHI catalog contains 4,305 total entries (2,967 papers, 763 proceedings, 575 theses). Of those, 2,083 have source material available, and 1,868 were successfully converted to markdown.

Metadata was harvested from INSPIRE-HEP and CERN CDS. Source PDFs and LaTeX were then downloaded. PDFs were converted to structured markdown using Nougat~\cite{blecher2023nougat} (a GPU-based neural OCR model) running on A100 80GB GPUs, while LaTeX sources were converted via Pandoc. The raw markdown then went through a reproducible postprocessing pipeline that normalizes math delimiters, strips missing-page markers and boilerplate, removes Pandoc div markers, and collapses excess whitespace.

When the analysis agent needs domain knowledge---for example, what charged-track quality cuts were used in previous ALEPH hadronic event selections, or what sources of systematic uncertainty were considered in a particular measurement---it queries the retrieval system and receives relevant passages from published papers, complete with the surrounding context needed to interpret them correctly.
This allows the agent to make informed decisions grounded in established experimental practice rather than guessing or hallucinating analysis choices.

The retrieval system is particularly important for three aspects of an analysis:
\begin{itemize}
    \item \textbf{Event selection}: The agent retrieves descriptions of selection criteria used in similar published analyses to inform its own selection design, adapting cuts to the specific measurement while maintaining consistency with established practice.
    \item \textbf{Systematic uncertainties}: Identifying the relevant sources of systematic uncertainty for a given measurement is one of the most expertise-intensive aspects of an analysis. The retrieval system provides the agent with descriptions of how systematic uncertainties were evaluated in prior work, including which variations were considered and how they were propagated.
    \item \textbf{Statistical methods}: The agent retrieves information about the statistical frameworks used in similar measurements (e.g., profile likelihood fits, template methods) to guide its choice of inference procedure.
\end{itemize}

The analyses on ALEPH Open Data presented here had access to this literature-based knowledge retrieval; those on DELPHI and CMS did not.

\subsection{Agent architecture}

The framework defines specialist agent profiles, each specified as a Markdown file in the framework's \texttt{agents/} directory (symlinked into each analysis) with a YAML frontmatter declaring the agent's name, description, available tools, and model tier.
These agents fall into four functional categories:

\begin{itemize}
    \item \textbf{Executor} (1): A single generalist \texttt{executor} agent handles all phases, with a deep domain prompt encoding HEP methodology (mandatory fit diagnostics, in-situ constraint strategies, the CL$_s$ exclusion procedure, etc.).
    The same agent profile adapts to each phase's requirements through the phase-specific instructions provided by the orchestrator.

    \item \textbf{Specialist agents} (4): \texttt{fixer} (makes targeted changes to resolve review findings during ITERATE cycles), \texttt{investigator} (traces regression triggers to their root cause across phases), \texttt{note-writer} (drafts and updates the analysis-note prose during the documentation steps), and \texttt{typesetter} (a LaTeX specialist that compiles the note to PDF and resolves rendering issues).

    \item \textbf{Reviewer agents} (6): \texttt{physics-reviewer}, \texttt{critical-reviewer}, \texttt{constructive-reviewer}, \texttt{rendering-reviewer}, \texttt{plot-validator}, and \texttt{bibtex-validator}, each detailed in Section~\ref{sec:review}.

    \item \textbf{Adjudication} (1): \texttt{arbiter} (synthesizes all reviewer findings into a PASS/ITERATE/ESCALATE verdict).
\end{itemize}

Every subagent is spawned as a fresh process with no memory of prior invocations, receiving only the specific artifacts and instructions relevant to its task.
This disposable-context architecture prevents context window exhaustion (Context Rot~\cite{hong2025context}) during long analyses and ensures that each agent operates from a clean state.

\subsection{Multi-agent review}
\label{sec:review}

Before any analysis phase is considered complete, it undergoes automated review by a panel of specialized reviewer agents.
This multi-agent review system is designed to mirror the internal review process of a real HEP collaboration, where different experts scrutinize different aspects of an analysis before it is approved for publication.

Seven distinct agents participate in the review process (the six reviewers above plus the arbiter), detailed below.

\paragraph{Physics reviewer (\texttt{physics-reviewer}).}
Operates as an independent senior collaboration member equivalent to an Analysis Review Committee (ARC) member or Level-2 convener.
Critically, this agent is \emph{deliberately denied access} to the methodology specification, conventions documents, review checklists, and previous review feedback.
It receives only the physics prompt and the artifact under review, and evaluates purely on the basis of physics knowledge: signal modeling, background identification and estimation, systematic uncertainty completeness, cross-check adequacy, and publication readiness.
This design choice ensures that the physics reviewer assesses the analysis as an external referee would, without being guided by the framework's own criteria.
Issues are classified as Category~A (blocking: ``would cause rejection''), B (important: ``weakens the analysis''), or C (minor: ``style or clarity'').

\paragraph{Critical reviewer (\texttt{critical-reviewer}).}
Performs the most comprehensive review.
Unlike the physics reviewer, the critical reviewer \emph{does} read the methodology specification and conventions documents, and applies them systematically.
Its protocol includes eight mandatory steps:
(1)~phase-specific review focus,
(2)~issue classification (A/B/C),
(3)~row-by-row conventions compliance check against the applicable conventions document,
(4)~reference analysis comparison (``what would a competing group have that we do not?''),
(5)~a 14-point figure and label checklist (verifying axis labels with units, luminosity stamps, experiment labels, ratio panels, uncertainty bands, and the absence of forbidden elements such as figure titles),
(6)~physics sanity checks on every plot (distribution shapes, yield magnitudes, data/MC agreement, uncertainty proportionality),
(7)~regression detection by comparing the current artifact against earlier phase outputs, and
(8)~upstream feedback identifying issues that originate in prior phases.
Each finding includes a file path or figure reference, an impact statement, and a suggested fix.

\paragraph{Constructive reviewer (\texttt{constructive-reviewer}).}
Complements the critical reviewer by focusing on improvements and alternatives rather than flaw detection.
It evaluates clarity, validation sufficiency, alternative approaches (additional signal regions, complementary cross-checks, improved statistical methodology), presentation quality, and notation consistency.
Each suggestion includes the current state, the improved state, a justification, and an effort estimate (low/medium/high).
The constructive reviewer also provides explicit positive feedback (marked with \texttt{[+]}) for genuinely strong aspects of the analysis, and is instructed not to duplicate findings already raised by the critical reviewer.
While it primarily produces Category~B and C findings, it can escalate to Category~A if a fundamental gap is discovered.

\paragraph{Rendering reviewer (\texttt{rendering-reviewer}).}
Activated only at the final documentation phase (Phase~5), where the typeset PDF is the deliverable; at Phases~4a and 4b the \texttt{typesetter}'s own compilation serves as the rendering check.
This agent compiles the analysis note to PDF, then inspects the compiled output across eight dimensions:
figure rendering integrity, LaTeX math compilation, page layout (orphaned text, page breaks, margins), cross-reference resolution (\texttt{@fig:}, \texttt{@tbl:}, \texttt{@eq:}, \texttt{@sec:} labels), citation resolution against the bibliography, table formatting and overflow, and page count assessment (targeting 50--100 pages for a complete analysis note).
It focuses exclusively on rendering quality and does not comment on physics content.

\paragraph{BibTeX validator (\texttt{bibtex-validator}).}
Activated at the phases where the analysis note exists and carries citations (Phases~4a, 4b, and~5).
This agent checks \texttt{references.bib} for formatting correctness and verifies that all citations point to real bibliographic records, cross-referencing against INSPIRE-HEP and other databases.

\paragraph{Plot validator (\texttt{plot-validator}).}
Unlike the other reviewers, the plot validator performs \emph{programmatic}, non-visual validation.
It runs in parallel with the other reviewers during every review cycle that involves figure-producing phases.
Its checks fall into four categories:
\begin{itemize}
    \item \textbf{Programmatic figure checks} (8 checks): Verifies that collaboration plotting styles (\texttt{mplhep}) are applied, figure sizes match the template, no forbidden \texttt{ax.set\_title()} calls exist, axis labels include units, no hardcoded font sizes appear, \texttt{bbox\_inches="tight"} is used at save time, both PDF and PNG formats are saved, and \texttt{plt.close()} is called after saving.

    \item \textbf{Physics sanity checks} (11 checks): Verifies that yields are non-negative, efficiencies lie in $[0, 1]$, data/MC ratios in control regions fall within $[0.5, 2.0]$, uncertainties scale as $\sqrt{N}$, $p_T$ and mass distributions fall off at high values, cutflow yields are monotonically non-increasing, and background composition fractions sum to $\sim$100\%.

    \item \textbf{Consistency checks} (6 checks): Cross-validates that the same process has consistent yields across different plots, pre-fit and post-fit yields are consistent with the fit, and nuisance parameter impact rankings match the uncertainty breakdown table.

    \item \textbf{Red flag detection} (10 checks): Automatic Category~A triggers including negative event yields, efficiencies outside $[0,1]$, data/MC ratios outside $[0.2, 5.0]$ in control regions, zero uncertainty on non-zero predictions, non-converged fits, nuisance parameter pulls $>3\sigma$, $\chi^2/\text{ndf} > 5.0$, and systematic variations exceeding 100\%.
\end{itemize}
Red flags from the plot validator are automatically classified as Category~A findings; the arbiter is explicitly forbidden from downgrading them.

\paragraph{Arbiter (\texttt{arbiter}).}
The arbiter is not a reviewer but an adjudicator: it reads all reviewer outputs and the original artifact, then synthesizes a single verdict.
Its adjudication follows a five-case decision framework:
(1)~both reviewers agree $\to$ accept at the higher severity;
(2)~reviewers disagree on severity $\to$ evaluate arguments and assign with documented reasoning;
(3)~only one reviewer raised the finding $\to$ assess validity independently;
(4)~reviewers contradict each other $\to$ examine the artifact directly to resolve;
(5)~both reviewers missed something $\to$ the arbiter adds its own finding.
The arbiter produces a structured adjudication table mapping each finding to its source reviewer(s), their assigned categories, and the final adjudicated category with rationale.
It then issues one of three decisions: \textbf{PASS} (no Category~A or unresolved B items remain), \textbf{ITERATE} (actionable Category~A items exist that can be fixed within the current phase), or \textbf{ESCALATE} (fundamental problem requiring human judgment or upstream phase changes).

This automated review layer serves two purposes.
First, it catches a class of errors---such as applying a selection cut that inadvertently removes signal, double-counting a systematic uncertainty, or using an inappropriate test statistic---before a human physicist ever needs to look at the analysis.
Second, it provides structured documentation of the analysis decisions and their justifications, making subsequent human review more efficient because the reviewer can focus on physics judgment rather than hunting for mechanical errors.

The review process is deliberately conservative: the analysis is not permitted to proceed to ``unblinding''---examining the real data in the signal region---until all reviewer agents have signed off. The review protocol is not advisory; it is \emph{binding}.
When the arbiter issues an \texttt{ITERATE} verdict, the orchestrator initiates a closed-loop revision cycle in which the executor agent must demonstrably address every Category~A finding before the analysis can advance.

In the documentation steps and the final documentation phase, the review system extends to the written analysis report itself: the analysis notes referenced throughout this paper, including all publication-quality plots, were drafted entirely by the agent system and refined through the same multi-agent review cycle.

\subsection{Pipeline state tracking}

The orchestrator tracks pipeline state through the append-only experiment log (\texttt{experiment\_log.md}) and the artifact history on disk.
Every phase transition, review verdict, and human gate decision is recorded with timestamps.
The experiment log also captures failed approaches with their failure mechanisms, guiding future agents to avoid re-trying approaches that have already been demonstrated not to work.
This enables the orchestrator to resume from the correct point after interruptions and provides a complete audit trail of the analysis pipeline's progression.

\subsection{Formal unblinding protocol}

The framework implements a structured, multi-stage blinding protocol that enforces a mandatory human gate before observed data in the signal region can be examined.

\paragraph{Blinding enforcement.}
From the start of the analysis, a blinding protocol defined in Phase~1 (Strategy) specifies the blinding variable and the signal region boundaries.
All agents operate under a standing constraint: ``never access signal region data until explicitly told unblinding is approved.''
During Phases~1--4a, the statistical analysis uses exclusively Asimov datasets (expected background, optionally with injected signal at $\mu=1$ for discovery projections).
Post-fit distributions show data points removed from the signal region or replaced with Asimov pseudodata.

\paragraph{Partial unblinding (Phase~4b).}
Phase~4b introduces a controlled partial unblinding: the \texttt{executor} agent runs the statistical fit on a 10\% random subsample of the signal region data, using a fixed random seed for reproducibility.
This limited exposure allows the analysis to verify that no unexpected pathologies appear (e.g., dramatic data/MC disagreement, fit non-convergence, anomalous nuisance parameter pulls) without revealing the full result.
The \texttt{note-writer} agent simultaneously produces a draft analysis note and an unblinding checklist covering seven criteria:
\begin{enumerate}
    \item Background model validated (closure tests pass in all validation regions),
    \item Systematic uncertainties evaluated and fit model stable,
    \item Expected results physically sensible,
    \item Signal injection tests confirm fit recovers injected signals,
    \item 10\% partial unblinding shows no unexpected pathologies,
    \item All agent review cycles resolved (arbiter PASS),
    \item Draft analysis note reviewed and considered publication-ready modulo full observed results.
\end{enumerate}

\paragraph{Human gate.}
After the Phase~4b review passes, the pipeline \emph{halts automatically}.
The orchestrator does \emph{not} proceed to Phase~4c under any circumstances without explicit human authorization.
The orchestrator presents a structured approval request to the human analyst containing:
\begin{itemize}
    \item A summary of the draft analysis note (physics process, methodology, 10\% results),
    \item The unblinding checklist with per-item pass/fail status,
    \item The review result (arbiter decision, iteration count, residual Category~B/C items),
    \item Key results from the 10\% data (expected vs.\ observed, goodness-of-fit, notable nuisance parameter pulls),
    \item Paths to all artifacts available for detailed human review.
\end{itemize}

The human responds with one of three decisions:

\begin{description}
    \item[\texttt{APPROVE}] The pipeline proceeds to Phase~4c (full unblinding).
    \item[\texttt{REQUEST\_CHANGES}] The pipeline re-enters Phase~4b execution with the human's change request as additional input. After the executor revises the artifacts, the review cycle runs again, and upon passing, the human gate is re-presented with the updated summary.
    \item[\texttt{HALT}] The pipeline is stopped entirely. The human's reason is recorded in the experiment log. The analysis can only resume after the human addresses the stated concerns.
\end{description}

\paragraph{Full unblinding (Phase~4c).}
Phase~4c executes only after the human has approved unblinding at the Phase~4b gate.
The executor runs the full statistical fit on the complete dataset.
The \texttt{executor} agent then validates consistency between the full observed results, the 10\% partial results, and the Asimov expected results.
The full results undergo a 1-bot review (critical reviewer plus plot-validator) before the analysis advances to the final documentation phase (Phase~5).

\paragraph{Design rationale.}
The three-stage sequence (Asimov $\to$ 10\% data $\to$ full data) with a mandatory human gate mirrors the blinding protocols of major HEP collaborations, and reflects the principle that while analysis execution and review can be automated, the scientific judgment to unblind a result should not be.

\section{Results}
\label{sec:results}

We demonstrate the \slop framework by reproducing \dminus{nine}\dplus{five} LEP measurements from archived ALEPH\dminus{ and DELPHI} data and one Higgs boson measurement from CMS Open Data~\cite{CERNOpenData}. Each was produced autonomously from a short natural-language prompt, with no human input prior to the unblinding gate.

\subsection{Analyses and pointers}
Table~\ref{tab:analyses} summarizes the analyses produced with \slop. Each analysis was initiated with a short natural-language prompt specifying only the measurement goal and the data location; the agent autonomously determined the full analysis strategy. All analyses presented here used a Claude Max subscription\footnote{\url{https://claude.ai}} with the Claude Opus~4.8 model. The wall-clock times reported in Table~\ref{tab:analyses} include idle periods spent waiting for Claude Code usage-limit resets, which can account for a substantial fraction of the total; the active compute time is significantly shorter (see Figure~\ref{fig:timings}).

The primary scientific output of each run is a full analysis note (AN), drafted by the agent system and refined through the multi-agent review cycle described in Section~\ref{sec:framework}. The complete ANs, all code, and all generated figures live in the per-analysis GitHub repositories linked in Table~\ref{tab:analyses}, and the prompts are reproduced verbatim in Appendix~\ref{app:prompts}. Reproducing all \dminus{ten}\dplus{six} ANs here would multiply the paper's length several-fold, so Appendices~\ref{app:tautau_pub} and~\ref{app:lund_pub} present two representative analyses --- the CMS $H\to \tau^+\tau^-$ measurement and the ALEPH Lund jet plane analysis --- each recast by the agent system into journal-letter format. These condensations are themselves agent-produced: a separate agent pass over the same analysis artifacts rather than a hand-edited summary, so the reader can compare what the framework writes as an AN with what it writes when asked to compress that note into a letter. The framework specification, agent definitions, and additional autonomously generated analyses are available from our GitHub organization (see Section~\ref{sec:availability}).

These results should be read as evidence of what the framework can do, not as measurements we put forward in their own right. The analyses surveyed below were produced autonomously and demonstrate current capability; \textbf{we have not vetted each to the standard we would apply to our own published work}. The two we have condensed into letters (Appendices~\ref{app:tautau_pub} and~\ref{app:lund_pub}) we examined more closely. The CMS $H\to\tau^+\tau^-$ measurement targets a well-established result that can be checked directly: it is consistent with the published CMS $\mu\tau_h$ measurement and rebuilds point-by-point in an independent \textsc{Combine} implementation. The ALEPH primary Lund jet plane density, by contrast, is in principle a genuine new result: to our knowledge it has not previously been measured in $e^+e^-$ collisions, and, obtained autonomously from a one-line prompt, it is also to our knowledge the first new HEP measurement produced autonomously by an AI agent --- with human input entering only at the unblinding gate, rather than the step-by-step human-in-the-loop guidance of Ref.~\cite{badea2026agentic}. Precisely because it has no prior counterpart to check against, we present it as a candidate measurement for community scrutiny rather than as a claim of discovery. As with any AI-assisted analysis, responsibility for validation rests with human experts, and every result here should be independently reproduced before it is treated as definitive.

\begin{table}[H]
\centering
\small
\begin{tabular}{p{4.5cm} p{7.0cm} c c}
\hline
\textbf{Analysis} & \textbf{Initial prompt (abridged)} & \textbf{Data} & \textbf{Repo} \\
\hline
$H\to \tau^+\tau^-$ & Measure the Higgs boson signal strength in the $\mu\tau_h$ final state at $\sqrt{s}=8$~TeV & CMS & \href{https://github.com/jfc-mit/analysis_cms_higgs_tautau}{Github} \\
Z lineshape & Measure Z lineshape parameters ($M_Z$, $\Gamma_Z$, $\sigma^0_\text{had}$) and $N_\nu$ from the invisible width & ALEPH & \href{https://github.com/jfc-mit/analysis_aleph_z_lineshape}{Github}  \\
Lund plane & Measure the primary Lund jet plane density in hadronic Z decays & ALEPH & \href{https://github.com/jfc-mit/analysis_aleph_lund_plane}{Github} \\
$R_b$, $R_c$, and $A_\text{FB}^b$ & Measure $R_b$, $R_c$, and $A_\text{FB}^b$ using signed impact parameter b-tagging and jet charge & ALEPH & \href{https://github.com/jfc-mit/analysis_aleph_z_heavy_flavour}{Github} \\
Energy-energy correlators & Measure the two- and three-point energy-energy correlators (E2C, E3C), their ratio, and the EEC asymmetry (AEEC) in hadronic Z decays & ALEPH & \href{https://github.com/jfc-mit/analysis_aleph_eec_correlators}{Github} \\
Event shapes + $\alpha_s$ & Measure six event shape distributions and extract $\alpha_s(M_Z)$ from NNLO+NLL QCD fits & ALEPH & \href{https://github.com/jfc-mit/analysis_aleph_eventshapes_alphas}{Github} \\
\ifdelphiprompt
\delphirow $N_\nu$ from $\Gamma_\text{inv}$ & Measure the number of light neutrino generations from the Z invisible width & DELPHI$^\dagger$ & \href{https://github.com/jfc-mit/archive_v0_analysis_delphi_z}{Github}\\
\delphirow Lund jet plane & Measure the primary Lund jet plane density in hadronic Z decays & DELPHI$^\dagger$ & \href{https://github.com/jfc-mit/archive_v0_analysis_delphi_lund}{Github} \\
\delphirow Energy-energy correlators & Measure the two-point energy-energy correlator in hadronic Z decays & DELPHI$^\dagger$ & \href{https://github.com/jfc-mit/archive_v0_analysis_delphi_eec_two_point}{Github} \\
\delphirow Event shapes + $\alpha_s$ & Measure six event shape distributions and extract $\alpha_s(M_Z)$ from NLO QCD fits & DELPHI$^\dagger$ & \href{https://github.com/jfc-mit/archive_v0_analysis_delphi_had_eventshape_alpha_s}{Github} \\
\fi
\hline
\end{tabular}
\caption{Summary of analyses produced with \slop. Each analysis was run autonomously from the initial prompt shown, with no human input prior to the unblinding gate. The per-analysis wall-clock time breakdown is given in Figure~\ref{fig:timings}.\ifdelphiprompt{} $^\dagger$The DELPHI analyses were earlier runs produced with Claude Opus~4.6; they are listed here for completeness, but their numerical results are not presented in this paper.\fi}
\label{tab:analyses}
\end{table}

The \dminus{ten}\dplus{six} analyses span a broad slice of LEP- and LHC-era measurement topics: Z-pole electroweak parameters ($M_Z$, $\Gamma_Z$, $\sigma^0_\text{had}$, $N_\nu$) and heavy-flavour observables ($R_b$, $R_c$, $A_\text{FB}^b$) from ALEPH\dminus{ and DELPHI}; QCD-substructure measurements (Lund jet plane, two- and three-point energy-energy correlators and the EEC asymmetry) at the $Z$ pole from \dminus{both LEP experiments}\dplus{ALEPH}; \dminus{extractions}\dplus{an extraction} of $\alpha_s(M_Z)$ from ALEPH\dminus{ and DELPHI} event shapes; and a Higgs-boson signal-strength measurement in $H\to \tau^+\tau^-$ from CMS Open Data.\ifdelphiprompt{} Table~\ref{tab:analyses} additionally lists four earlier DELPHI analyses (run on Claude Opus~4.6) whose numerical results we do not present here.\fi{} They cover the full range of standard analysis ingredients --- event classification, lifetime-based b/c-tagging, jet reconstruction and declustering, iterative Bayesian unfolding, profile-likelihood fits with nuisance parameters --- so the framework is exercised across rather than along a single methodological axis.

As a worked example, the prompt issued for the ALEPH Lund jet plane analysis is reproduced verbatim below; the remaining \dminus{nine}\dplus{five} follow the same shape.

\begin{quote}
\footnotesize
\color{black}
Measure the primary Lund jet plane density in hadronic Z decays using archived ALEPH data at $\sqrt{s} = 91.2$~GeV.

Decluster each thrust hemisphere with the Cambridge/Aachen algorithm and map primary splittings to Lund coordinates $(\ln 1/\Delta_\theta,\, \ln k_t/\mathrm{GeV})$. Correct to charged-particle level.

This is the first Lund plane measurement in $e^+e^-$ collisions.

Data: [...]
\end{quote}

\subsection{Quantitative and qualitative assessment}

\begin{figure}[ht!]
  \centering
  \ifdelphi\else\includegraphics[width=0.85\linewidth]{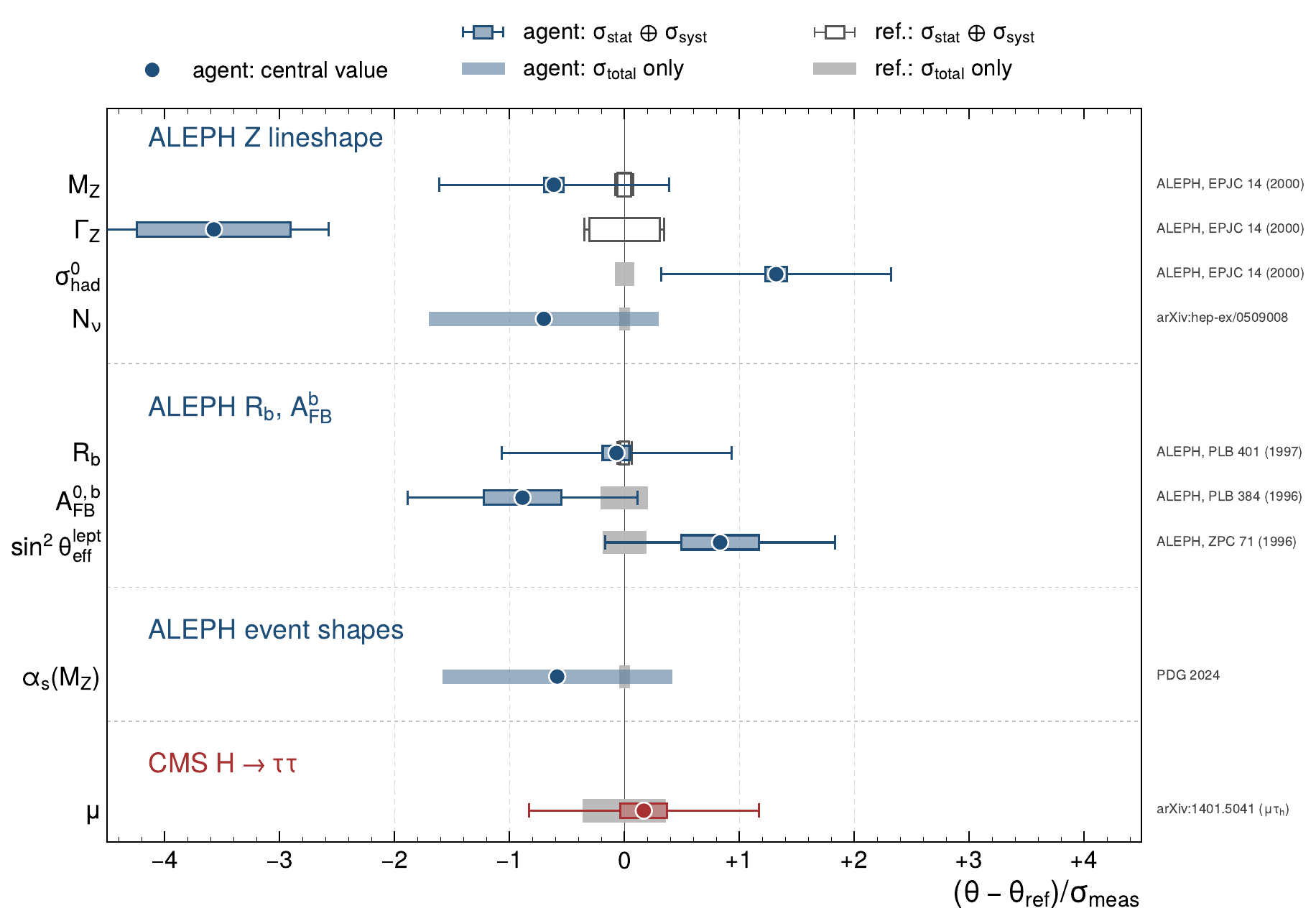}\fi
  \caption{Agent-produced scalar measurements versus their published reference values, shown as the deviation in units of the agent's own total uncertainty, $(\theta - \theta_\mathrm{ref})/\sigma_\mathrm{meas}$, with $\pm1\sigma$ and $\pm2\sigma$ guides (dashed). For the lineshape and heavy-flavour observables ($M_Z$, $\Gamma_Z$, $\sigma^0_\mathrm{had}$, $R_b$, $A_\mathrm{FB}^{b}$, $\sin^2\theta_\mathrm{eff}$) we compare against the measuring experiment's own published value rather than the LEP+SLD combination; the derived quantities $N_\nu$ and $\alpha_s(M_Z)$ are compared against the LEP electroweak combination and the PDG world average, respectively. \dminus{Greyed rows are superseded runs pending a re-run. }The plotted values, their published references, and the reference sources are listed in Table~\ref{tab:results}; pure shape/density measurements (Lund plane, EEC) have no scalar reference and are omitted.}
  \label{fig:summary}
\end{figure}

Figure~\ref{fig:summary} compares the agent-produced scalar observables against their published reference values; the numerical values and per-row pulls are tabulated in Table~\ref{tab:results} (Appendix~\ref{app:results}). Of the \dminus{thirteen}\dplus{nine} scalar comparisons, \dminus{nine}\dplus{eight} sit within $|\mathrm{pull}|<2$ of the reference. The ALEPH $\Gamma_Z$ row sits $-3.4\sigma$ low: the lineshape note traces this to a residual energy-dependent selection (``aftercut'') efficiency at the off-peak scan points, which---because $\Gamma_Z$ is fixed largely by the ratio of off-peak to peak cross-sections---is imperfectly corrected and pulls the fitted width down.\dminus{ The remaining outliers (the DELPHI $\Gamma_Z$ at $-3.4\sigma$, the DELPHI $N_\nu$ at $+2.8\sigma$, and the NLO-only $\alpha_s$ extraction from DELPHI event shapes at $+2.5\sigma$) are likewise flagged by the agent in the corresponding analysis notes.} Because the pull weights agent and reference uncertainties symmetrically, a small pull can reflect a wide agent uncertainty rather than genuine agreement: the CMS $H\to \tau^+\tau^-$ signal strength (agent uncertainty nearly three times the published one) and the theory-dominated ALEPH $\alpha_s$ extraction ($\sim$17\% uncertainty absorbing a real $-9.8\%$ offset) are both of this kind and should not be over-read as precision results.

\begin{figure}[t]
  \centering
  \includegraphics[height=7.5cm]{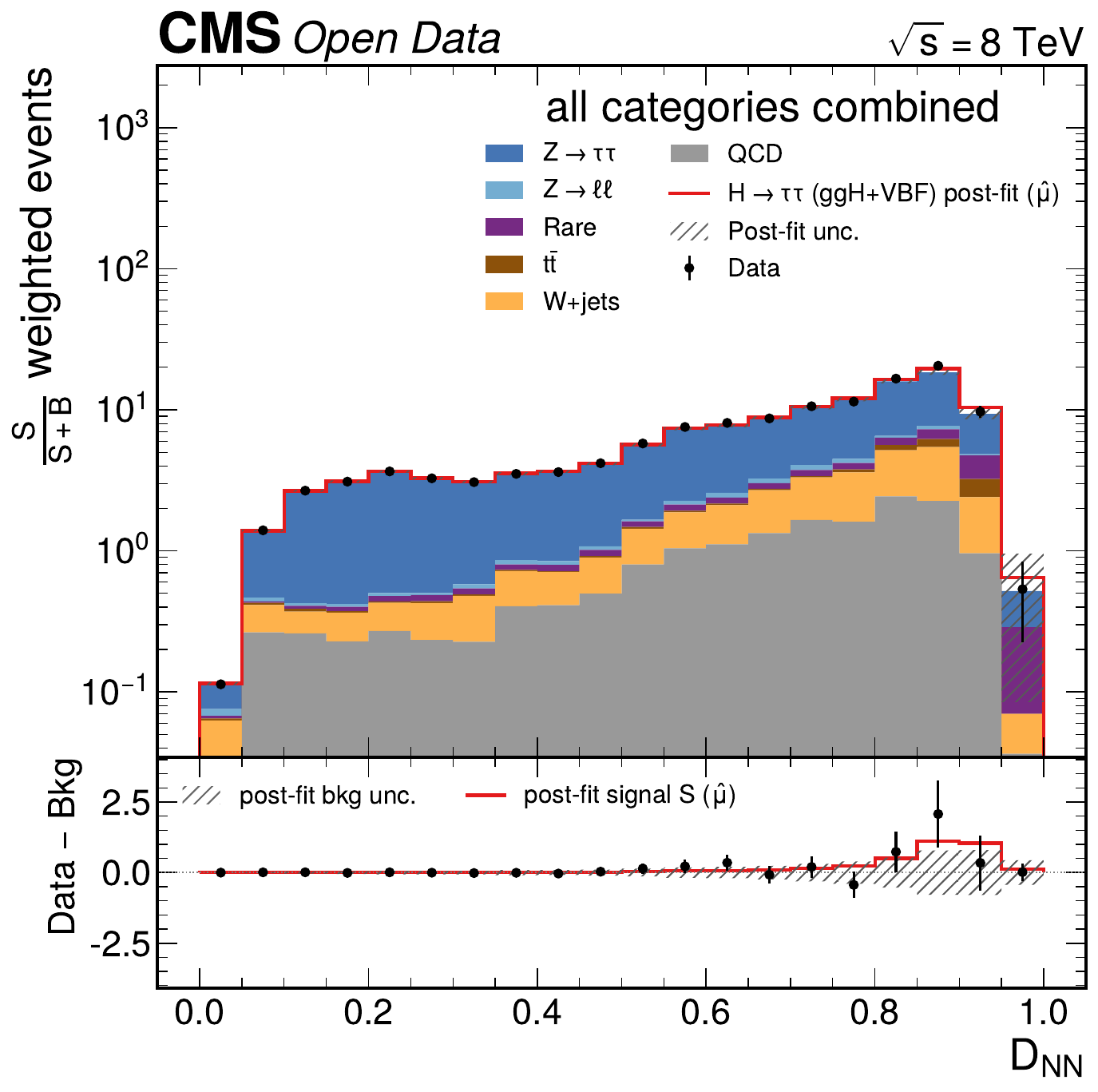}%
  \hspace{0.04\linewidth}%
  \includegraphics[height=7.5cm]{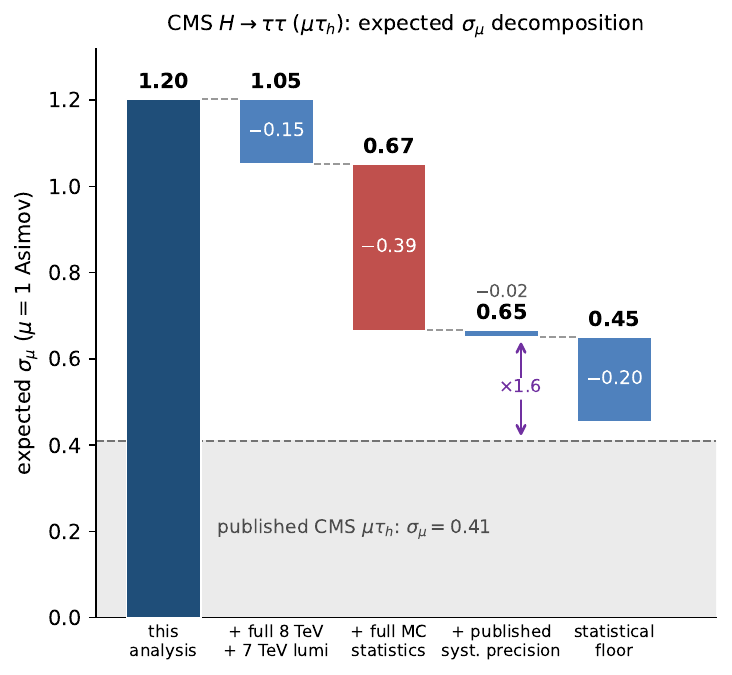}
  \caption{Headline CMS $H\to \tau^+\tau^-$ result, reproduced from
  Appendix~\ref{app:tautau_pub}. The $S/(S+B)$-weighted post-fit $D_{NN}$
  distribution of the CMS $H\to \tau^+\tau^-$ measurement (left), combining the 0-jet,
  boosted, and VBF categories, with the post-fit signal at $\hat{\mu}=1.20$
  (red) overlaid on the background stack and the weighted residual in the lower
  panel. Decomposition of the expected signal-strength uncertainty
  $\sigma_\mu$ (right), obtained by sequential Asimov re-fits of the analysis workspace:
  starting from this analysis ($\sigma_\mu=1.20$ on the partial
  $11.5~\mathrm{fb}^{-1}$ 8~TeV open-data sample), successively bringing the
  luminosity (full Run-1, 8~TeV${+}$7~TeV), simulation statistics, and
  systematic precision to CMS scale lowers the expected uncertainty to
  $\sigma_\mu\approx0.65$ --- within a factor $\times1.6$ of the published CMS
  $\mu\tau_h$ value of $0.41$, with the residual reflecting analysis method and
  design rather than inputs; removing systematics entirely leaves a statistical
  floor of $0.45$.}
  \label{fig:headline_cms}
\end{figure}

The two analyses condensed into journal-letter form (Appendices~\ref{app:tautau_pub} and~\ref{app:lund_pub}) bracket the range of what the framework produces: a reasonable reproduction of a well-established LHC measurement and a candidate first measurement of a previously unpublished $e^+e^-$ observable. Figure~\ref{fig:headline_cms} shows the headline CMS $H\to \tau^+\tau^-$ result.

The $H\to \tau^+\tau^-$ signal strength, $\hat{\mu}=1.20\pm1.13$, lands within $0.16\sigma$ of the published CMS $\mu\tau_h$-channel value ($\mu=1.01\pm0.41$) and its full statistical model reproduces point-by-point in an independent CMS \textsc{Combine} rebuild, but its uncertainty is nearly three times the published one --- set by the single $\mu\tau_h$ channel and the partial 2012 open-data sample, as the sensitivity decomposition in Figure~\ref{fig:headline_cms} (right) makes explicit --- so it stands as a correctness and reproducibility demonstration rather than a competitive measurement.

\begin{figure}[t]
  \centering
  \includegraphics[width=0.6\linewidth]{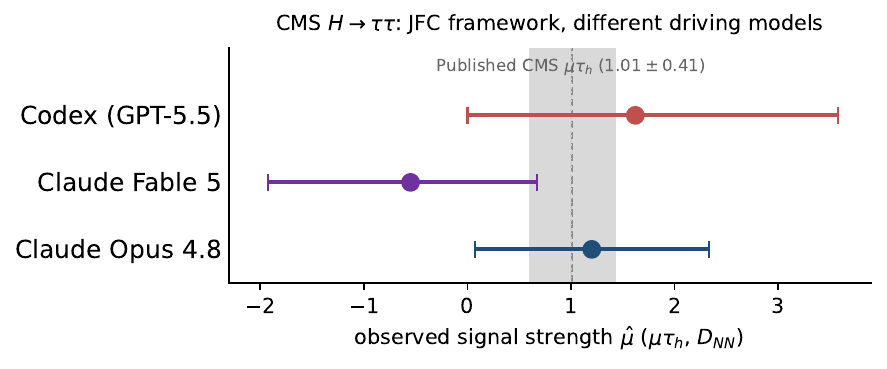}
  \caption{Indicative cross-model comparison for the CMS $H\to \tau^+\tau^-$ ($\mu\tau_h$) measurement: the observed signal strength $\hat{\mu}$ on the $D_{NN}$ discriminant from our JFC framework run with three different driving models --- Claude Opus~4.8 (used for the analyses throughout this paper), a Codex-based agent (GPT-5.5), and Claude Fable~5 --- against the published CMS value (grey band, $1.01\pm0.41$; dashed line). Best-fit values are shown as observed (the Fable~5 fit prefers a negative $\hat{\mu}$). The three runs share the identical prompt, framework, single $\mu\tau_h$ channel, and partial 2012 CMS Open Data sample, and differ only in the underlying model. This is a single run per model with no repeat trials --- an illustrative hint at model-to-model variation, not a statistically controlled benchmark.}
  \label{fig:model_comparison}
\end{figure}

As a preliminary probe of how much the underlying model matters, we re-ran the identical CMS $H\to \tau^+\tau^-$ prompt and framework with the driving model swapped from Claude Opus~4.8 (used for every analysis above) to a Codex-based agent (GPT-5.5) and to Claude Fable~5, recovering compatible but distinct results. Figure~\ref{fig:model_comparison} compares the observed $\mu\tau_h$ signal strength from the three JFC runs against the published CMS value: the central values span a range --- Claude Opus~4.8 and the Codex agent land near or somewhat above the Standard Model expectation, while the Fable~5 run prefers a slight deficit (a negative best fit, still consistent with $\mu=0$ and within roughly one standard deviation of the Standard Model) --- all with large, single-channel uncertainties. We emphasize that this is a single run per model, with no repeat trials or ensemble statistics; token-budget constraints preclude the repeated testing a controlled benchmark would require, so we present it only as an indicative hint at model-to-model variation rather than a quantitative comparison.

\begin{figure}[t]
  \centering
  \includegraphics[height=7.5cm]{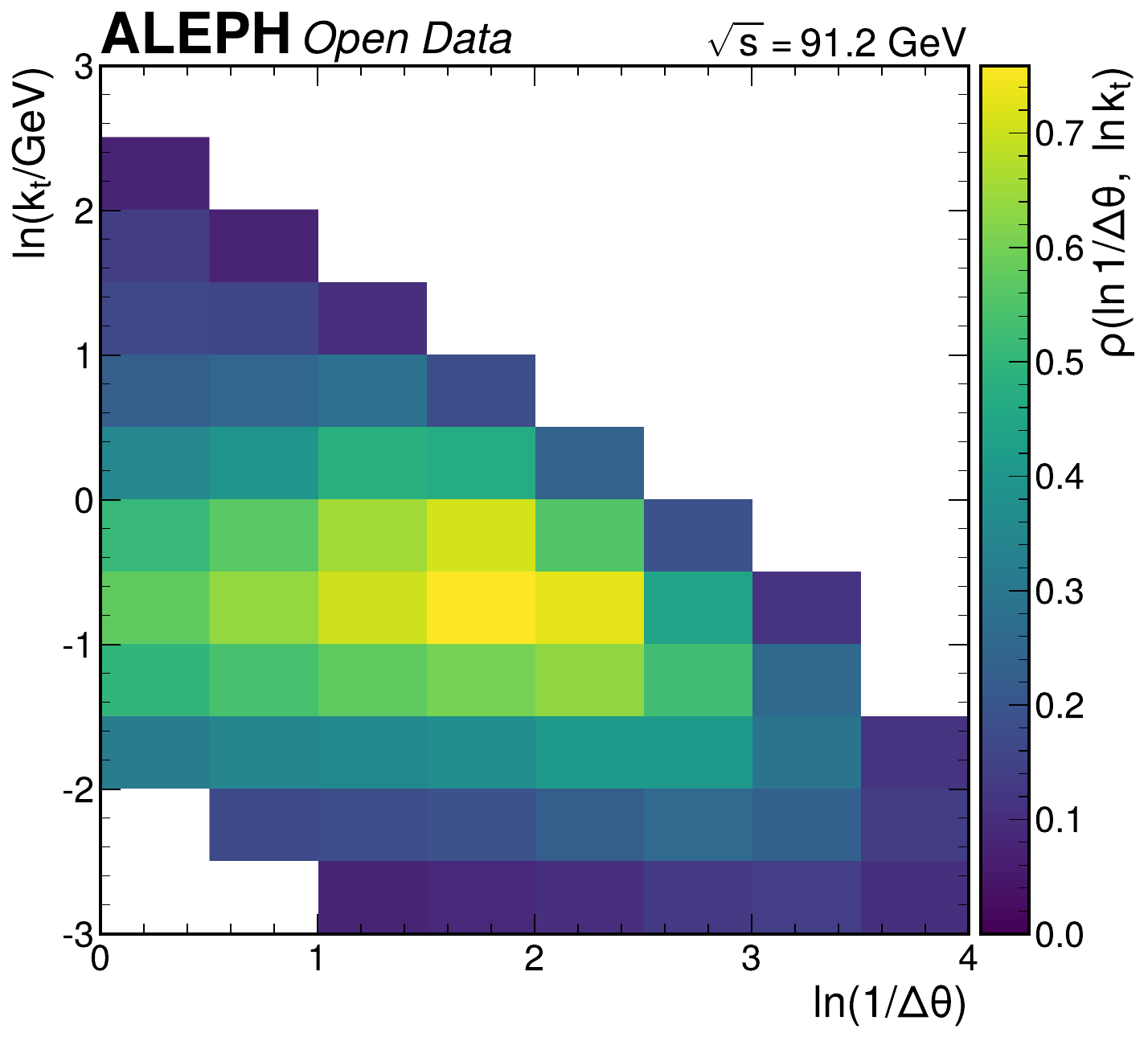}
  \caption{Headline ALEPH Lund jet plane result, reproduced from
  Appendix~\ref{app:lund_pub}. Unfolded primary Lund jet plane density
  $\rho(\ln 1/\Delta\theta,\,\ln k_t)$ at charged-particle level, measured on
  archived ALEPH LEP1 data --- to our knowledge the first measurement of this
  density in $e^+e^-$ collisions.}
  \label{fig:headline_lund}
\end{figure}

Figure~\ref{fig:headline_lund} shows the ALEPH primary Lund jet plane density, a candidate first measurement of this observable in $e^+e^-$ collisions: the unfolded charged-particle-level density measured on archived ALEPH LEP1 data.

For the Lund jet plane density the agent reports a mean primary-emission multiplicity $\langle N\rangle = 4.751\pm0.224$ (systematics-limited). The unfolded density shows the qualitative structure expected of a primary Lund jet plane, lies somewhat above the modern parton-shower tunes it is compared against, and is reported to be stable under the unfolding and correction variations the agent tested. As an observable with no prior $e^+e^-$ counterpart, none of these features can be cross-checked against an existing measurement, so the result should be treated as a candidate to be independently verified rather than as an established one.
Beyond the numerical comparisons, the agent-produced analyses demonstrate that an LLM-driven system can design and execute particle physics analyses that are methodologically standard and honestly documented.
The event selections are sensible, and correct systematic uncertainty sources are identified, even when not all can be fully evaluated.
The analysis strategies align closely with published approaches, with deviations that are well-motivated and clearly documented.

In all cases, the agent's analysis strategy is recognizably similar to a corresponding published analysis. The agent consistently makes the same high-level choices that the original experimental collaborations made --- for example, iterative Bayesian unfolding (or bin-by-bin correction where appropriate) for the QCD measurements, HistFactory likelihood for the Higgs search, and Breit--Wigner lineshape fitting for $N_\nu$. The selection of observables, binning strategies, and correction procedures are all standard and well-motivated. The treatment of systematic uncertainties is one of the strongest aspects: every note includes a completeness table that explicitly compares the sources considered against those in the corresponding published analyses (\dminus{DELPHI, }ALEPH, CMS, ATLAS, or ALICE). This similarity is expected and in fact desired --- the agents consult the literature and are asked to repeat measurements that have been done before --- but autonomously reproducing full analysis workflows is itself a significant demonstration of capability.

The analyses presented here all rely on open-data releases, which carry intrinsic constraints relative to the data the original experiments worked with. Despite the best efforts of large HEP experiments, releasing and publishing sufficient information to fully replicate a physics analysis remains difficult: calibrations may require niche dedicated runs (luminosity at LEP) or sub-analyses (jet-energy corrections at the LHC); the sets of required inputs are extensive and often not well tracked or documented, so some elements are absent from open-data releases (in part because open data is often intended for educational rather than analysis-replication use); and the released sample is typically only a subset of the total, inflating statistical uncertainties on the replicated measurements. These constraints are inherited by every result above and limit what the agent-produced analyses can claim independently of the framework's behavior.

Within these open-data constraints, the remaining weaknesses of the analyses largely reflect the prototype scope rather than errors in the agent's physics reasoning. The most significant genuine errors (DY template contamination, rank-deficient covariance, NLO-only $\alpha_s$ bias) are the kinds of issues that standard review processes are designed to catch, and the agent itself flags most of them in its ``limitations'' discussions. The agent tends to be conservative with selection cuts, preferring high purity over high efficiency.
This is a reasonable default for a first-pass analysis, but in several cases, it substantially hurts sensitivity. A more experienced analyst would likely investigate and relax or work around such cuts earlier in the analysis cycle.

Where the agent falls short relative to an experienced human analyst is in the iterative refinement loop.
The agent produces thorough first drafts that identify their own problems but does not always close the loop on fixing them. Some limitations that are correctly identified would effectively require a sub-analysis of a similar extent, which is not something that the current framework can compel the agents to do. 
This pattern of correct diagnosis but deferred treatment is the single most characteristic feature of the agent's working style across all analyses, and it suggests that the review-and-iterate cycle is where the most value would come from further improvements in the framework.

Otherwise, the review rounds worked as intended at catching mistakes: each reviewer agent independently read its analysis note and produced a critical assessment.

\ifreviewfindings
For the two analyses we examined most closely, the most instructive reviewer findings --- and the points where human inspection would probe further --- are collected below.

\paragraph{CMS $H\to\tau^+\tau^-$.}
\begin{itemize}
    \item The first unblinded fit returned a low signal strength, $\hat\mu = 0.34\pm1.20$, because the dominant $t\bar t$ background carried an arbitrary $\pm35\%$ normalization prior; replacing it with a direct in-situ measurement of $t\bar t$ in a dedicated b-tag control region (which the agent performed, but did not wire into the fit at first) moved the result onto the Standard Model, $\hat\mu = 1.20\pm1.13$. The agent identified and corrected this itself, and the sensitivity of a headline value to a background prior is exactly what a human reviewer should check.
    \item The primary discriminant fit passes its goodness-of-fit only narrowly ($p=0.065$); separately, an earlier in-house goodness-of-fit test was mis-implemented and made every fit look near-perfect ($p\approx0.99$) until the agent replaced it with the standard saturated-model test --- a reminder to scrutinize not only the fit but the tools used to judge it.
\end{itemize}

\paragraph{ALEPH Lund jet plane.}
\begin{itemize}
    \item The automated review caught the first analysis-note draft over-stating its result --- claiming the data excluded the parton-shower generators ``at high significance'' --- traced the apparent significance to an over-simplified treatment of the correlated systematic uncertainties, and forced a revision; the final comparison adopts a more conservative metric and makes no such exclusion claim.
    \item The largest single systematic --- a $\sim3\%$ modelling uncertainty in the unfolding --- cannot be cross-checked in the usual way, by repeating the correction with a different simulated generator, because only one ALEPH sample carries a full detector simulation; it is instead estimated indirectly, a limitation the note flags plainly.
\end{itemize}
\fi

\subsection{Throughput, cost, and pipeline telemetry}
\label{sec:cost}

The Claude Max subscription used throughout costs approximately \$200/month and provides sufficient throughput for the full pipeline, including all review iterations. Each end-to-end analysis completes in roughly 10 hours of wall-clock time, with phases typically passing review after one to two iterations under soft and hard caps that prevent runaway loops.

\begin{figure}[ht!]
  \centering
  \includegraphics[width=0.95\linewidth]{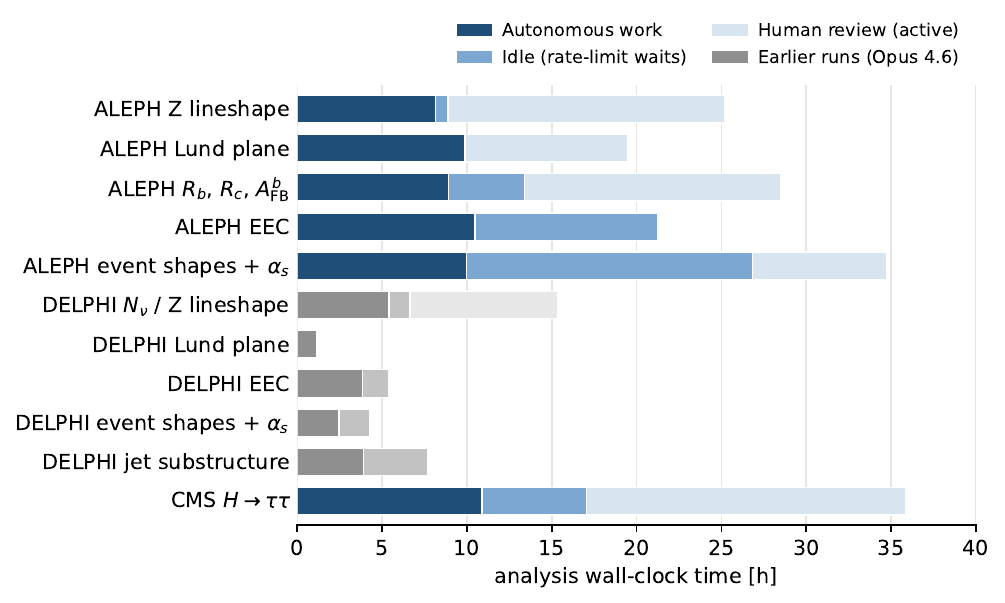}
  \caption{Per-analysis decomposition of the wall-clock times reported in Table~\ref{tab:analyses}, measured from \texttt{git log} of the per-analysis repositories. Dark segments are autonomous agent work before the unblinding gate (real compute); mid segments are the idle gaps (rate-limit waits) within that pre-gate phase; pale segments are the post-gate human-review/iteration time with idle gaps likewise removed, measured up to the point at which paper writing begins (the first commit touching the letter/paper sources): the analysis note is the deliverable, so once the letter is being drafted the analysis is effectively complete and any later edits are errata-grade rather than review.\dminus{ Greyed rows are results from superseded runs still pending a re-run to the current pipeline.}\dplus{ The greyed rows are earlier DELPHI analyses produced with Claude Opus~4.6; they are not among the results presented in this paper and are shown only to illustrate the wall-clock profile of those earlier runs.}}
  \label{fig:timings}
\end{figure}

Substantial cost savings could in principle be achieved by delegating narrowly scoped roles to lighter-weight models---for example, Claude Sonnet for certain reviews, and Claude Haiku for mechanical checks and one-off figure inspection.
Conversely, additional review iterations (beyond the current default) represent an opportunity to improve analysis quality at the cost of additional model invocations and wall-clock time --- a quality--cost trade-off whose systematic exploration we return to in Section~\ref{sec:discussion}.

Figure~\ref{fig:timings} separates each run's wall-clock time into autonomous pre-gate compute, idle usage-limit waits, and post-gate human-review iteration, with idle gaps removed from the active segments; the human-review segment is measured only up to the first commit that touches the letter/paper sources, since the analysis note is the deliverable and the subsequent letter writing is downstream of the analysis itself.
The same per-analysis git logs offer a coarse qualitative read of where effort goes: review iterations concentrate at the strategy (Phase~1) and inference-setup (Phase~4a) steps rather than the mechanical execution phases, and note-writing and rendering review consume a substantial share of the total iteration budget once the physics work is complete.

\dminus{The clearest single piece of case-study evidence that the unblinding gates do useful work is the DELPHI Z lineshape Phase~4b: the agent's 10\% partial-data fit was deferred outright after the agent diagnosed that the cross-section computation was algebraically circular --- the per-file luminosity was being derived from the same event yields that fed the cross section, so the resulting fit would have contained zero information from the data --- and the analysis advanced to Phase~4c only with a corrected procedure. Without the partial-unblinding gate, this bug would have propagated into the full-data inference and produced a fit that looked statistically clean but was tautological by construction.}

\subsection{Limitations and risks}
\label{sec:limitations}

We do not wish to overstate what current systems can do.
Several important limitations must be acknowledged.

\paragraph{Physics}
\begin{itemize}
    \item \textbf{Subtle physics errors}: Agents can produce analyses that are superficially correct but contain subtle physics errors---for example, applying a selection criterion that biases the measurement in a non-obvious way, or omitting a systematic uncertainty that an experienced analyst would know to include.
    \item \textbf{Complexity ceiling}: The analyses we have demonstrated---Z lineshape, event shapes, $\alpha_s$---are relatively standard measurements. More complex analyses involving advanced multivariate techniques, data-driven background estimation in multiple control regions, or simultaneous fits across many channels remain to be demonstrated. That said, the primary Lund jet plane density has not previously been measured in $e^+e^-$ collisions, and these substructure measurements (Lund plane, EEC) required full unfolding---suggesting that the complexity ceiling for current state-of-the-art models is already quite high.
    \item \textbf{Novelty}: Current agents excel at reproducing established analysis strategies retrieved from the literature but are not necessarily reliable when asked to develop genuinely novel approaches. For analyses that require creative new techniques, human-driven design remains essential.
    \item \textbf{Missing niche methods}: Agents are unlikely to employ analysis techniques that are not widespread in the literature, even when they would be appropriate. This represents an opportunity for domain-specific fine-tuning.
    \item \textbf{Limited knowledge of review practices}: Internal collaboration review processes are not publicly available and are unlikely to appear in training data. Agents can reason about physics reasonably well, but many analyses have characteristic failure modes that experienced reviewers would catch immediately. Building internal retrieval-augmented generation systems over review documentation could help address this.
    \item \textbf{Verification burden}: The need for thorough human verification of agent-produced results cannot be shortcut. An analysis that looks correct but contains a subtle error is arguably more dangerous than no analysis at all, because it carries an unearned sense of confidence. The community must develop robust practices for validating AI-produced analyses.
\end{itemize}
\paragraph{Execution}
\begin{itemize}
    \item \textbf{Instruction compliance}: Agents are prone to ignoring instructions when they become buried in a large context window. For example, an agent may repeatedly use absolute font sizes in plots despite explicit instructions to the contrary.
    \item \textbf{Prompt nondeterminism}: Agent outputs are inherently stochastic, making prompt optimization difficult. The same prompt can yield markedly different analysis strategies or code quality across runs.
    \item \textbf{Stale tool knowledge}: Agents tend to rely on APIs and interfaces from their training data, and convincing them to consult up-to-date documentation for newer package versions is surprisingly difficult.
    \item \textbf{Poor dependency management}: Agents are inclined to write ad hoc patches or workarounds rather than pulling an updated dependency or adding a lightweight new one, even when the latter would be the correct solution.
    \item \textbf{Context and prompt engineering trade-offs}: There is an inherent tension between providing detailed instructions (which improve specificity but risk context bloat and instruction-following degradation) and keeping prompts concise (which preserves context but may leave important conventions unspecified).
\end{itemize}
\paragraph{Style}
\begin{itemize}
    \item \textbf{Figures}: Much of the scientific (review) process is driven by reading and understanding figures to analyze correctness and provide feedback. This leads to an informally established ``style'' for each subfield that makes this process easy for humans. However, since conventions likely differ widely across the model training data, enforcing extremely strict rules in the specification becomes necessary. Moreover, current agents seem to struggle with visual inspection, particularly at scale, and leave a number of figures that are not only poorly styled but in fact fully nonsensical (see the per-analysis repositories linked in Table~\ref{tab:analyses}). 
    \item \textbf{Note formatting}: While LaTeX has reasonable defaults, there are many options and packages to choose from, and there is no uniform style across fields and journals. It is necessary to be extremely specific when describing and encoding what the final analysis note should look like. 
\end{itemize}

\section{Discussion}
\label{sec:discussion}
\subsection{Implications for analysis workflows}

The results presented here suggest that a significant fraction of the technical work in a standard HEP analysis can be automated with current AI agent technology.
This does not mean that human physicists are no longer needed --- quite the contrary.
The role of the physicist shifts from implementer to architect and critic: defining what should be measured and why, assessing whether the agent's approach is physically sensible, and catching the subtle errors that automated review may miss.

From this perspective, AI coding agents join a long lineage of time-saving innovations that unburden scientists from technical drudgery, shifting the primary occupation --- at every career stage --- toward analysis design, physical reasoning, and judgment (the skills sought in faculty candidates and foundational to career success).

This shift has the potential to dramatically increase the throughput of an experimental program.
Consider the situation facing the LHC experiments today: thousands of potential measurements that could be performed with existing data, but only a finite number of students and postdocs to carry them out.
Each analysis can occupy a person for years, so if the implementation phase can be compressed from years to months, the bottleneck moves from coding capacity to physics ideas and human review bandwidth---a much more desirable constraint, because it means the limiting factor is intellectual rather than mechanical.
Measurements that might take a year to complete could potentially be produced in days, reviewed by a suite of specialized AI agents before any human even looks at the result, and iterated rapidly in response to feedback.
Follow-up studies, closure checks, systematics investigations, and many other burdensome tasks could be implemented within minutes at the level of code and bottlenecked only by runtime on large datasets, as they are now anyway.
Studies that are currently skipped because they are ``too tedious'' become tractable when the cost of implementation drops by orders of magnitude.
Even with LEP data, there are significant prospective improvements, such as event-level analyses of all Standard Model parameters, tunes of the hadronic shower for every analysis, and optimized particle reconstruction for core Standard Model parameters. With these toolkits, the burden of analysis is lifted, encouraging all of us to think more broadly about the scope of research that can be conducted.

Furthermore, large collaboration reviews, which involve 3--4 levels of human review (see Figure~\ref{fig:vision}), can be expedited. Agents have a role at both the review and testing stages, providing immediate feedback and immediate results from cross-checks. The role of the human reviewer then focuses on the technically challenging elements of the review process, largely paralleling the ``approvals'' and ``collaboration-wide reviews'' performed in the later stages.


While a reduced coding burden would come as a relief to most in the field, \textbf{technical proficiency in programming must remain a critical and carefully developed skill for all HEP practitioners}. If an AI agent writes a sophisticated analysis pipeline, a physicist must, at minimum, be able to read and understand the codebase to check for correctness. Modern coding assistants are incredibly capable but still routinely make fundamental errors, particularly when performing the kinds of specialized tasks required in HEP (e.g.\ complicated fits and statistical hypothesis testing). The reliability of these systems will continue to improve, but the conceptual basis of all software---what the code should \textit{do}---will remain human-driven.

\subsection{Legacy data and reanalysis}

One particularly compelling application is the reanalysis of legacy datasets.
Enormous volumes of data from past experiments, including LEP, the Tevatron, and the B-factories, sit in archives, potentially containing information relevant to current physics questions\footnote{\url{https://ee-alliance.org/} is one such organization pursuing these questions}. Previous analysis codes are built on legacy software frameworks using compilers, such as FORTRAN, and toolkits, such as PAW, that have fallen out of the scope of expertise of practicing high energy physicists. 
Autonomous analysis agents could systematically work through these datasets, producing first-pass results, with modern software, that human physicists can then evaluate and refine.
Our work with ALEPH and DELPHI data is a proof of concept for exactly this kind of application.
Moreover, legacy datasets become accessible to reanalysis with modern computational tools---including deep learning techniques that were unavailable when the data were originally collected.

Beyond new analyses, this paradigm also opens the door to automatic reproducibility of published results, where AI agents could systematically re-execute the computational pipelines of existing papers, flagging inconsistencies or confirming findings at a scale that is currently impractical due to the resource and personnel constraints needed for large-scale verification\footnote{Coding assistants will also prove helpful for understanding legacy codebases and data formats.}.

\subsection{Toward robust AI-assisted analysis}

The path forward is not to deploy AI agents as unsupervised analysis machines, but to develop them as tools within a framework that preserves the rigor and scrutiny that the HEP community rightly demands.
This requires investment in several concrete areas:

\begin{enumerate}
    \item \textbf{Benchmarks for physics analysis.} The community is only beginning to develop standardized benchmarks for evaluating agent performance on realistic, end-to-end physics analysis tasks. Existing benchmarks such as CelloAI~\cite{Atif:2026eju} focus on code generation and documentation, and Collider-Bench~\cite{colliderbench} evaluates agents on reproducing LHC searches as simulation pipelines; what is still needed are benchmarks that test the full analysis chain---from event selection design through statistical inference---on real datasets with known ground truth. The LHC Olympics~\cite{Kasieczka:2021xcg} anomaly detection challenge offers a partial template, but benchmarks targeting standard measurements (cross-sections, coupling extractions, event shape variables) with realistic systematic uncertainty structures would be far more informative.

    \item \textbf{Standardized review protocols for AI-produced analyses.} Collaborations should develop explicit protocols for reviewing analyses in which substantial portions of the code and strategy were produced by AI agents. These protocols should specify what additional scrutiny is warranted---for example, mandatory independent re-implementation of key results, systematic comparison of agent-chosen selections against published baselines, or automated regression testing against known-good analysis outputs. The multi-agent review system we describe in this paper is a step in this direction, but it complements rather than replaces human review.

    \item \textbf{Training for AI-assisted workflows.} Physicists need practical training in how to effectively supervise and validate AI-assisted work. This includes developing the ability to read and critically evaluate code they did not write, understanding the failure modes of LLM-based systems (hallucination, instruction drift, stale API knowledge), and learning to formulate analysis specifications that are precise enough to guide an agent while remaining flexible enough to allow it to exercise judgment. Graduate curricula should evolve to treat AI literacy as a core competency alongside statistical methods, detector physics, and programming.

    \item \textbf{Institutional adaptation.} Collaborations will need to update their authorship, review, and accountability norms to accommodate AI-assisted work. Clear standards for disclosure of AI involvement, attribution of responsibility for AI-produced results, and documentation requirements for AI-assisted analyses will be essential for maintaining the trust and rigor that underpin the field's scientific credibility.
\end{enumerate}

The HEP community has a long tradition of rigorous internal review before publication.
That tradition should be extended, not abandoned, as AI tools become more capable.

\subsection{Future work on the framework}
Several directions for systematic improvement of the framework remain to be explored. A natural next step is A/B testing of different specification variants, comparing the effect of more prescriptive methodology documents against minimal-guidance prompts on analysis quality, completeness, and error rates. Equally important is characterizing the stochastic variability of agent outputs: running the same prompt and specification multiple times and quantifying the spread in analysis strategies, numerical results, and documentation quality would establish a baseline for reproducibility and identify which aspects of the pipeline are most sensitive to prompt nondeterminism. 

The review system itself offers a rich optimization target; the current iteration caps and reviewer panel composition were chosen heuristically, and a systematic sweep over the maximum number of review rounds, reviewer agent combinations, and arbiter thresholds could reveal favorable operating points on the quality--cost frontier. 

Beyond Claude Code, the framework's architecture is in principle backend-agnostic, and benchmarking against other agentic coding systems such as OpenAI's Codex, Google's Jules, or open-weight agent frameworks would clarify which capabilities are specific to the underlying model and which emerge from the specification and orchestration layer. 

Finally, extending the framework to more complex analysis topologies, including multi-channel simultaneous fits, data-driven background estimation with transfer factors, and analyses requiring custom reconstruction or machine learning components, would map the current boundary between what agents can handle autonomously and what still demands sustained human involvement.

\subsection{Rethinking graduate training}

The implications for graduate education deserve particular attention. As computer technology and deep learning have co-evolved with HEP's ``big data'' era, software engineering has, perhaps unintentionally, become a significant component of a graduate student's training. This is a valuable technical foundation and prepares students well for futures inside and outside academia\footnote{The value of software engineering in industry may drop in the coming years, depending on the continued evolution of AI coding agents. At the time of writing, this matter is hotly debated.}, but inevitably distracts from core training as a \textit{physicist}.
Writing code has become synonymous with HEP graduate student work, and technical implementation is the largest bottleneck to the realization of an idea.

If AI agents can handle the implementation, graduate training can be restructured to emphasize the aspects of research that require human intelligence: developing physical intuition, learning to ask good questions, understanding the theoretical context of measurements, and practicing the kind of critical thinking needed to evaluate whether an analysis result is trustworthy.
Students would still need to understand their analysis code and be able to validate anything an agent contributes, but they would not need to write every line of it from scratch. This focus on high-level design and evaluation rather than implementation mirrors how faculty and postdocs work, contributing primarily through scientific judgment rather than writing code. AI agents would simply extend this training to earlier career stages.


\section{Conclusion}
\label{sec:conclusion}

AI agents based on large language models, combined with a relatively small set of structured prompts, can already autonomously execute substantial portions of a standard HEP analysis pipeline.
We have demonstrated this concretely by reproducing published ALEPH\dminus{ and DELPHI} measurements using archived LEP data and CMS measurements using CMS Open Data, with agents that plan their own analysis strategy, retrieve domain knowledge from the literature, execute the full analysis chain, undergo automated multi-agent review, and produce complete written reports. Beyond reproducing known results, one of these analyses --- the primary Lund jet plane density in hadronic $Z$ decays --- is to our knowledge both the first measurement of that observable in $e^+e^-$ collisions and the first new HEP measurement produced autonomously by an AI agent, indicating that such systems can already reach genuinely novel measurements and not only re-derive established ones (subject, as always, to independent verification). We stress that the toolkit for these analyses is largely composed of structured prompts and guidance for the LLM; it is by no means sophisticated, and we encourage the development of related approaches given the low barrier to entry.

The technology to perform analysis at the graduate student level is here and warrants serious attention from the experimental HEP community.
We are not suggesting that AI agents should replace physicists.
We are suggesting that they can take over much of the technically demanding, often tedious, implementation work that currently consumes the majority of an analyst's time, freeing physicists to focus on what they do best: developing physical insight, asking creative questions, and exercising the expert judgment that no AI system can yet replicate. Moreover, we see this as a path towards more sophisticated physics analyses. Data can be more thoroughly combed, details and features within the data can be explored, and there is significant potential for new, comprehensive analyses to be built on these ideas.

The community should begin experimenting with these tools now, developing the workflows, benchmarks, and review practices needed to use them responsibly.
The question is not whether AI agents will become part of the HEP analysis toolkit, but how quickly the community adapts to these tools.

\section{Code and Data Availability}
\label{sec:availability}

The \slop framework specification---including the methodology document, conventions directory, agent profile definitions, and orchestrator prompts---is publicly available at \url{https://github.com/jfc-mit/jfc} and archived on Zenodo~\cite{jfc_zenodo}.
The agent-produced analyses presented in this paper, including complete analysis notes, all code, and generated figures, are available in dedicated repositories linked from the main framework repository --- \url{https://github.com/jfc-mit} --- and archived on Zenodo~\cite{analyses_holding_zenodo}. The two analyses presented as condensed letters (Appendices~\ref{app:tautau_pub} and~\ref{app:lund_pub}) are additionally archived on Zenodo~\cite{cms_htautau_zenodo,aleph_lund_zenodo}.
The archived ALEPH, DELPHI, CMS data and Monte Carlo samples used in this work are publicly available through the CERN Open Data portal and are cited in the specific ANs.

\section{Acknowledgments}
We thank Yen-Jie Lee for providing us with preprocessed ALEPH Open Data and Monte Carlo for use in agentic analysis, and we thank the Electron-Positron Alliance\footnote{\href{https://ee-alliance.org/home/}{https://ee-alliance.org/home/}} for their ongoing work to preserve and re-analyze archival LEP data. We also thank CERN for providing the open data and MC from the ALEPH, DELPHI, and CMS experiments that we use in this paper\footnote{\href{https://opendata.cern.ch/}{https://opendata.cern.ch/}}.  Lastly, while preparing this document, we became aware of~\cite{badea2026agentic}. We have appropriately qualified the differences and conclusions within the paper.

This work is supported by the National Science Foundation under Cooperative Agreement PHY-2019786 (The NSF AI Institute for Artificial Intelligence and Fundamental Interactions, IAIFI, \href{http://iaifi.org/}{http://iaifi.org/}). Computing resources for the SciTreeRAG literature extraction were provided by IAIFI on the FASRC Cannon cluster supported by the FAS Division of Science Research Computing Group at Harvard University. A. N. is supported by the NSF-funded A3D3 Institute (NSF-PHY-2117997), and a DOE Early Career award FY2021, ``Harnessing the Large Hadron Collider with New Insights in
Real-Time Data Processing and Artificial Intelligence''. E. A. M. acknowledges support from the National Science Foundation with Grant No. GRFP2141064.

\paragraph{Competing interests.}
The authors declare no competing interests.

\paragraph{Ethical statement.}
This work raises no ethical concerns: it involves no human or animal subjects and uses only publicly released open data, so no ethical approval was required.

\printbibliography

\appendix
\numberwithin{figure}{section}
\numberwithin{table}{section}

\newpage
\section{Numerical results}\label{app:results}
Table~\ref{tab:results} lists the agent-produced scalar measurements shown in Fig.~\ref{fig:summary}, together with the published reference value used for each comparison and a link to the public code repository for each analysis.

\begin{table}[!htbp]
\centering
\small
\begin{tabular*}{\textwidth}{@{\extracolsep{\fill}} l c c c c c c c c r @{}}
\toprule
                & \multicolumn{4}{c}{Reference}                                                              & \multicolumn{4}{c}{Agent measurement}                                                  &        \\
\cmidrule(lr){2-5} \cmidrule(lr){6-9}
Observable      & Value      & $\sigma_\mathrm{stat}$ & $\sigma_\mathrm{syst}$ & $\sigma_\mathrm{tot}$      & Value      & $\sigma_\mathrm{stat}$ & $\sigma_\mathrm{syst}$ & $\sigma_\mathrm{tot}$ & Pull   \\
\midrule
\multicolumn{10}{@{}l}{\textit{ALEPH Z lineshape~\cite{ALEPH:1999dsx,LEPEWWG:2006}}~\href{https://github.com/jfc-mit/analysis_aleph_z_lineshape}{[code]}} \\
$M_Z$ [GeV]                 & $91.1885$   & $0.0024$ & $0.0017$ & $0.0031$ & $91.1638$ & $0.003$  & $0.040$  & $0.0402$ & $-0.6$           \\
$\Gamma_Z$ [GeV]            & $2.4951$    & $0.0038$ & $0.0016$ & $0.0043$ & $2.4508$  & $0.008$  & $0.009$  & $0.0124$ & $\mathbf{-3.4}$  \\
$\sigma^0_\mathrm{had}$ [nb]& $41.559$    & ---      & ---      & $0.058$  & $42.47$   & $0.065$  & $0.68$   & $0.69$   & $+1.3$           \\
$N_\nu$                     & $2.9840$    & ---      & ---      & $0.0082$ & $2.864$   & ---      & ---      & $0.171$  & $-0.7$           \\
\midrule
\multicolumn{10}{@{}l}{\textit{ALEPH $R_b$, $A_\mathrm{FB}^b$~\cite{ALEPH:1996nz,ALEPH:1996we,ALEPH:1996zz}}~\href{https://github.com/jfc-mit/analysis_aleph_z_heavy_flavour}{[code]}} \\
$R_b$                                     & $0.2158$ & $0.0009$ & $0.0011$ & $0.0014$ & $0.21421$ & $0.0027$ & $0.0227$ & $0.0228$ & $-0.1$           \\
$A_\mathrm{FB}^{0,b}$                     & $0.0927$ & ---      & ---      & $0.0052$ & $0.07025$ & $0.0085$ & $0.0239$ & $0.0253$ & $-0.9$           \\
$\sin^2\theta_\mathrm{eff}^\mathrm{lept}$ & $0.2330$ & ---      & ---      & $0.0009$ & $0.23691$ & $0.0016$ & $0.0044$ & $0.0047$ & $+0.8$           \\
\midrule
\multicolumn{10}{@{}l}{\textit{ALEPH event shapes and $\alpha_s$ extraction~\cite{ParticleDataGroup:2024cfk}}~\href{https://github.com/jfc-mit/analysis_aleph_eventshapes_alphas}{[code]}} \\
$\alpha_s(M_Z)$             & $0.1180$    & ---      & ---      & $0.0009$ & $0.1064$  & ---      & ---      & $0.0198$ & $-0.6$           \\
\midrule
\multicolumn{10}{@{}l}{\textit{CMS $H\to \tau^+\tau^-$ ($\mu\tau_h$ channel)~\cite{CMS:2014wdm}}~(App.~\ref{app:tautau_pub})~\href{https://github.com/jfc-mit/analysis_cms_higgs_tautau}{[code]}} \\
$\mu$                       & $1.01$      & ---    & ---    & $0.41$ & $1.20$    & $0.23$   & $1.10$   & $1.13$   & $+0.2$           \\
\bottomrule
\end{tabular*}
\caption{Numerical values plotted in Fig.~\ref{fig:summary}: agent-produced scalar measurements alongside the published reference value used for each comparison, grouped by analysis. Both the Reference and Agent blocks split $\sigma_\mathrm{stat}$ and $\sigma_\mathrm{syst}$ where the source separates the two; $\sigma_\mathrm{tot}$ is the corresponding quadrature sum or the single total when no split was provided. Em-dashes mark missing splits. The compatibility pull is $(\theta_\mathrm{agent} - \theta_\mathrm{ref})\big/\sqrt{\sigma_\mathrm{agent,\,tot}^2 + \sigma_\mathrm{ref,\,tot}^2}$, treating reference and agent uncertainties as uncorrelated. Bold pulls flag $|\mathrm{pull}|\geq 2$. Each analysis group links to its public code repository (\texttt{[code]}); the CMS measurement is additionally detailed in Appendix~\ref{app:tautau_pub}. The numerical content is read directly from the same \texttt{measurements.yaml} and \texttt{references.yaml} that feed Fig.~\ref{fig:summary}.\dminus{ Shaded rows marked $^\dagger$ are from superseded v0 runs still pending a re-run (greyed in Fig.~\ref{fig:summary}); all other rows use the current v1 results.}}
\label{tab:results}
\end{table}

\newpage
\begin{refsection}[appendices_pub/cms_higgs_tautau/references.bib]
\section{An Open and Reproducible Measurement of $H\to \tau^+\tau^-$ in the $\mu\tau_h$ Final State with CMS Open Data}\label{app:tautau_pub}

\graphicspath{{appendices_pub/cms_higgs_tautau/figures/}}

\tikzset{
  gluon/.style={decorate, decoration={coil, aspect=0.5, segment length=2mm, amplitude=1.2mm}},
  boson/.style={decorate, decoration={snake, segment length=2.4mm, amplitude=0.7mm}},
  ferm/.style={postaction={decorate}, decoration={markings,
    mark=at position 0.55 with {\arrow{Stealth[length=2mm]}}}},
}


\begingroup
\leftskip=2em \rightskip=2em
\small
\noindent\textbf{Abstract.}\enspace
We present a complete, systematics-aware, categorized measurement of
Standard Model Higgs-boson production in the $H\to \tau^+\tau^-$
muon--hadronic-tau ($\mu\tau_{h}$) final state, carried out end to end on
public CMS Open Data at $\sqrt{s}=8$~TeV ($11.467~\mathrm{fb}^{-1}$, 2012)
with open-source tools. Events are split into 0-jet, boosted, and
vector-boson-fusion (VBF) categories---the latter using forward jets out to
$|\eta|<4.7$---with the $W$+jets and QCD multijet backgrounds estimated from
data and the $t\bar{t}$ background constrained in situ by a dedicated b-tag
control region included in the simultaneous fit. A multivariate
discriminant $D_{NN}$ is fit across the three categories in a binned
maximum-likelihood fit with roughly twenty systematic uncertainties
profiled as nuisance parameters. The full statistical model is published
and independently validated: a direct reimplementation in the CMS
\textsc{Combine} framework reproduces the \textsc{pyhf} likelihood
point-by-point and every fitted quantity. The fit yields a signal strength
$\hat{\mu}=1.20\pm1.13$ relative to the Standard Model expectation, with an
observed (expected) 95\% confidence-level upper limit $\mu<3.72$
($\mu<2.58$), consistent with the Standard Model ($\mu=1$). The result
lands directly on the published CMS $\mu\tau_{h}$-channel measurement
$\mu=1.01\pm0.41$, agreeing to $0.16\sigma$, and on the published combined
measurements $\mu=0.78\pm0.27$ and $\mu=1.09^{+0.27}_{-0.26}$. The
precision is set by the use of a single final state and the partial 2012
dataset. This Letter provides a transparent, dual-engine-validated
reference implementation of a categorized profile-likelihood Higgs
measurement from public data.
\par
\endgroup

\subsection{Introduction}
The Yukawa coupling of the Higgs boson to fermions is not predicted by the
gauge structure of the Standard Model but is fixed by the requirement that
the same field that breaks electroweak symmetry also generate the fermion
masses. The decay of the 125~GeV Higgs boson to a pair of tau
leptons~\cite{Ellis1988} is the most accessible direct probe of this
coupling to leptons: the tau is the heaviest lepton, its Yukawa coupling is
the largest in the lepton sector, and the $H\to \tau^+\tau^-$ branching fraction
of roughly $6\%$~\cite{LHCHXSWG2013} is large enough to be measured at the
LHC despite the substantial backgrounds. Following the discovery of the
boson by the ATLAS and CMS collaborations in
2012~\cite{ATLAS2012disc,CMS2012disc}, evidence for and then observation of
the $H\to \tau^+\tau^-$ decay were reported by CMS using the 7 and 8~TeV
data~\cite{CMS2014htautau} and the combined 7, 8, and 13~TeV
data~\cite{CMS2018htautau}, with corresponding ATLAS
results~\cite{ATLAS2015htautau}. These frontier measurements combine
several di-tau final states, exploit the distinct kinematics of the
gluon-fusion and vector-boson-fusion production modes through event
categorization, employ multivariate discriminants, and extract the signal
from a per-category profile-likelihood fit over the full recorded
luminosity.

The methods that drive such measurements are now reproducible outside the
collaborations. The release of LHC collision data through the CERN Open
Data Portal under the CC0 waiver~\cite{CMSopendata2012B,CMSopendata2012C,CMSopendatapolicy},
the growing practice of publishing full statistical
likelihoods~\cite{ATLAS2019likelihood}, and the availability of mature
open-source modeling and inference tools~\cite{pyhf,Cranmer2012,CMSCombine}
together make it possible to reconstruct a complete Higgs-boson measurement
transparently---and, crucially, to verify it with a statistical engine
independent of the one used to obtain it. In this Letter we provide such a
reference implementation: a categorized $H\to \tau^+\tau^-$ measurement in the
$\mu\tau_{h}$ final state, built entirely from public 2012 CMS data, whose
full statistical model is released and cross-validated against the CMS
\textsc{Combine} framework. We report the signal strength $\mu$, the ratio
of the measured to the Standard-Model-predicted $H\to \tau^+\tau^-$ cross
section times branching fraction, with a single parameter scaling the
gluon-fusion (ggH) and VBF production modes coherently. The two signal
processes and the $\mu\tau_{h}$ decay topology targeted here are shown in
Fig.~\ref{tpub:fig:feyn}.

\begin{figure}[t]
\centering
\resizebox{\linewidth}{!}{%
\begin{tikzpicture}[line width=0.7pt]

\begin{scope}[shift={(0,0)}]
  \coordinate (H) at (1.6,0.0);
  \coordinate (g1) at (0,0.9);
  \coordinate (g2) at (0,-0.9);
  \draw[gluon] (g1) -- (H);
  \draw[gluon] (g2) -- (H);
  \fill (H) circle (3.2pt);              
  \coordinate (D) at (3.2,0.0);
  \draw[dashed] (H) -- (D) node[midway,above] {$H$};
  \coordinate (ta) at (4.4,0.95);
  \coordinate (tb) at (4.4,-0.95);
  \draw[ferm] (D) -- (ta);
  \draw[ferm] (tb) -- (D);
  \draw[ferm] (ta) -- (5.6,1.5) node[right] {$\tau_h$};
  \draw[ferm] (ta) -- (5.6,0.55) node[right] {$\nu$};
  \draw[ferm] (tb) -- (5.6,-1.5) node[right] {$\mu$};
  \draw[ferm] (tb) -- (5.6,-0.55) node[right] {$\nu\bar\nu$};
  \node[left] at (g1) {$g$};
  \node[left] at (g2) {$g$};
  \node[above left] at (ta) {$\tau^{+}$};
  \node[below left] at (tb) {$\tau^{-}$};
  \node at (2.8,-1.85) {gluon--gluon fusion};
\end{scope}

\begin{scope}[shift={(8.6,0)}]
  \coordinate (qi1) at (0,1.6);
  \coordinate (qi2) at (0,-1.6);
  \coordinate (v1) at (1.4,1.1);
  \coordinate (v2) at (1.4,-1.1);
  \coordinate (qo1) at (2.8,1.6);
  \coordinate (qo2) at (2.8,-1.6);
  \draw[ferm] (qi1) -- (v1);
  \draw[ferm] (v1) -- (qo1) node[right] {$q$};
  \draw[ferm] (qi2) -- (v2);
  \draw[ferm] (v2) -- (qo2) node[right] {$q$};
  \node[left] at (qi1) {$q$};
  \node[left] at (qi2) {$q$};
  \coordinate (H) at (1.4,0.0);
  \draw[boson] (v1) -- (H);
  \draw[boson] (v2) -- (H) node[midway,right=1pt] {$W/Z$};
  \coordinate (D) at (3.0,0.0);
  \draw[dashed] (H) -- (D) node[midway,above] {$H$};
  \coordinate (ta) at (4.2,0.85);
  \coordinate (tb) at (4.2,-0.85);
  \draw[ferm] (D) -- (ta);
  \draw[ferm] (tb) -- (D);
  \draw[ferm] (ta) -- (5.4,1.4) node[right] {$\tau_h$};
  \draw[ferm] (ta) -- (5.4,0.45) node[right] {$\nu$};
  \draw[ferm] (tb) -- (5.4,-1.4) node[right] {$\mu$};
  \draw[ferm] (tb) -- (5.4,-0.45) node[right] {$\nu\bar\nu$};
  \node[above left] at (ta) {$\tau^{+}$};
  \node[below left] at (tb) {$\tau^{-}$};
  \node at (2.7,-2.05) {vector-boson fusion};
\end{scope}

\end{tikzpicture}%
}
\caption{Representative leading-order diagrams for the two Higgs-boson
production modes scaled by the signal strength in this analysis: gluon--gluon fusion (left), $gg\to H$, proceeding through a heavy-quark loop here
drawn as an effective $ggH$ vertex, and vector-boson fusion (right),
$qq\to qqH$, through $t$-channel $W$ or $Z$ exchange with two
forward-tagging quark jets. In both cases the Higgs boson decays to a tau
pair, $H\to\tau^{+}\tau^{-}$, with one tau decaying leptonically to a muon,
$\tau\to\mu\,\nu\bar\nu$, and the other hadronically,
$\tau\to\tau_{h}\,\nu$, defining the $\mu\tau_{h}$ final state.}
\label{tpub:fig:feyn}
\end{figure}

\subsection{Data and event selection}
The analysis uses the \textsc{TauPlusX} primary dataset from the 2012 CMS
run periods B and C at $\sqrt{s}=8$~TeV, corresponding to an integrated
luminosity of $11.467~\mathrm{fb}^{-1}$ with a 2.6\% normalization
uncertainty~\cite{CMSlumi2012}. The dataset is collected with a set of
triggers requiring the coincidence of an isolated muon and a hadronic tau
candidate, matching the $\mu\tau_{h}$ topology of the signal. Signal and
background processes are modeled with the simulated samples provided
alongside the data on the open data portal, normalized to
$\sigma\mathcal{L}/N_{\mathrm{gen}}$ with the recommended Higgs and
electroweak cross sections~\cite{LHCHXSWG2013} and reweighted to the
distribution of reconstructed primary vertices to match the in-time pileup
of the data.

Events are selected in the $\mu\tau_{h}$ final state by requiring an
isolated muon with transverse momentum $p_{T}>20$~GeV, within the tracker
acceptance $|\eta|<2.1$, and passing the tight identification criteria,
together with a hadronically decaying tau candidate reconstructed with the
hadron-plus-strips algorithm~\cite{CMStau2016}, satisfying the medium
isolation working point and a tight discriminator against muons
misidentified as taus. The muon and the tau candidate are required to carry
opposite electric charge, as expected for the two taus from a Higgs-boson
decay. A large fraction of the residual background is $W$+jets production,
in which a genuine muon from the $W$ decay is accompanied by a jet
misidentified as a hadronic tau; because the muon and the neutrino from the
$W$ are nearly back-to-back in the transverse plane, this background is
suppressed by requiring the transverse mass of the muon and the missing
transverse momentum to satisfy
$m_{T}(\mu,p_{T}^{\mathrm{miss}})<30$~GeV. The resulting opposite-sign
signal region contains $27{,}240$ events.

\subsection{Event categories}
Selected events are partitioned into three mutually exclusive categories
that isolate the production modes of Fig.~\ref{tpub:fig:feyn} and order the
sample by signal-to-background ratio. The VBF category requires two jets
with invariant mass $m_{jj}>500$~GeV and pseudorapidity separation
$|\Delta\eta_{jj}|>3.5$, the characteristic signature of the two forward
quark jets recoiling against the Higgs boson in vector-boson fusion; the
jets are reconstructed out to $|\eta|<4.7$ to retain the most forward
tags, which carry much of the discriminating power of this category. The
boosted category collects the remaining events with a reconstructed di-tau
transverse momentum $p_{T}^{\tau\tau}>100$~GeV, enriched in gluon-fusion
events recoiling against initial-state radiation and benefiting from the
improved angular resolution of a boosted topology. All remaining events
form the 0-jet category, which dominates the event count but has the lowest
signal-to-background ratio. A veto on b-tagged jets is applied throughout to
suppress top-quark backgrounds; inverting this veto defines the $t\bar{t}$
control region used below. The three signal-region categories contain
$26{,}020$, $1{,}149$, and $71$ events, respectively.

\subsection{Backgrounds}
\begin{figure}[t]
\centering
\includegraphics[width=0.49\linewidth]{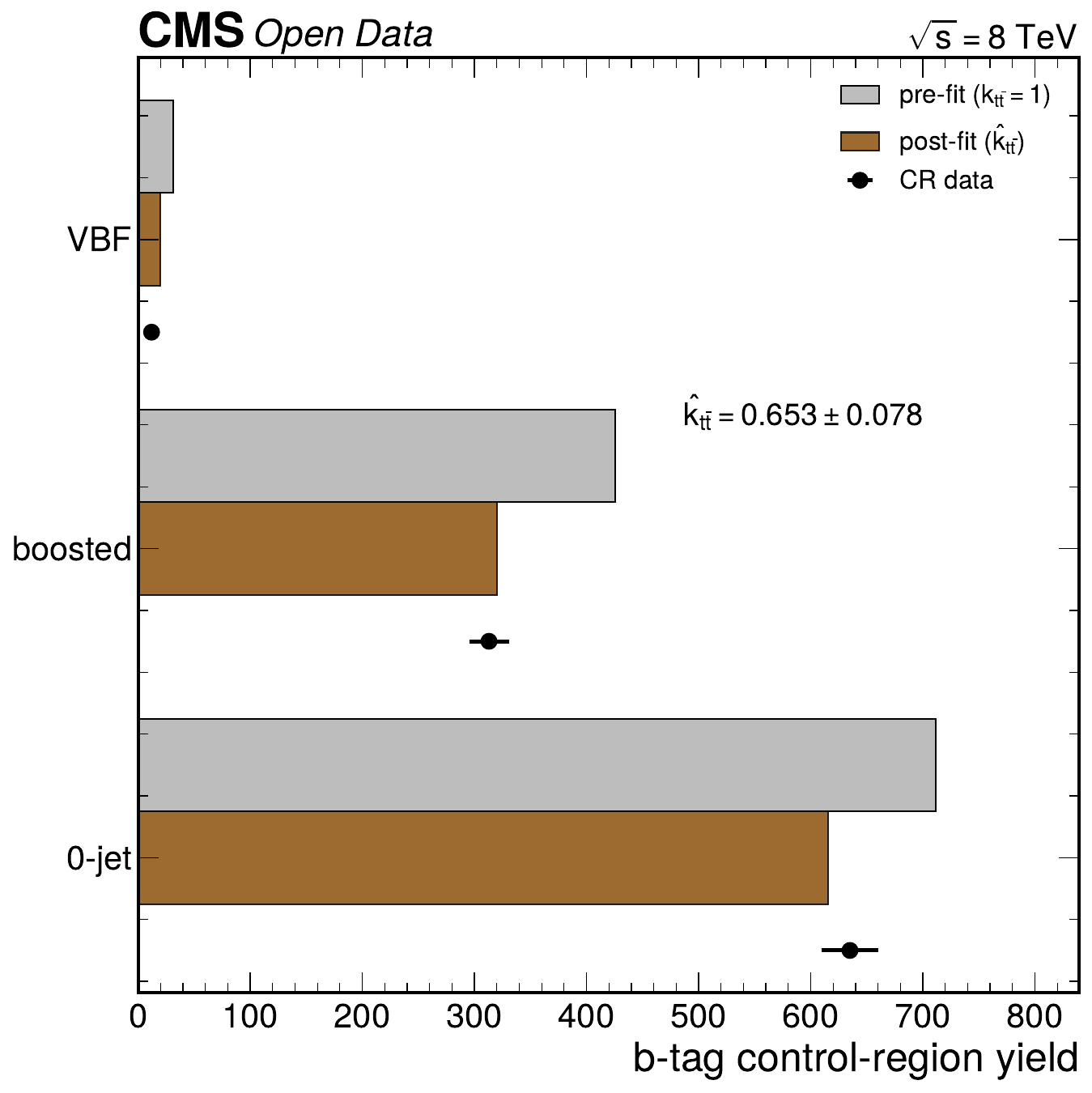}\hfill
\includegraphics[width=0.49\linewidth]{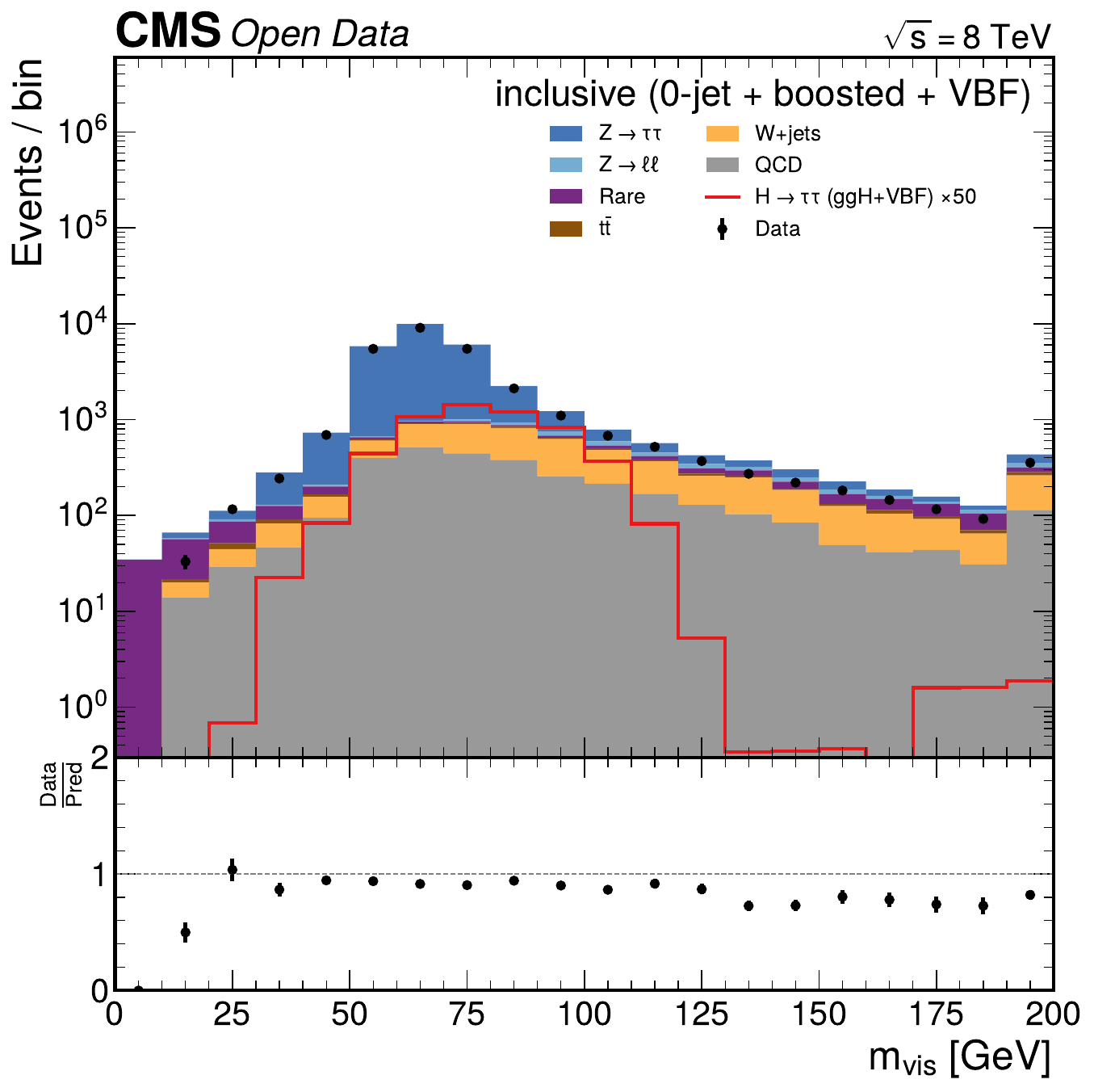}
\caption{Yields in the three b-tagged $t\bar{t}$ control regions (0-jet,
boosted, VBF) entering the simultaneous fit (left): for each region the pre-fit
prediction ($k_{t\bar{t}}=1$, grey) and the post-fit prediction (scaled by the
fitted $\hat{k}_{t\bar{t}}=0.653\pm0.078$, brown) are compared with the
observed control-region data (points); the pre-fit prediction overshoots the
data in all three regions, and the single shared normalization $k_{t\bar{t}}$
brings the post-fit prediction into agreement, providing the in-situ
$t\bar{t}$ constraint that is propagated into the signal regions.
Distribution of the visible di-tau mass $m_{\mathrm{vis}}$, the invariant
mass of the muon and the visible hadronic-tau decay products, for the inclusive
$\mu\tau_{h}$ selection (right): the data are compared with the stack of background
processes, the $W$+jets and QCD multijet contributions estimated from data; the
spectrum is broad and peaks well below the Higgs-boson mass because the
neutrinos from both tau decays escape detection, and the absence of a narrow
mass peak motivates the use of a multivariate discriminant rather than a simple
mass window to separate signal from background.}
\label{tpub:fig:ttcr}
\label{tpub:fig:mvis}
\end{figure}

The dominant and irreducible background is $Z\to \tau^+\tau^-$ production, which
shares the di-tau final state of the signal and is separated from it only
kinematically; it is taken from simulation with a normalization uncertainty
of 10--15\%. Smaller contributions from $Z\to\ell\ell$ (with a lepton or
jet misidentified as the hadronic tau) and a combined
diboson-plus-single-top (``rare'') category are likewise taken from
simulation. The two backgrounds that are most difficult to model from
simulation are estimated directly from data: the $W$+jets background is
normalized in a high-$m_{T}$ control region orthogonal to the signal
selection and extrapolated into the signal region, and the QCD multijet
background is estimated from a same-sign $\mu\tau_{h}$ control region, with
the opposite-sign-to-same-sign transfer factor measured in data. These
data-driven estimates avoid reliance on the simulation of jet
misidentification and hadronic activity, and their normalizations enter the
fit as free parameters constrained by the control-region statistics. The
visible di-tau mass of the inclusive sample is shown in
Fig.~\ref{tpub:fig:mvis} (right).

The $t\bar{t}$ background is constrained in situ rather than fixed to a
prior. Inverting the b-jet veto---requiring at least one b-tagged jet---in
each of the three categories defines a set of $t\bar{t}$-enriched control
regions, with a $t\bar{t}$ purity ranging from $56\%$ in the 0-jet region
to $80\%$ in the VBF region. The three control regions are entered as
single-bin counting channels in the simultaneous fit, and a single freely
floating normalization $k_{t\bar{t}}$, shared between the control regions
and the signal regions, scales the $t\bar{t}$ yield everywhere; the
b-tagging efficiency nuisance parameter is correlated between the two
region sets so that the extrapolation across the b-tag boundary is
self-consistent. The control-region yields, observed and post-fit, are
shown in Fig.~\ref{tpub:fig:ttcr} (left). The fit determines
\begin{equation}
k_{t\bar{t}} = 0.653 \pm 0.078,
\label{tpub:eq:kttbar}
\end{equation}
a $12\%$ in-situ constraint that replaces any external $t\bar{t}$
normalization assumption; the data prefer a $t\bar{t}$ yield about $35\%$
below the simulation, well within the spread of $t\bar{t}$ normalizations
seen in the published analysis, and the constraint is propagated coherently
into the signal regions.

\subsection{Discriminant}
Because the escaping neutrinos smear the reconstructed di-tau mass into a
broad distribution that overlaps the dominant $Z\to \tau^+\tau^-$ background
(Fig.~\ref{tpub:fig:mvis} (right)), the signal is separated from the total background by
a multivariate discriminant, $D_{NN}$, rather than by a mass window. The
discriminant is a single gradient-boosted decision-tree
classifier~\cite{Chen2016}\footnote{The discriminant is denoted $D_{NN}$ throughout this Letter, including in the figure axes; despite the ``NN'' subscript it is the gradient-boosted decision-tree output described here, not a neural network.}, trained once on signal-region events to
distinguish $H\to \tau^+\tau^-$ from the summed background and applied unchanged
in all three categories, using a fixed set of fifteen kinematic and
angular inputs: the visible and collinear-approximation di-tau masses, the
transverse momenta of the muon, the tau candidate, and the di-tau system,
the muon and tau pseudorapidities, the angular separations
$\Delta R$ and $\Delta\phi$ between the muon and the tau candidate, the
missing transverse momentum and its significance, the transverse masses of
the muon, the tau candidate, and the full system with the missing momentum,
and the raw tau isolation. The classifier reaches an area under the
receiver-operating-characteristic curve of approximately $0.82$, and its
output is the observable fit to extract the signal.

\subsection{Statistical model}
The signal strength is extracted from a simultaneous binned
maximum-likelihood fit of the $D_{NN}$ distributions in the three signal
categories together with the three $t\bar{t}$ control regions. The
likelihood is built with \textsc{pyhf}~\cite{pyhf}, a
pure-Python implementation of the \textsc{HistFactory}
template~\cite{Cranmer2012}, with a single parameter of interest $\mu$
scaling the ggH and VBF yields coherently and a single freely floating
$t\bar{t}$ normalization $k_{t\bar{t}}$ shared between the control and
signal regions. Roughly twenty systematic
uncertainties enter as nuisance parameters, each motivated by a measured or
published uncertainty:
the tau energy scale~\cite{CMStau2016}, which shifts the $D_{NN}$ inputs
built from the hadronic tau;
the jet energy scale and resolution~\cite{CMSjes2017}, which affect the
category boundaries and the VBF dijet variables;
the b-tagging efficiency, entering through the b-jet veto and correlated
between the control and signal regions;
the unclustered missing transverse momentum~\cite{CMSmet2015}, which
propagates to $m_{T}$ and the missing-energy inputs;
the data-driven $W$+jets and QCD normalizations and their extrapolation
shapes, sized by the control-region statistics;
the $Z\to \tau^+\tau^-$ and rare normalizations and a $5\%$ uncertainty on the
$t\bar{t}$ control-to-signal-region extrapolation;
the signal theory uncertainties from the renormalization and factorization
scales, the parton distribution functions, and the parton shower;
and the 2.6\% luminosity uncertainty~\cite{CMSlumi2012}. The finite size of
the simulated samples is accounted for with per-bin Barlow--Beeston
statistical parameters~\cite{BarlowBeeston1993}, and the Poisson
likelihood for binned histograms follows the standard
treatment~\cite{Baker1984}. The signal strength is obtained by profiling
the likelihood; the discovery significance is computed from the one-sided
$q_{0}$ test statistic and the upper limit from the $\mathrm{CL}_{s}$
prescription~\cite{Read2002}, both in the asymptotic
approximation~\cite{Cowan2011}.

\subsection{Results}
\begin{figure}[t]
\centering
\includegraphics[width=0.32\linewidth]{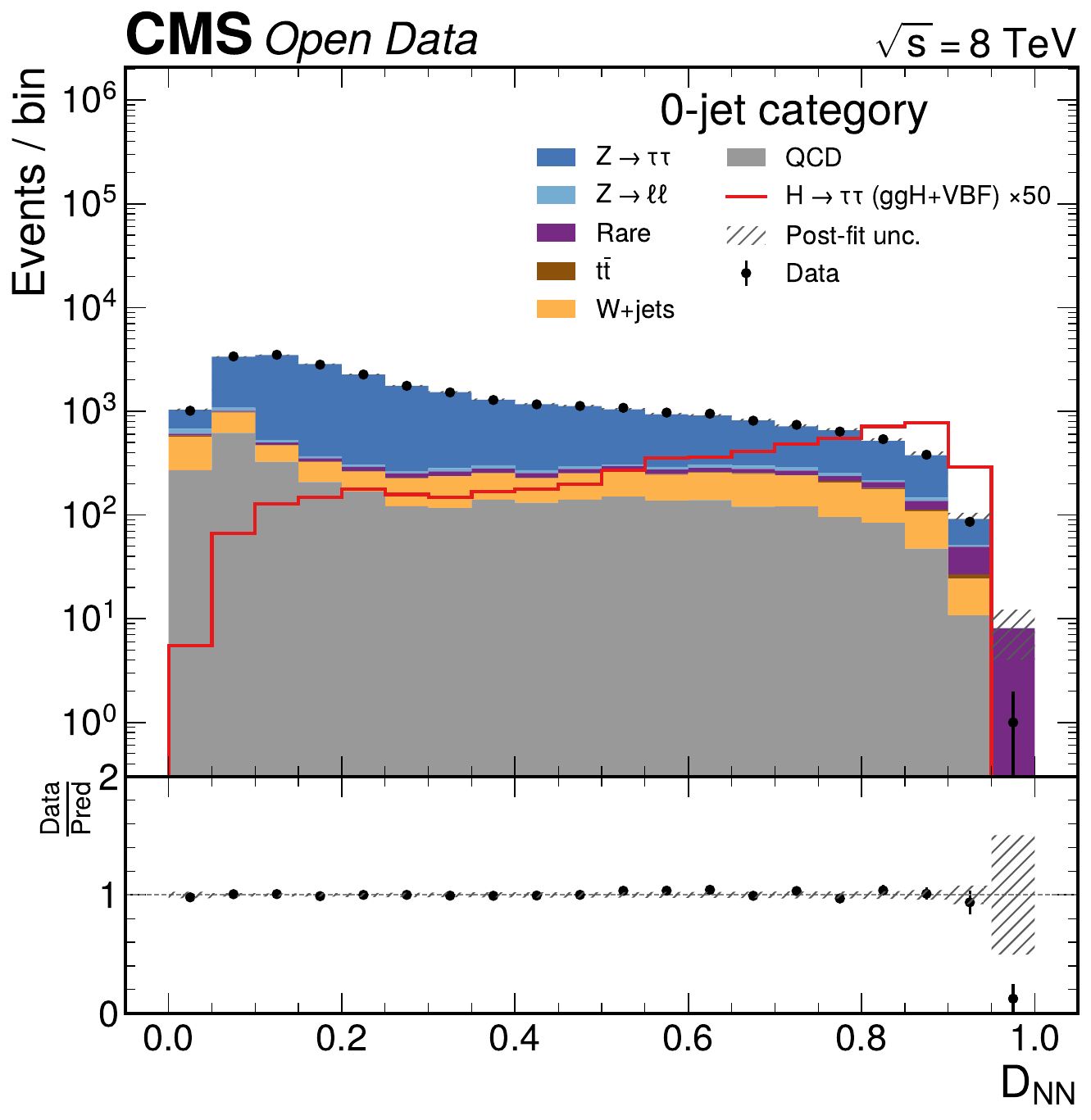}\hfill
\includegraphics[width=0.32\linewidth]{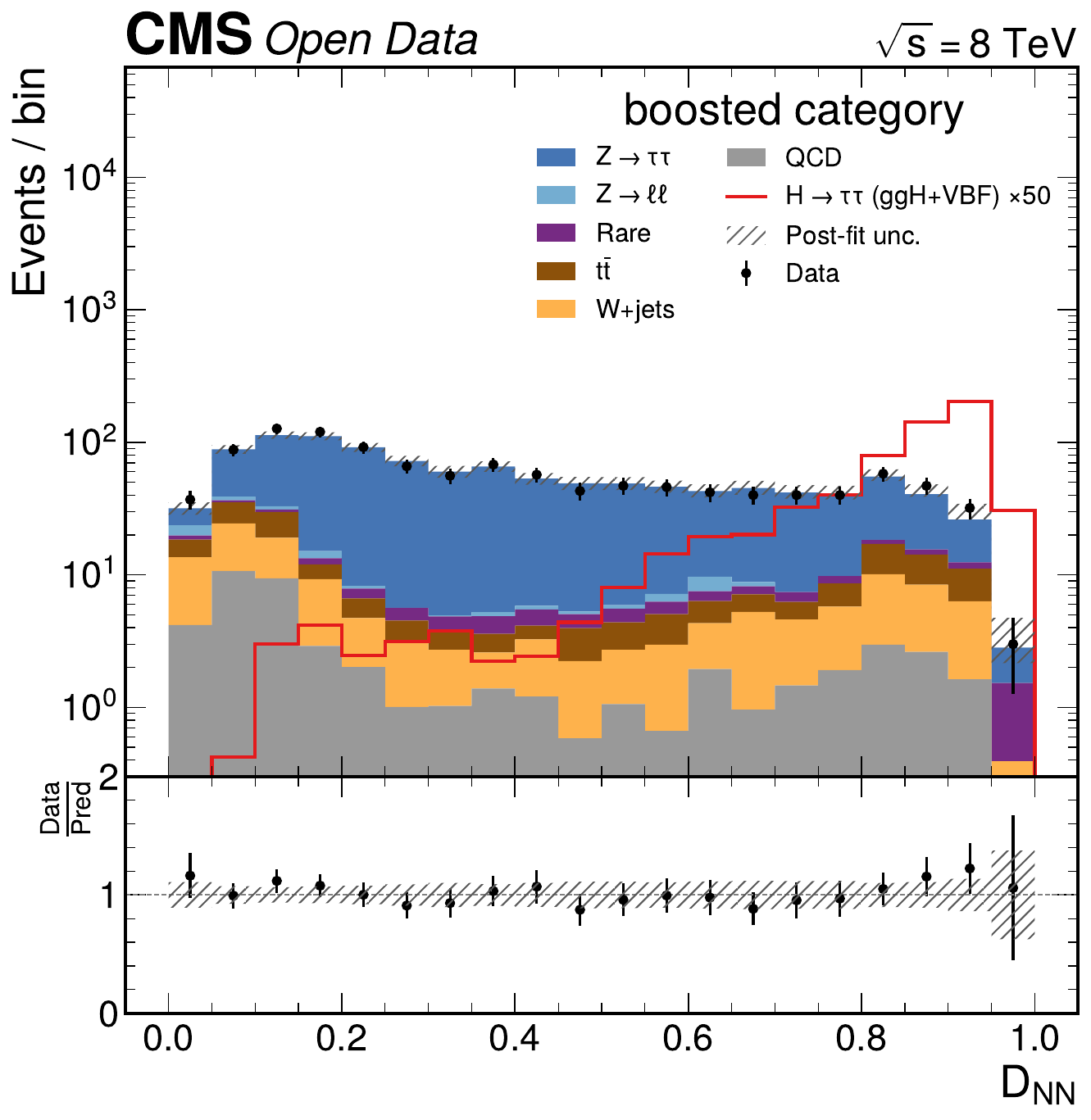}\hfill
\includegraphics[width=0.32\linewidth]{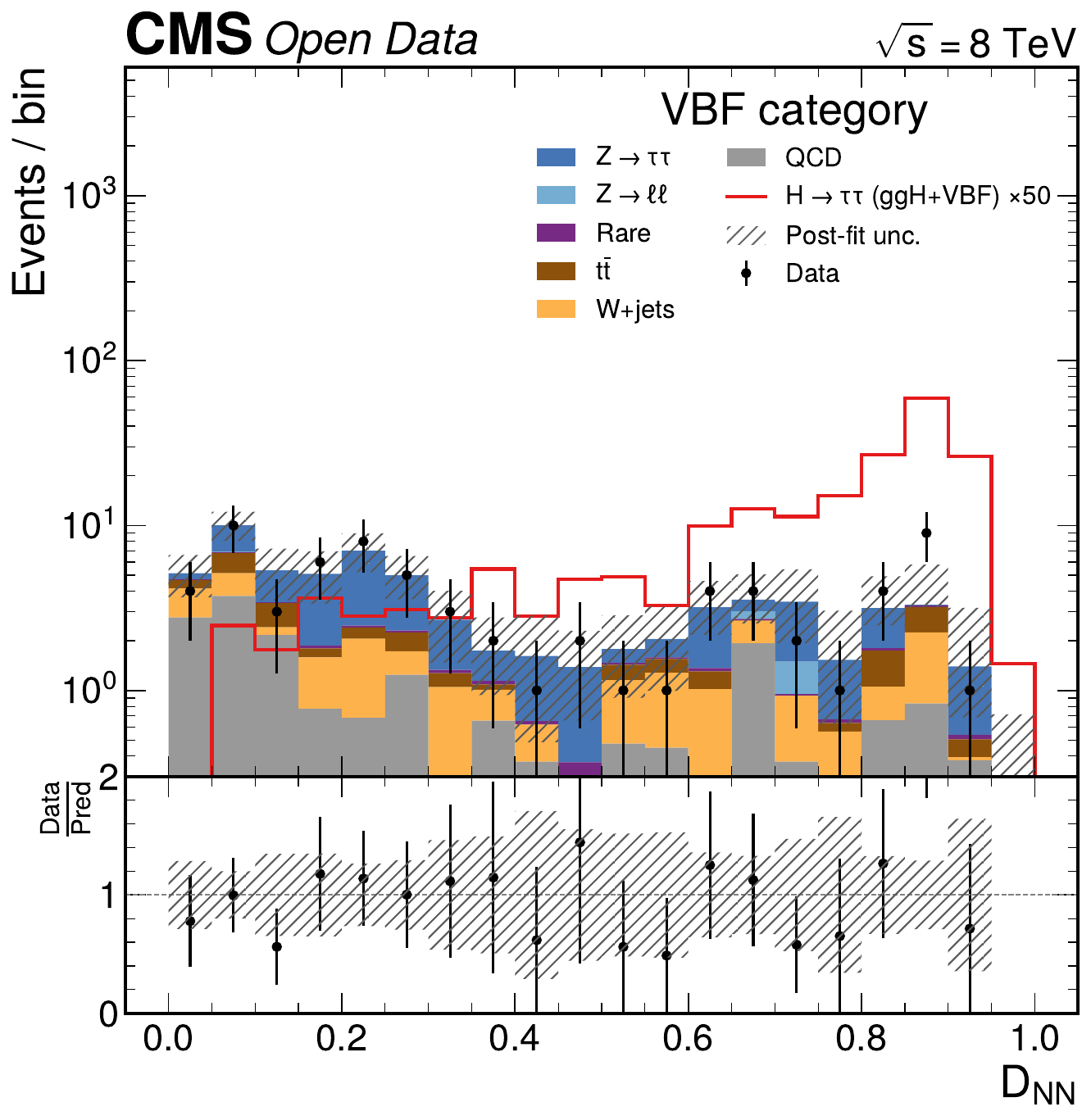}
\caption{Post-fit $D_{NN}$ distributions in the three event categories: the
data are overlaid on the post-fit background stack with its uncertainty band
and the post-fit signal (red) scaled to $\hat{\mu}$, with the data-to-model
ratio shown below each panel. 0-jet category (left), the highest-yield and lowest
signal-to-background category. Boosted category
(middle, $p_{T}^{\tau\tau}>100$~GeV), enriched in gluon-fusion events recoiling against
initial-state radiation. VBF category (right, $m_{jj}>500$~GeV,
$|\Delta\eta_{jj}|>3.5$), the highest signal-to-background but lowest-yield
category, whose few tens of events are well described by the post-fit model.
The post-fit model reproduces the data in all three categories without
localized discrepancies, across the three orders of magnitude in event count
that separate them.}
\label{tpub:fig:cat_zerojet}
\label{tpub:fig:cat_boosted}
\label{tpub:fig:cat_vbf}
\end{figure}

The simultaneous fit to the observed $D_{NN}$ distributions and the
$t\bar{t}$ control regions yields a signal strength
\begin{equation}
\hat{\mu} = 1.20 \pm 1.13,
\label{tpub:eq:muhat}
\end{equation}
relative to the Standard Model expectation, where the uncertainty is
dominated by its systematic component ($\pm1.10$, versus $\pm0.23$
statistical). The per-category post-fit $D_{NN}$ distributions are shown in
Fig.~\ref{tpub:fig:cat_zerojet}(a--c): the model reproduces the
data in all three categories with no localized structure, across the three
orders of magnitude in event count and the order of magnitude in
signal-to-background ratio that separate the 0-jet and VBF categories. The
data-driven $W$+jets and QCD backgrounds and the simulated $Z\to \tau^+\tau^-$
are each pulled within their pre-fit uncertainties to absorb a uniform
few-percent normalization offset, with no nuisance parameter constrained
beyond two standard deviations.

\begin{figure}[t]
\centering
\includegraphics[height=7.6cm]{fig_money.pdf}\hfill
\includegraphics[height=7.6cm]{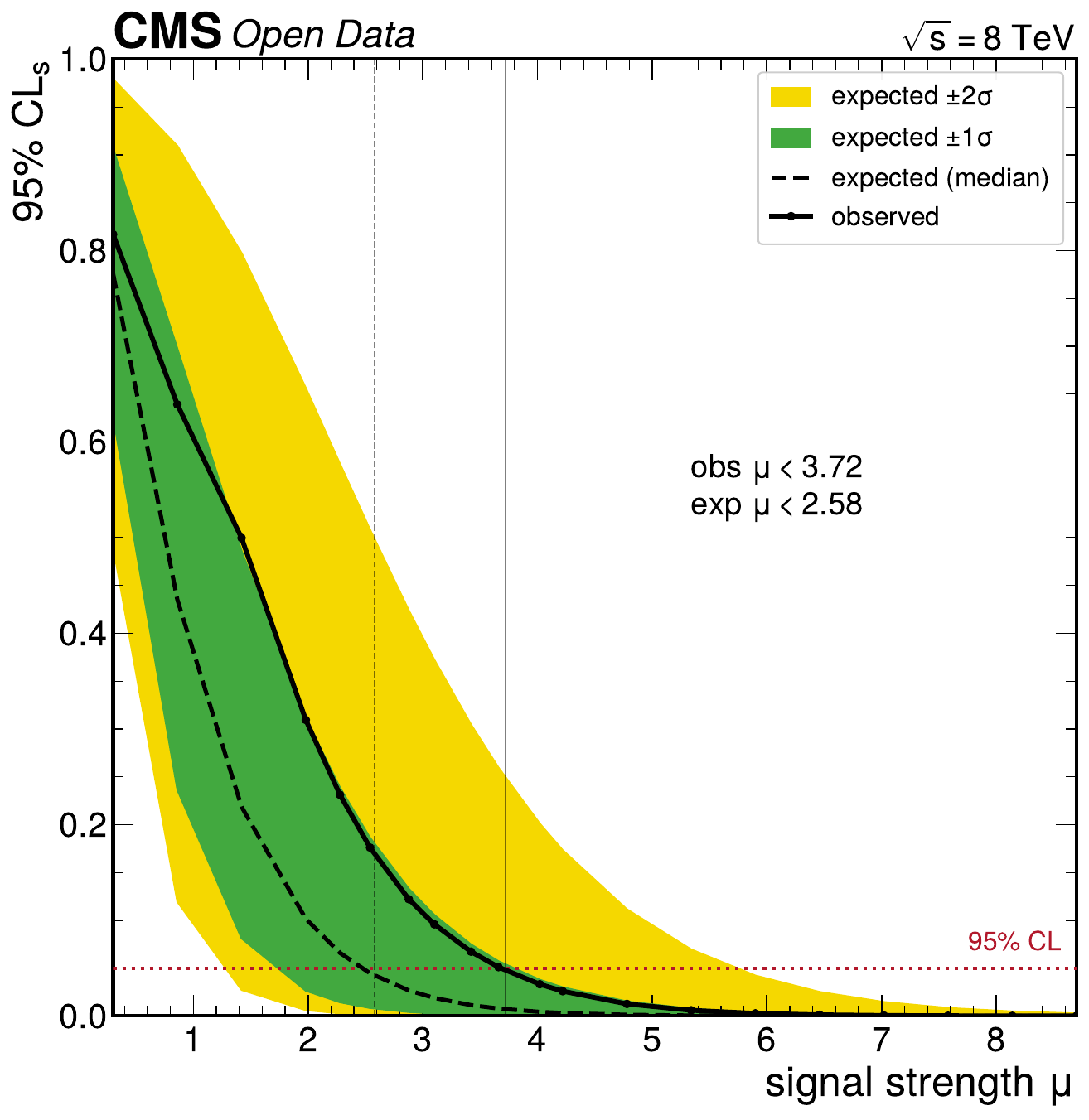}
\caption{$S/(S+B)$-weighted post-fit $D_{NN}$ distribution combining the
0-jet, boosted, and VBF categories (left): each (category, bin) is weighted by its
post-fit $S/(S+B)$ and the weighted yields are accumulated on the common
$D_{NN}$ axis. The points show the data; the filled histogram is the post-fit
per-process background stack ($Z\to \tau^+\tau^-$, $Z\to\ell\ell$, rare, $t\bar{t}$,
$W$+jets, QCD), with the hatched band showing its total post-fit uncertainty
propagated from the fit covariance; the solid red line is the post-fit
signal-plus-background at $\hat{\mu}=1.20$. The lower panel shows the weighted
residual (data minus background) as points, with the post-fit signal $S$
overlaid as a red step and the post-fit background uncertainty as a hatched band
about zero. Observed and expected 95\% confidence-level $\mathrm{CL}_{s}$
upper limits on the signal strength $\mu$ (right): the observed limit is $\mu<3.72$ and
the expected median $\mu<2.58$, with the green and yellow bands showing the
$\pm1\sigma$ and $\pm2\sigma$ ranges of the expectation and the Standard Model
value $\mu=1$ indicated; the observed limit lies within the $\pm1\sigma$ band of
the expectation. The fitted signal is consistent with both the background-only
and the Standard Model hypotheses.}
\label{tpub:fig:result}
\label{tpub:fig:limit}
\end{figure}

The three per-category distributions are summarized in
Fig.~\ref{tpub:fig:result} (left), which shows the $D_{NN}$ distribution summed
over the three categories with each event weighted by its category- and
bin-dependent $S/(S+B)$ ratio, so that the most signal-like bins are
emphasized. The
post-fit signal, scaled to $\hat{\mu}$, sits within the
background uncertainty, and the weighted residual shows a mild positive
excess consistent with a Standard-Model-sized signal. The fit is compatible
with both the absence of signal and the Standard
Model: the observed discovery significance is $Z\approx1.15$ standard
deviations. The goodness of fit is consistent with the model describing the data,
with a corrected frequentist saturated-model test giving a $p$-value of
$0.065$, independently reproduced with the CMS
\textsc{Combine}~\cite{CMSCombine} tool.

Since the data show no significant excess, the result is also expressed as
an upper limit on the signal strength. The observed and expected 95\%
confidence-level $\mathrm{CL}_{s}$ limits are shown in
Fig.~\ref{tpub:fig:limit} (right); the observed limit is
\begin{equation}
\mu < 3.72,
\label{tpub:eq:limit}
\end{equation}
compared with an expected median limit of $\mu<2.58$ under the
background-only hypothesis, the observed value lying within the
$\pm1\sigma$ band of the expectation. The measured signal strength of
Eq.~(\ref{tpub:eq:muhat}) lands directly on the published CMS
$\mu\tau_{h}$-channel result, $\mu=1.01\pm0.41$ (Ref.~\cite{CMS2014htautau},
Fig.~16a)---the like-for-like comparison, the same final state extracted
here---agreeing to $0.16$ standard deviations, where the uncertainties are
combined in quadrature. It is likewise consistent with the published CMS
combined $H\to \tau^+\tau^-$ measurements
$\mu=0.78\pm0.27$~\cite{CMS2014htautau} and
$\mu=1.09^{+0.27}_{-0.26}$~\cite{CMS2018htautau}, within $0.36$ and $0.09$
standard deviations, respectively. The total uncertainty of $\pm1.13$, set
by the single $\mu\tau_{h}$ channel and the partial 2012 dataset
($11.5~\mathrm{fb}^{-1}$, versus the $19.7~\mathrm{fb}^{-1}$ of the
published 8~TeV analysis), makes the upper limit of Eq.~(\ref{tpub:eq:limit})
the most informative statement of the result.

\subsection{Independent validation}
The complete statistical model is published as a \textsc{HistFactory}
workspace and is independently validated with the CMS \textsc{Combine}
framework~\cite{CMSCombine}, which uses the
\textsc{RooFit}/\textsc{RooStats} engine, entirely separate from
\textsc{pyhf}. An independent rebuild of the model as a direct
\textsc{Combine} datacard, cross-checked against a build produced with
\textsc{rhalphalib}~\cite{rhalphalib}, reproduces the \textsc{pyhf}
likelihood point-by-point: the two likelihoods agree to $\le0.005$ at every
$\pm1\sigma$ template knot of the seven shape systematics and to
$10^{-6}$ along the signal-strength ladder, and the fully profiled
$-2\Delta\ln\mathcal{L}$ scans overlay across the entire physical range with
a maximum difference of $0.055$. The agreement was established at the
likelihood level, so the two engines provably fit the identical model: all
60 signal-region bins and the three control regions, the per-bin yields, the
14 normalization ($\ln\mathcal{N}$) systematics, the 7 shape systematics,
the per-bin Barlow--Beeston statistical parameters, the parameter of
interest, and the shared $t\bar{t}$ normalization are reproduced with no
systematic missing or mis-correlated. As summarized in
Table~\ref{tpub:tab:xcheck}, every fitted quantity matches: the signal strength
($\hat{\mu}=1.200\pm1.128$, with a like-for-like profiled $68\%$ interval
$\hat{\mu}=1.19^{+1.28}_{-1.04}$, versus $\hat{r}=1.17^{+1.27}_{-1.02}$ from
\textsc{Combine}'s profile-likelihood scan), the in-situ
$t\bar{t}$ normalization ($k_{t\bar{t}}=0.653\pm0.078$ versus $0.647\pm0.080$),
and the observed discovery significance ($Z=1.152$ in both engines, agreeing
to four decimal places). Both engines independently confirm that the fit is
normal and the data are well described.

\begin{table}[t]
\centering
\caption{Independent validation of the $D_{NN}$ fit: the \textsc{pyhf}
result and its reproduction in the CMS \textsc{Combine} framework on the
point-by-point identical statistical model. The best-fit signal strength
$\hat{\mu}$, the in-situ $t\bar{t}$ normalization $k_{t\bar{t}}$, and the
observed discovery significance $Z$ are listed; the signal-strength
intervals are the like-for-like profiled $68\%$ ($-2\Delta\ln\mathcal{L}=1$)
intervals from each engine's profile-likelihood scan, quoted as offsets from
the best fit (the \textsc{pyhf} headline $\hat{\mu}=1.200\pm1.128$ is the
symmetric Hessian uncertainty). The
profiled $-2\Delta\ln\mathcal{L}(\mu)$ scans overlay to a maximum difference
of $0.055$ across the physical range, and the likelihoods agree to
$\le0.005$ at every template knot.}
\label{tpub:tab:xcheck}
\begin{tabular}{lcc}
\hline
Quantity & \textsc{pyhf} & \textsc{Combine} \\
\hline
$\hat{\mu}$ (profile-scan $68\%$) & $1.19^{+1.28}_{-1.04}$ & $1.17^{+1.27}_{-1.02}$ \\
$k_{t\bar{t}}$         & $0.653\pm0.078$ & $0.647\pm0.080$ \\
$Z$ (observed)        & 1.152           & 1.152                     \\
\hline
\end{tabular}
\end{table}

\subsection{Reproducibility}
The full \textsc{pyhf} statistical model (the workspace in JSON form), the
analysis code, and the intermediate artifacts are openly available, so that
both the measurement and its independent \textsc{Combine} validation can be
rerun end to end. Combined with the public input data, this makes every step
of the analysis---object selection, data-driven background estimation, the
multivariate discriminant, and the categorized profile-likelihood
fit---fully reproducible and independently verifiable.

\subsection{Conclusion}
We have measured the $H\to \tau^+\tau^-$ signal strength in the $\mu\tau_{h}$
final state from CMS Open Data, with the $t\bar{t}$ background constrained
in situ by a b-tag control region in the simultaneous fit, obtaining
$\hat{\mu}=1.20\pm1.13$ and an observed 95\% confidence-level upper limit of
$\mu<3.72$, in agreement with the Standard Model. The analysis is a
transparent, fully reproducible, dual-engine-validated reference
implementation of a categorized profile-likelihood Higgs measurement built
entirely from public data with open-source tools~\cite{pyhf,CMSCombine}.
Its precision is set by the single channel and the partial dataset rather
than by the method; the same pipeline applied to additional final states
and the full luminosity would scale directly toward the published
sensitivity, while remaining open and independently verifiable at every
step. That an independent open-data reproduction lands within $0.16$
standard deviations of the published CMS $\mu\tau_{h}$-channel signal
strength, $\mu=1.01\pm0.41$, illustrates how open data and published
likelihoods turn the reproduction of frontier measurements into a routine
and transparent exercise.

\medskip
\noindent\textbf{Acknowledgments.}\enspace
We thank the CMS Collaboration for releasing the 2012 collision data and
the associated simulated samples through the CERN Open Data Portal under the
CC0 waiver. This is an independent analysis of CMS Open Data and is not
endorsed by or affiliated with the CMS Collaboration. We acknowledge the
developers of the \textsc{pyhf}, \textsc{Combine}, and \textsc{rhalphalib}
tools.

\printbibliography[heading=subbibliography,title={References}]
\end{refsection}

\newpage
\begin{refsection}[appendices_pub/aleph_lund_plane/references.bib]
\section{First Measurement of the Primary Lund Jet Plane Density in Hadronic $Z$ Decays}\label{app:lund_pub}

\graphicspath{{appendices_pub/aleph_lund_plane/figures/}}


\begingroup
\leftskip=2em \rightskip=2em
\small
\noindent\textbf{Abstract.}\enspace
We report the first measurement of the primary Lund jet plane density in
$e^+e^-$ collisions, using archived ALEPH data at the $Z$ pole
($\sqrt{s}=91.2$~GeV). Thrust hemispheres are reclustered with the $e^+e^-$
Cambridge/Aachen algorithm, declustered along the harder branch, and corrected
to charged-particle level by two-dimensional iterative Bayesian unfolding. At
the $Z$ pole the density is free of the underlying event, multiple-parton
interactions, and pileup, isolating pure quark-initiated ($C_F$) QCD radiation
and hadronization, and the perturbative bulk traces the
expected $(2/\pi)\,C_F\,\alpha_s(k_t)$ running-coupling form. The mean
primary-emission multiplicity per hemisphere is
$\langle N\rangle = 4.751\pm0.224$, a systematics-limited $4.7\%$. The result is
robust against the unfolding prior: re-unfolding with modern-generator priors
moves $\langle N\rangle$ by $<0.02\%$, a fully self-consistent operator rebuild
by $<0.7\%$, and an independent singular-value-decomposition unfolding agrees to
$0.19\%$. Used as a differential benchmark, the density lies $9$--$18\%$ above
all modern parton-shower plus hadronization predictions; this offset is
\emph{predominantly an overall normalization}, as a single rescaling
$k_g=1.10$--$1.17$ per generator removes the entire integrated excess (consistent
with a density-level multiplicity retune). A residual differential structure---a
shallow U-shape in $\ln k_t$---is resolved by the data but lies within current
generator-theory uncertainties: after the normalization and an
analytically motivated per-bin theory band, $\chi^2/\mathrm{ndf}\simeq0.4$--$0.6$.
With realistic prediction uncertainties the PYTHIA~8 tunes are statistically
consistent with the data, while the Sherpa cluster-hadronization model remains
the most discrepant. The measurement provides a contamination-free, fixed-$C_F$
quark reference and a differential target for the soft, hadronization-dominated
region of the plane.
\par
\endgroup

\subsection{Introduction}
The Lund jet plane (LJP) turns the radiation pattern
inside a jet into a near-uniform map of QCD splittings, separating onto a single
plane the perturbative parton shower, the running of the strong coupling, and
nonperturbative hadronization---mechanisms that are otherwise entangled in
inclusive observables~\cite{dreyer2018}. It is a two-dimensional representation
of the phase space of $1\to2$ QCD splittings inside a jet. Reclustering a jet's
constituents with the angular-ordered Cambridge/Aachen (C/A) algorithm and
reversing the clustering history maps each primary splitting to a point whose
coordinates encode the transverse momentum $k_t$ of the softer prong relative to
the harder one and the opening angle $\Delta\theta$ between them. Because QCD is
approximately scale invariant, in the soft-and-collinear (eikonal) limit the
primary emissions populate the plane nearly uniformly, and for independent
soft-collinear emissions the density is directly proportional to the running
coupling,
\begin{equation}
\rho(k_t,\Delta\theta) \;\approx\; \frac{2}{\pi}\,C_R\,\alpha_s(k_t),
\label{lpub:eq:rho-alphas}
\end{equation}
with $\alpha_s(k_t)$ the strong coupling evaluated at the emission scale and
$C_R$ the color factor of the emitter ($C_F=4/3$ for a
quark)~\cite{cms2024ljp}. Different regions of the plane are thus governed by
distinct physics: hard wide-angle radiation, the running of $\alpha_s$,
hadronization at low $k_t$, and---at hadron colliders only---the underlying
event (UE), multiple-parton interactions (MPI), initial-state radiation, and
pileup near the jet boundary~\cite{dreyer2018,atlas2020ljp,cms2024ljp}.

Hadronic $Z$ decays at the $Z$ pole provide an exceptionally clean environment
for this observable. There are no beam remnants and no parton distribution
functions; on the resonance peak initial-state radiation is strongly suppressed;
and there is no pileup. The entire soft, wide-angle corner that the UE
contaminates in $pp$ collisions is, in $e^+e^-$, populated by genuine QCD
radiation and hadronization alone. The color content is equally distinctive: the
$e^+e^-\to Z\to q\bar q$ hemispheres are quark dominated, so the density samples
almost pure $C_F$ radiation. The measurement is therefore the cleanest available
differential, fixed-$C_F$ reference for the soft, hadronization-dominated region
of the plane---precisely where modern parton-shower and hadronization models,
tuned on LEP/SLD integrated event shapes and fragmentation
spectra~\cite{monash2014}, are least directly constrained by existing $e^+e^-$
data.

All published primary-LJP-density measurements use $pp$ collisions: ALICE at
intermediate jet $p_T$, reaching $k_t\approx5$~GeV~\cite{alice2021ljp}; CMS at
high $p_T$ in the same $\ln k_t$ plane~\cite{cms2024ljp}; and ATLAS, differential
in $\ln(1/z)$ rather than $\ln k_t$~\cite{atlas2020ljp}. No prior $e^+e^-$
LJP-\textit{density} measurement exists. The same archived ALEPH dataset
underlies a published SoftDrop substructure
measurement~\cite{aleph2022softdrop} and an energy-energy-correlator
analysis~\cite{bossi2025eec}; sharing the identical detector and Monte Carlo,
the latter provides our correction-chain and systematic-magnitude template. The
generator-independent analytic benchmark is the Lifson--Salam--Soyez (LSS)
next-to-leading-log (NLL) plus next-to-leading-order calculation of the primary
LJP density~\cite{lifson2020ljp}.

\subsection{Data and method}
The analysis uses the archived ALEPH LEP1 data and
the corresponding detector-simulated Monte Carlo from the ALEPH open-data
archive. Only the 1994 run carries a full detector simulation---a single sample
of hadronic $Z$ decays generated with PYTHIA~6.1 in the ALEPH tune under 1994
conditions---so, following the same-dataset EEC analysis~\cite{bossi2025eec}, the
baseline corrected measurement uses the 1994 peak data. After the peak energy
window ($|E-M_Z|<0.5$~GeV, $M_Z=91.188$~GeV~\cite{pdg2024z}) and the standard
ALEPH hadronic-$Z$ selection, the sample comprises $1{,}293{,}167$ events
($2{,}586{,}334$ hemispheres). Non-$q\bar q$ contamination is below
$0.6\%$~\cite{aleph_ew1999}, dominated by $\tau^+\tau^-$ and two-photon events.

For each event the charged-particle thrust axis defines the plane that splits the
event into two hemispheres. The charged particles of each hemisphere are
reclustered with the $e^+e^-$ generalized-$k_t$ algorithm with $p=0$ (the C/A,
angular-ordered metric) in
FastJet~\cite{cacciari2012fastjet,dokshitzer1997,wobisch1999}. The clustering
history is reversed; at each step the harder prong $j_1$ is followed and the
softer prong $j_2$ is recorded as a primary
emission~\cite{dreyer2018,atlas2020ljp,cms2024ljp} at coordinates
\begin{equation}
\Delta\theta = \angle(j_1,j_2), \qquad k_t = p_{j_2}\,\sin\Delta\theta ,
\label{lpub:eq:coords}
\end{equation}
i.e.\ the emitter-relative $k_t$ of the original Lund proposal and its analytic
calculation~\cite{lifson2020ljp,dreyer2018}. The primary LJP density is
\begin{equation}
\rho(k_t,\Delta\theta) \;\equiv\; \frac{1}{N_{\rm hem}}
\frac{\mathrm{d}^2 N_{\rm emissions}}{\mathrm{d}\ln(k_t/\mathrm{GeV})\,\mathrm{d}\ln(1/\Delta\theta)},
\label{lpub:eq:density}
\end{equation}
normalized to the number of hemispheres $N_{\rm
hem}$~\cite{atlas2020ljp,cms2024ljp}: $\rho$ is the average number of primary
emissions per hemisphere per unit area of the $(\ln 1/\Delta\theta,\,\ln k_t)$
plane, an inclusive count over \emph{all} primary splittings. In the perturbative
bulk it follows the running-coupling form $(2/\pi)\,C_F\,\alpha_s(k_t)$ of
Eq.~(\ref{lpub:eq:rho-alphas}), rising toward lower $k_t$ as $\alpha_s(k_t)$ grows and
turning over at the hadronization scale $k_t\sim\Lambda_{\rm QCD}$, where the
perturbative picture breaks down; its integral over the plane is the average
number of primary emissions per hemisphere,
$\langle N\rangle\equiv\langle N_{\rm emissions}\rangle$. The measurement
is corrected to a charged-particle-level definition---stable hadrons and leptons
with $c\tau>10$~mm, following the ATLAS
convention~\cite{atlas2020ljp,atlas2019frag}.

The plane is binned uniformly in $\ln(1/\Delta\theta)\in[0,5.5]$ and
$\ln(k_t/\mathrm{GeV})\in[-4,3]$ with width $0.5$. The detector-level emission
spectrum is corrected with two-dimensional iterative Bayesian (D'Agostini)
unfolding~\cite{dagostini1995}, using a response matrix, efficiency, purity, and
prior built from the PYTHIA~6.1 simulation; a bin-by-bin split-sample correction
serves as an independent cross-check. Because the response is strongly diagonal
(global diagonal fraction $0.944$), a single iteration is used. The chain is
validated by a split-sample closure test, a graded stress test that demonstrates
few-percent resolving power, and the bin-by-bin cross-check, which agrees with
the unfolding baseline to $0.10\%$ in $\langle N\rangle$ on the full data. We
report a $57$-bin fiducial region---bins with genuine
density ($\rho>0.05$) and reliable precision (total relative uncertainty below
$25\%$)---which also keeps the total covariance well conditioned (condition
number $2.4\times10^5$).

\subsection{Systematic uncertainties}
Each of ten systematic sources is propagated
through the full correction chain as a bin-dependent shift rather than a flat
percentage, a systematic being defined as a change in the correction operator
applied to fixed pseudo-data. The dominant source is the prior/model dependence
of the unfolding ($3.0\%$, density-weighted), evaluated from a data-driven
reconstruction-to-data reweighting together with a generator-bracketed gen-level
envelope spanning the PYTHIA~8 string variants and the Sherpa cluster model. It
is large because only one detector-simulated generator exists and the data
fragment slightly harder than PYTHIA~6.1. Next come the archived per-track
energy-flow weight ($2.7\%$) and the binning/projection non-closure ($2.2\%$);
detector effects (tracking, resolution, thrust-axis definition, matching) each
remain below $2\%$, reflecting the clean $e^+e^-$ environment. There is
\textit{no} UE, MPI, or pileup systematic---these do not apply at the $Z$ pole.
The marginal split-sample closure ($\chi^2/\mathrm{ndf}=2.44$, density-weighted
residual $\sim0.4\%$, below the $\chi^2/\mathrm{ndf}>3$ hard alarm) is carried
honestly as a $0.4\%$ correction-bias systematic, following ATLAS/CMS
practice~\cite{atlas2020ljp,cms2024ljp}. The total density-weighted uncertainty
on the reported region is $6.0\%$, and the integrated $\langle N\rangle$
uncertainty is $0.224$ ($4.7\%$); the prior/model term alone accounts for about
a quarter of the total variance, with no single detector source dominant. The
precision is modest and prior-dominated: this is a systematics-limited shape
measurement, and we state it as such.

\subsection{Results}
\begin{figure}[t]
  \centering
  \includegraphics[width=0.48\linewidth]{fig1a_density_2d.pdf}\hfill
  \includegraphics[width=0.48\linewidth]{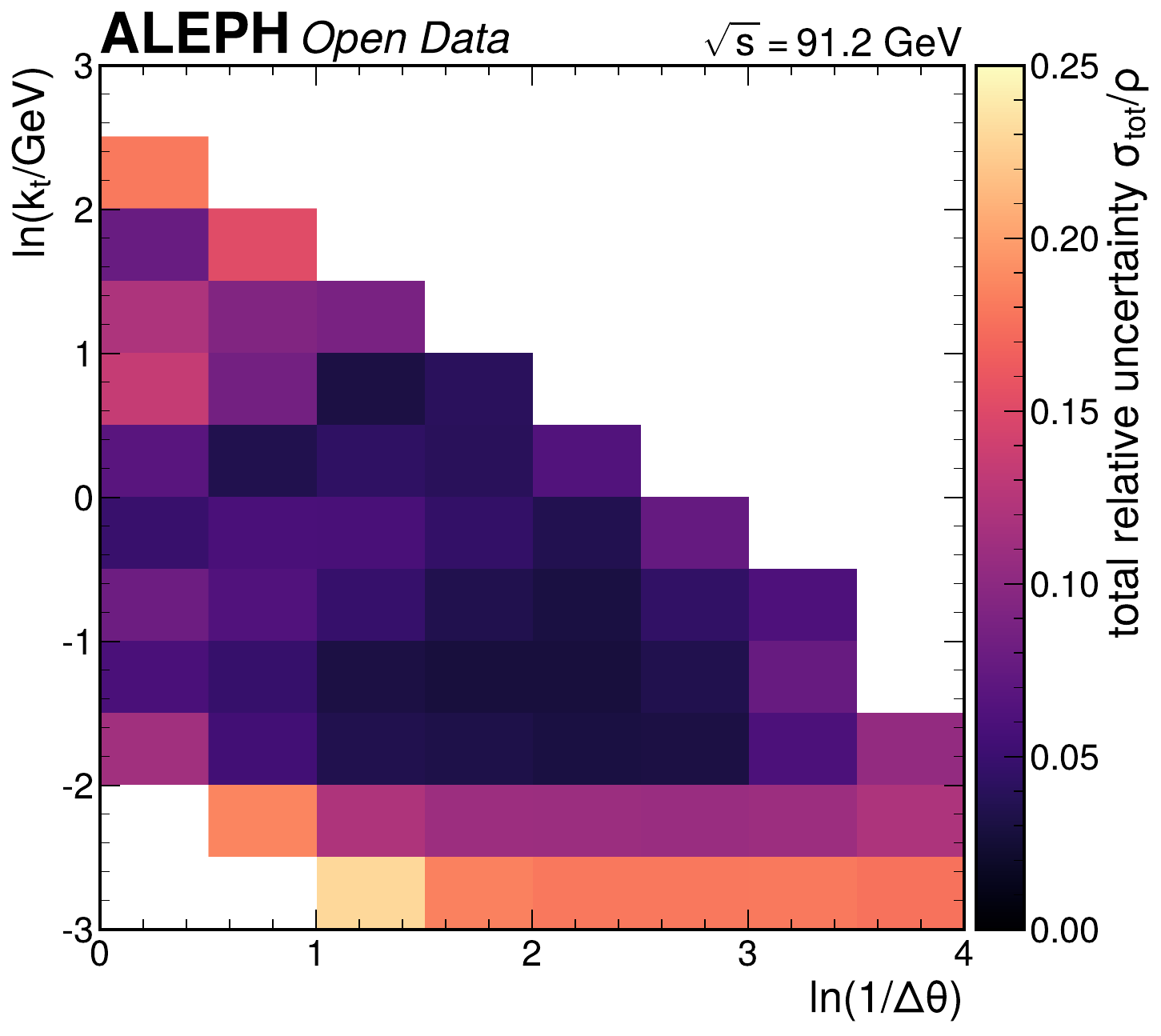}
  \caption{Corrected primary Lund jet plane density
  $\rho(\ln 1/\Delta\theta,\,\ln k_t)$ at charged-particle level (left), measured on the
  1994 ALEPH peak data, and the matching map of the total relative
  uncertainty $\sigma_{\rm tot}/\rho$ over the same bins (right). The density displays the
  canonical triangular structure; the central perturbative region is measured to
  a few percent, while the larger uncertainties are confined to the kinematic
  edges. The soft, wide-angle corner (upper left of the populated region) carries
  no underlying-event, MPI, or pileup contamination and is therefore pure QCD
  radiation and hadronization.}
  \label{lpub:fig:money}
\end{figure}

The corrected charged-particle-level density is shown in
Fig.~\ref{lpub:fig:money} (left), with the matching map of the total relative uncertainty
in Fig.~\ref{lpub:fig:money} (right). It exhibits the canonical triangular Lund structure,
the kinematically forbidden collinear-hard corner empty. Across the bulk of the
plane near $\ln k_t\approx-0.7$ ($k_t\approx0.5$~GeV) the eikonal approximation
holds and the density follows the running coupling of
Eq.~(\ref{lpub:eq:rho-alphas}), rising toward lower $k_t$ as $\alpha_s(k_t)$ grows and
turning over below $k_t\sim\Lambda_{\rm QCD}$. The soft, wide-angle corner that
is UE/MPI/pileup-contaminated in every $pp$ measurement is here pure QCD
radiation and hadronization. The uncertainty map confirms that the central
perturbative region is measured to a few percent, the larger uncertainties
confined to the kinematic edges.

The average number of primary emissions per hemisphere over the reported region
is
\begin{equation}
\langle N\rangle = 4.751 \pm 0.224 \quad (4.7\%),
\label{lpub:eq:nemi}
\end{equation}
with the uncertainty decomposed as
$0.0014\,(\mathrm{stat})\oplus0.224\,(\mathrm{syst})$: the measurement is
overwhelmingly systematics limited, as expected for a shape corrected with a
single detector-simulated generator. The median per-bin total relative
uncertainty is $6.4\%$. Nothing is tuned to the data---the correction operator
and the systematic shifts are the values fixed before unblinding, and only the
input spectrum and statistical covariance change. The corrected density is
compatible with the expectation evaluated on the correction Monte Carlo
(diagonal $\chi^2/\mathrm{ndf}=0.33$; worst per-bin pull $1.10\sigma$; zero bins
above $2\sigma$). It sits $1.33\%$ below the PYTHIA~6.1 charged-particle truth
($4.815$)---the harder-fragmentation difference of the legacy ALEPH tune,
covered by the prior/model systematic.

The result is robust against the unfolding methodology. Re-unfolding the same
data with the prior swapped from PYTHIA~6.1 to each modern-generator truth shape
moves $\langle N\rangle$ by at most $0.02\%$; rebuilding the \emph{entire}
correction operator---response, efficiency, and purity---self-consistently to
each generator's particle-level truth moves it by at most $0.67\%$, an order of
magnitude below the $9$--$18\%$ data-vs-generator offset and below the existing
$3.0\%$ prior/model systematic. The deficit is therefore not an artifact of the
single legacy PYTHIA~6.1 prior. An independent singular-value-decomposition
unfolding~\cite{hocker1996svd} of the same response and input reproduces the
baseline iterative result to $0.19\%$ in $\langle N\rangle$ and a $0.10\%$ median
per bin---a third unfolding leg, alongside the bin-by-bin split-sample
cross-check that agrees to $0.10\%$. The corrected density is thus stable under
the choice of prior, the construction of the correction operator, and the
unfolding algorithm itself.

\subsection{Generator benchmark}
\begin{figure}[t]
  \centering
  \includegraphics[width=0.49\linewidth]{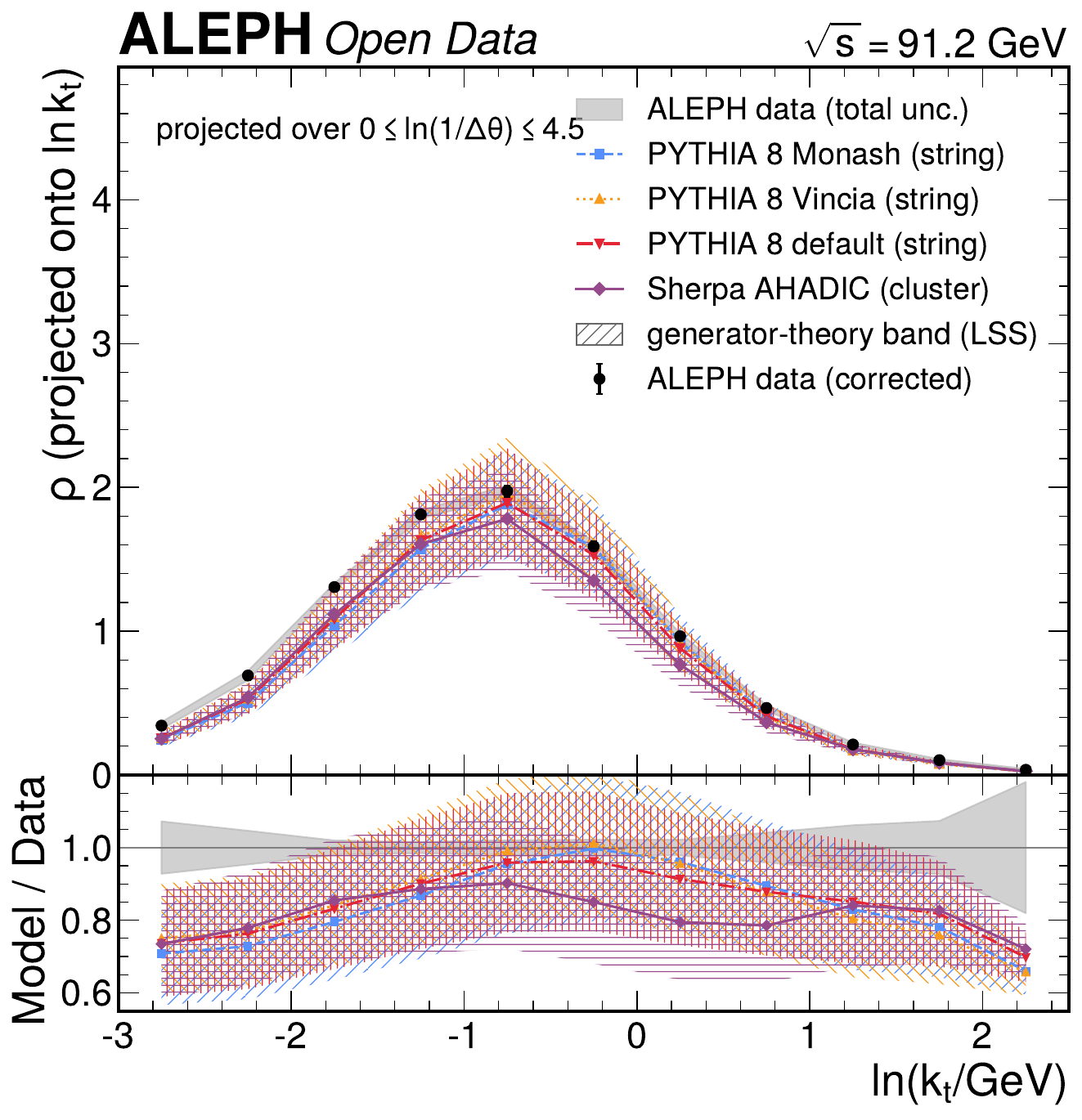}\hfill
  \includegraphics[width=0.49\linewidth]{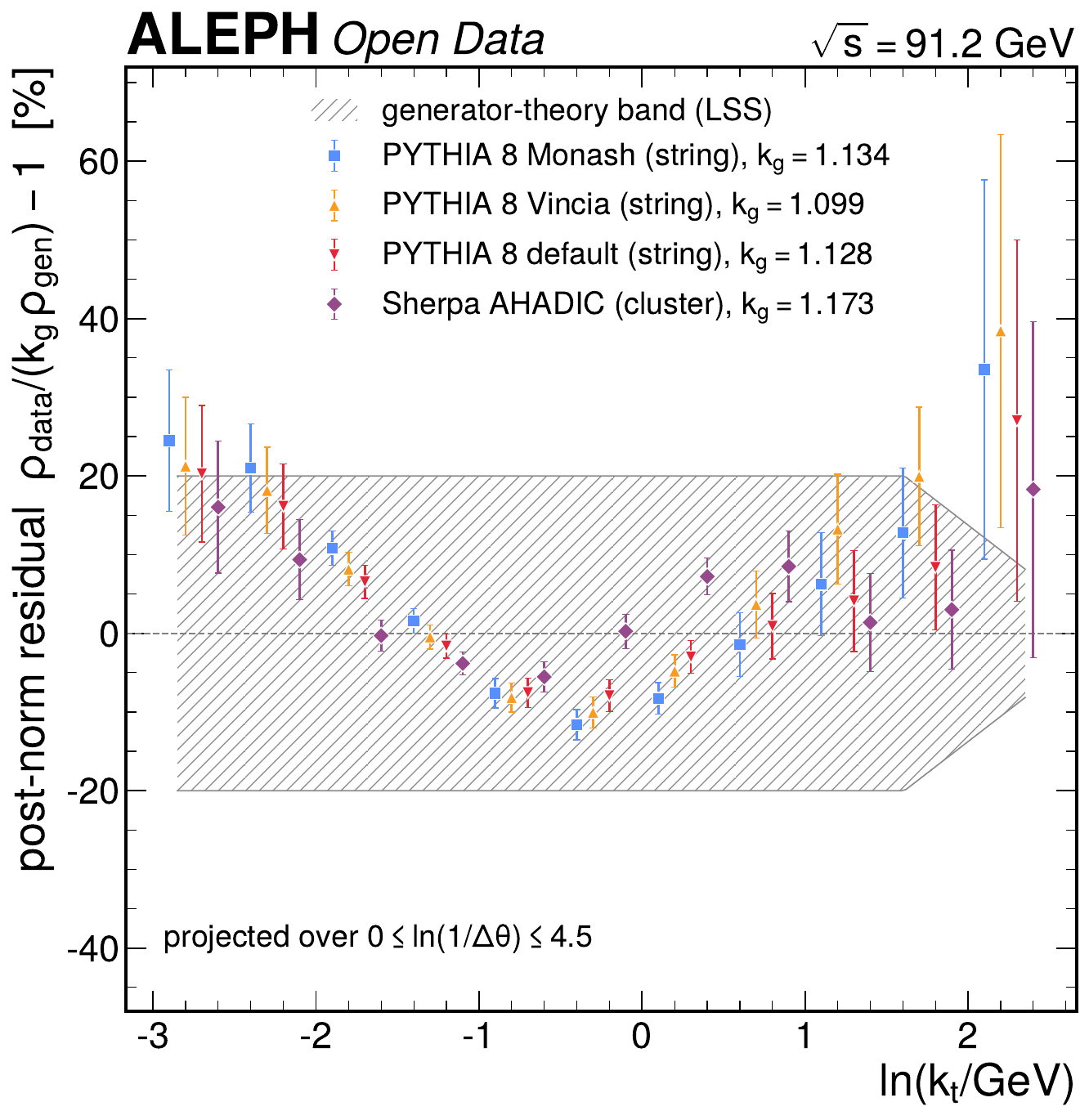}
  \caption{Corrected data (left; points, grey total-uncertainty band) versus the
  standalone PYTHIA~8 Monash, Vincia, and default string tunes and the Sherpa
  AHADIC cluster prediction, with the density projected onto $\ln k_t$ over the
  fiducial $\ln(1/\Delta\theta)$ range; the lower panel shows the model/data
  ratio, the grey band giving the data total uncertainty about unity. The shaded
  colored bands are the assumed per-bin generator-theory uncertainty, scaling from
  $\sim6\%$ at high $k_t$ to $\sim20\%$ at the soft edge following the
  next-to-leading-log LSS calculation~\cite{lifson2020ljp}. All modern tunes lie
  $9$--$18\%$ below the data (occupancy-weighted, like-for-like), the deficit
  largest for the Sherpa cluster model, and the rise toward lower $k_t$ reflects
  the running of $\alpha_s(k_t)$. Post-normalization residual
  $\rho_{\rm data}/(k_g\,\rho_{\rm gen})-1$ (right, in percent), projected onto $\ln k_t$
  over the same range after rescaling each generator by its single best-fit
  normalization $k_g$ (legend); error bars are the projected data total
  uncertainty and points are staggered horizontally for clarity, and the grey band
  is the same per-bin generator-theory uncertainty centered on zero. After the
  normalization the residual is a shallow U-shape rising toward both $k_t$ ends,
  resolved in the data covariance ($\chi^2/\mathrm{ndf}=3.4$--$6.7$) but contained
  within the theory band ($\chi^2/\mathrm{ndf}=0.4$--$0.6$ after the band,
  $p\simeq1$): the offset is predominantly an overall normalization plus a
  differential shape within current generator-theory uncertainty, within which the
  PYTHIA~8 tunes are statistically consistent with the data and Sherpa remains the
  most discrepant.}
  \label{lpub:fig:generators}
  \label{lpub:fig:postnorm}
\end{figure}

As a differential benchmark, we compare the
corrected density, projected onto $\ln k_t$, with standalone particle-level
predictions: the PYTHIA~8 Monash~\cite{monash2014}, Vincia, and default
tunes~\cite{pythia8}, all using Lund-string hadronization (Vincia is an
alternative antenna shower), and Sherpa~2.2.16 with the AHADIC
cluster-hadronization model~\cite{sherpa}---the same cluster-model family ATLAS
used as its string-versus-cluster anchor~\cite{atlas2020ljp}.
Figure~\ref{lpub:fig:generators} (left) shows that all modern tunes lie below the data across
the spectrum. The integrated, occupancy-weighted excess over the measured region
is $+12.4\%$ (Monash), $+9.0\%$ (Vincia), $+11.7\%$ (default), and $+17.7\%$
(Sherpa); the mean unweighted per-bin excess is larger, $\sim20$--$24\%$, because
it up-weights the lower-occupancy soft and hard edges relative to the central
bins. We quote the occupancy-weighted, like-for-like figure (data and each
generator over the \emph{same} fiducial region) as the benchmark. The excess
comfortably exceeds the dominant prior/model systematic ($3.0\%$
density-weighted, $6$--$9\%$ in the perturbative region), whose magnitude is
corroborated by the same-detector energy-energy-correlator
analysis~\cite{bossi2025eec}; together with the prior-independence checks above,
this establishes the offset as a genuine feature of the data rather than an
unfolding artifact.

This offset is \emph{predominantly an overall normalization}. A single best-fit
rescaling $k_g$ of each prediction---the inverse-variance-weighted estimator,
equal to the integral ratio to better than $1\%$---takes the value
$k_g=1.134$ (Monash), $1.099$ (Vincia), $1.128$ (default), and $1.173$ (Sherpa),
and removes the \emph{entire} integrated excess and $44$--$60\%$ of the diagonal
$\chi^2$. This is the behavior expected if the generators share a single
density-level multiplicity offset, retunable without altering the differential
shape---e.g.\ a global fragmentation-multiplicity adjustment. What remains after
the rescaling is a genuine differential residual: a shallow U-shape in $\ln k_t$,
the post-norm excess rising toward both the soft ($k_t\lesssim0.2$~GeV) and the
hard ($k_t\gtrsim2$~GeV) ends and passing through zero in the perturbative
plateau ($k_t\sim1$~GeV). This residual is resolved in the measurement's own
covariance (diagonal $\chi^2/\mathrm{ndf}=3.4$--$6.7$ after the norm) but is
small relative to the predictions' theory uncertainty:
Fig.~\ref{lpub:fig:postnorm} (right) overlays the post-norm residual with an
analytically motivated per-bin generator-theory band, scaling from $\sim6\%$ at
high $k_t$ to $\sim20\%$ at the soft edge following the next-to-leading-log LSS
calculation~\cite{lifson2020ljp}, and the residual U-shape sits largely within
it. With the normalization absorbed and this band applied, the residual-shape
$\chi^2/\mathrm{ndf}=0.48$ (Monash), $0.45$ (Vincia), $0.35$ (default), and
$0.56$ (Sherpa), all with $p\simeq1$. We therefore classify the comparison as
norm-plus-shape: a coherent multiplicity offset plus a differential structure
that the measurement resolves but that lies within present generator-theory
uncertainty.

The same band controls the goodness of fit before any normalization is removed.
Compared as central values, the nominal diagonal $\chi^2/\mathrm{ndf}=9.9$
(Monash), $6.9$ (Vincia), $8.2$ (default), and $14.7$ (Sherpa); these are upper
bounds, computed without the predictions' theory uncertainties. Adding an
incoherent $10\%$ per-bin band reduces them to $2.0$--$4.0$, and the LSS
$k_t$-dependent band to $0.9$--$1.5$. At the LSS band the three PYTHIA~8 tunes are
statistically consistent with the data ($p=0.16$--$0.67$), while Sherpa AHADIC
remains marginally inconsistent ($p\simeq0.01$). The ordering---Sherpa most
discrepant---survives at every band level. The Sherpa deficit is the largest
single feature, but the comparison varies generator, shower, and hadronization
together; a clean isolation of the hadronization model requires a same-generator
string-versus-cluster swap, which we discuss below. We do \emph{not} read the
comparison as excluding any generator at present precision: it is a differential
target, and a more precise soft-corner measurement or a tighter theoretical
prediction would sharpen it.

The systematic covariance must be handled with care here. Built from the shifts
of a single detector-simulation generator, it is dominated by a coherent
normalization mode: the top eigenvector is fully same-sign, carries $66\%$ of the
total variance, and overlaps the density-normalization direction with
$|\cos|=0.93$ (condition number $2.4\times10^5$). A full-covariance
generalized-least-squares one-parameter rescaling is then biased---it returns
$k_g=0.32$--$0.68$, pushing the rescaled prediction \emph{below} the data, the
wrong direction---a textbook instance of Peelle's Pertinent
Puzzle~\cite{peelle1987,dagostini1994}; the same fit on the statistical
covariance alone recovers the physical $k_g=1.10$--$1.16$. The robust comparison
object is therefore the diagonal, inverse-variance metric used above. We release
the full bin-to-bin covariance as machine-readable supplemental material so the
comparison can be recomputed.

The $k_t$ dependence of the same density traces the running-coupling behavior of
Eq.~(\ref{lpub:eq:rho-alphas}): the density rises toward lower $k_t$ as $\alpha_s(k_t)$
grows, the expected $(2/\pi)C_F\alpha_s(k_t)$ shape, visible here in $e^+e^-$ as
it is in $pp$. We report this as a shape consistency and do not attempt an
absolute $\alpha_s$ extraction, for which a dedicated $e^+e^-$ thrust-hemisphere
LJP-density calculation---not yet available---would be required.

\subsection{The residual differential structure}
Figure~\ref{lpub:fig:postnorm} (right) shows
the post-normalization residual $\rho_{\rm data}/(k_g\,\rho_{\rm gen})-1$ projected
onto $\ln k_t$, after the single best-fit rescaling $k_g$ of each generator. Once
the multiplicity offset is absorbed, what remains is a coherent differential
structure rather than a flat zero: a shallow U-shape, the residual rising toward
both the soft and the hard $k_t$ ends and dipping slightly below zero in the
perturbative plateau, a single multiplicative constant cannot reproduce it. The
shape is generator-robust in this sense, though its detail varies between tunes,
and we caution against over-reading the highest-$k_t$ bins, which are one to three
emissions wide with $\sim25\%$ uncertainties. The defensible statement is the
U-shape itself, not a localization in any single corner. Crucially, the residual
sits largely within the analytically motivated generator-theory band (grey):
after the normalization and this band the residual-shape
$\chi^2/\mathrm{ndf}=0.4$--$0.6$ with $p\simeq1$, so the measurement
\emph{resolves} the differential structure---its median per-bin precision of
$6.4\%$ is finer than the band in the bulk---but does not significantly disfavor
the generators relative to theory. The soft, hadronization-dominated region that
is contamination-free in $e^+e^-$ but swamped by the underlying event,
multiple-parton interactions, and pileup near the jet boundary in every $pp$
measurement is exactly where this residual and the predictions' theory band are
largest, and is where a more precise measurement would have the most leverage.

\subsection{Discussion}
The Sherpa AHADIC cluster model is the most discrepant
generator at every level of the comparison, which raises whether this isolates a
hadronization-model (string-versus-cluster) effect. It does not: the comparison
varies the generator, the parton shower, and the hadronization model together
(Sherpa versus PYTHIA). A clean test requires a same-generator hadronization swap
with the shower held fixed, infeasible in-house as the available Sherpa~2.2.16
build lacks the Lund-string interface. The published clean swap---ATLAS,
Sherpa~2.2.5 with the dipole shower fixed, AHADIC versus Lund
string~\cite{atlas2020ljp}---finds a hadronization-only effect of only
$\sim5.5\%$ in the mean primary-emission multiplicity, comparable to both the
confounded Sherpa-versus-PYTHIA gap here ($3.5$--$6.4\%$ in $\langle N\rangle$)
and to the intra-string shower/tune spread ($3.1\%$). Hadronization is therefore
not cleanly separable from the shower and tune at this level: we report Sherpa
AHADIC as the most discrepant generator, with the confound stated, and do not
claim a string-versus-cluster discrimination.

A data-driven heavy-versus-light-flavor cross-check, using a displaced-track
lifetime tag applied identically to data and Monte Carlo, finds the $b$-enriched
hemispheres $32.4\%$ higher in reconstruction-level $\langle N\rangle$ ($4.948$
versus $3.739$), with a localized dead-cone shape effect. This is a tagged-sample
comparison, not a flavor-corrected density, and does not affect the baseline. The
1992, 1993, and 1995 corrected $\langle N\rangle$ agree with 1994 to within
$0.11\%$, confirming stability across the LEP1 periods; the 1994 peak remains the
baseline under the Monte-Carlo-coverage caveat.

A quantitative comparison to the published $pp$ Lund-plane
densities~\cite{atlas2020ljp,cms2024ljp,alice2021ljp} is precluded by their
differing jet definitions, energies, and quark/gluon mixtures: the
quark-dominated $e^+e^-$ density is expected to lie below the gluon-enriched $pp$
plateau by a color factor approaching $C_A/C_F=9/4\approx2.25$~\cite{dreyer2022qg},
so the two cannot be overlaid as a consistency test. The present measurement cedes
high-$k_t$ reach and substructure/tagging applications to the LHC; its unique
contribution is the environment and color content---a contamination-free,
fixed-$C_F$ quark reference for the soft, hadronization-dominated region.

\subsection{Conclusions}
We have presented the first measurement of the primary
Lund jet plane density in $e^+e^-$ collisions, using archived ALEPH LEP1 data at
$\sqrt{s}=91.2$~GeV, with $\langle N\rangle = 4.751\pm0.224$ ($4.7\%$,
systematics limited). The density shows the canonical triangular Lund structure
and, in the perturbative bulk, the expected $(2/\pi)\,C_F\,\alpha_s(k_t)$
running-coupling form---a clean, contamination-free, fixed-$C_F$ quark reference
for the soft, hadronization-dominated corner that the underlying event obscures
in $pp$. It is robust against the unfolding methodology: prior swaps move
$\langle N\rangle$ by $<0.02\%$, a full self-consistent operator rebuild by
$<0.7\%$, and an independent SVD unfolding agrees to $0.19\%$. Used as a
differential benchmark, the density lies $9$--$18\%$ above all modern
parton-shower plus hadronization predictions, but this offset is predominantly an
overall normalization: a single rescaling $k_g=1.10$--$1.17$ per generator removes
the entire integrated excess, consistent with a density-level multiplicity retune.
The residual differential structure---a shallow U-shape in $\ln k_t$---is resolved
by the data yet lies within current generator-theory uncertainty
($\chi^2/\mathrm{ndf}\simeq0.4$--$0.6$ after normalization and an LSS-style theory
band), so the PYTHIA~8 tunes are statistically consistent with the data while
Sherpa AHADIC remains the most discrepant. We frame this as a benchmark, not a
generator-excluding constraint: a differential target for the soft,
non-perturbative Lund-plane region that a more precise soft-corner measurement or
a tighter $e^+e^-$ thrust-hemisphere LJP-density calculation would sharpen.

\medskip
\noindent\textbf{Acknowledgments.}\enspace
We thank the ALEPH Collaboration for making the LEP1 data publicly available,
and the maintainers of the ALEPH open-data archive on which this analysis is
based.

\printbibliography[heading=subbibliography,title={References}]
\end{refsection}

\newpage
\section{Initial agent prompts}\label{app:prompts}

The natural-language prompts used to initiate each agent run listed in Table~\ref{tab:analyses} are reproduced verbatim below. Each was issued once to the autonomous pipeline, with no human input prior to the unblinding gate. The prompt for the ALEPH Lund jet plane analysis is reproduced in Section~\ref{sec:results} as a worked example and is not repeated here.

\paragraph{CMS Open Data $H\to \tau^+\tau^-$ (\href{https://github.com/jfc-mit/analysis_cms_higgs_tautau}{Github}).}
\begin{quote}
    \footnotesize
You are performing a Higgs boson search in the $\tau^+\tau^-$ decay channel using CMS Open Data from 2012 at $\sqrt{s}$ = 8 TeV.

The final state is one muon and one hadronically decaying tau lepton ($\mu \tau_h$). Your goal is to produce distributions of key observables --- particularly the visible di-tau mass --- showing the Higgs signal contribution on top of Standard Model backgrounds. This loosely follows the official CMS publication (Phys. Lett. B 779 (2018) 283 and JHEP 05 (2014) 104).

To optimize the Higgs signal selection, do some categorization, in particular add a VBF category, and be sure to fit all categories simultaneously. Our dataset is missing the full trigger and tau efficiency scale factors, so loosen the tau efficiency selection to $10$--$15\%$ to allow for good agreement of the Drell-Yan peak.

In addition to a baseline analysis, perform three more approaches that focus on fitting a different final observable: (a) fit an NN discriminator to find the Higgs, propagating the systematic uncertainties in the fit; (b) train an NN to regress the direction of the genMET and genMET $\phi$ in the final state and use this to construct a combined mass with the visible objects as the final observable; and (c) fit the mass after adding the missing energy to the mass distribution.

Also, make sure to apply a tight anti-muon veto, and to normalize the W+jets sample from data using the high $m_T$ region. Make sure you put a tight anti-muon veto on the hadronic tau ID. Make sure your $Z$ normalization uncertainty is larger than the MC prediction, at a level of $10$--$15\%$, to reflect missing trigger turn-on scale factors and a larger tau efficiency scale factor.

Data: [...]
\end{quote}

\paragraph{Z lineshape and $\alpha_s$ extraction, ALEPH (\href{https://github.com/jfc-mit/analysis_aleph_z_lineshape}{Github}).}
\begin{quote}
    \footnotesize
Measure the Z lineshape parameters ($M_Z$, $\Gamma_Z$, $\sigma^0_\text{had}$), the number of light neutrino species $N_\nu$ from the invisible width, and extract $\alpha_s(M_Z)$ from the hadronic-to-leptonic width ratio $R_\ell$, using archived ALEPH data at $\sqrt{s} \sim 91.2$ GeV.

Classify events as hadronic, $e^+e^-$, $\mu^+\mu^-$, or $\tau^+\tau^-$. $R_\ell$ is a counting ratio (luminosity cancels). $N_\nu$ comes from the invisible width. $\alpha_s$ comes from $R_\ell$ via the N4LO QCD correction. If scan data at multiple $\sqrt{s}$ points is available, fit the lineshape.

Data: [...]
\end{quote}

\paragraph{$R_b$, $R_c$, and $A_\text{FB}^b$, ALEPH (\href{https://github.com/jfc-mit/analysis_aleph_z_heavy_flavour}{Github}).}
\begin{quote}
    \footnotesize
Scaffold and run a measurement analysis of the partial width ratios $R_b$ = $\Gamma(Z\to b\bar{b})/\Gamma(Z\to \mathrm{hadrons})$ and $R_c$ = $\Gamma(Z\to c\bar{c})/\Gamma(Z\to \mathrm{hadrons})$ in hadronic Z decays using archived ALEPH data at $\sqrt{s} \sim 91.2$~GeV.

Setup: scaffold analyses/rb\_rc as a measurement, set data\_dir=[...] in .analysis\_config, install the pixi environment, then begin orchestrating.

Observable: $R_b$ and $R_c$ are the fractions of hadronic Z decays to $b\bar{b}$ and $c\bar{c}$ respectively. Tag b and c events using lifetime-based methods: signed impact parameter significance of charged tracks (\texttt{pwflag} == 0, \texttt{highPurity} == 1) relative to the primary vertex, and/or secondary vertex reconstruction. Use a double-tag method to reduce dependence on MC tagging efficiency.

Deliverables:
\begin{enumerate}
   \item $R_b$ and $R_c$ with statistical and systematic uncertainties.
   \item b-tagging and c-tagging performance (efficiency, purity, mistag rates) validated against MC.
   \item Double-tag consistency checks.
   \item Comparison to published ALEPH $R_b$/$R_c$ values and the SM prediction.
   \item Machine-readable results.
\end{enumerate}

Data: [...]
\end{quote}

\paragraph{Energy-energy correlator, ALEPH (\href{https://github.com/jfc-mit/analysis_aleph_eec_correlators}{Github}).}
\begin{quote}
    \footnotesize
Measure the projected two-point energy-energy correlator (E2C), the projected three-point energy correlator (E3C), and the energy-energy correlator asymmetry (AEEC) in hadronic $Z$ decays using archived ALEPH data at $\sqrt{s} = 91.2$~GeV.

The EEC is the energy-weighted angular correlation of all final-state particle pairs per event; the E3C extends this to energy-weighted triplets projected onto the largest pairwise angle; the AEEC is the forward-backward asymmetry of the EEC, whose leading hadronization correction cancels in the difference. Correct the detector-level distributions to particle level using the ALEPH full-simulation MC, and present the EECs both as a function of the opening angle $\chi$ and in the $z = (1-\cos\theta)/2$ representation. Where the observables support it, extract $\alpha_s(M_Z)$ --- classically from the EEC and AEEC, and via the modern $E3C/E2C$ ratio, which benefits from cancellation of normalization and reduced hadronization sensitivity.

Data: [...]
\end{quote}

\paragraph{Event shapes and $\alpha_s$ extraction, ALEPH (\href{https://github.com/jfc-mit/analysis_aleph_eventshapes_alphas}{Github}).}
\begin{quote}
    \footnotesize
Perform a state-of-the-art determination of the strong coupling constant $\alpha_s(M_Z)$ using hadronic Z decay data from ALEPH at $\sqrt{s} = 91.2$~GeV. Measure the distributions of classic event shape variables --- thrust ($T$), heavy jet mass ($\rho_H$), wide and total jet broadenings ($B_W$, $B_T$), $C$-parameter, and the Durham $y_{23}$ jet resolution parameter --- using the full LEP1 dataset. Correct the data to particle level using the ALEPH simulation chain and unfold using modern techniques (e.g., iterative Bayesian or OmniFold).

Fit the unfolded distributions with NNLO QCD predictions matched to NNLL (or N$^3$LL where available) resummation, extracting $\alpha_s$ and non-perturbative power corrections simultaneously. Compare FOPT and CIPT prescriptions. Combine the individual extractions into a single ALEPH $\alpha_s$ result and compare with the current world average.

The theoretical landscape has improved dramatically since the original ALEPH publications: NNLO corrections are now available for all six event shapes, and resummation has been pushed to N$^3$LL for thrust. This reanalysis could yield one of the most precise $\alpha_s$ determinations from $e^+e^-$ data, directly relevant to the ongoing global $\alpha_s$ tension.

Data: [...]
\end{quote}

\ifdelphiprompt
\paragraph{$N_\nu$ from $\Gamma_\text{inv}$, DELPHI (\href{https://github.com/jfc-mit/archive_v0_analysis_delphi_z}{Github}).}
\begin{quote}
    \footnotesize
Reproduce the classic LEP measurement of the number of light neutrino generations $N_\nu$ from the Z boson invisible decay width using DELPHI open data.

Measure the Z lineshape (cross section vs $\sqrt{s}$) in the hadronic and leptonic channels from the LEP1 energy scan runs. Extract $\Gamma_Z$ (total width), $\Gamma_{had}$, and $\Gamma_\ell$. Compute the invisible width $\Gamma_{inv} = \Gamma_Z - \Gamma_{had} - 3\Gamma_\ell$ and determine $N_\nu = \Gamma_{inv} / \Gamma^{(SM)}_{\nu\bar{\nu}}$. The original LEP result is $N_\nu = 2.9840 \pm 0.0082$, consistent with exactly 3 generations.

Data: [...]
\end{quote}

\paragraph{Lund jet plane, DELPHI (\href{https://github.com/jfc-mit/archive_v0_analysis_delphi_lund}{Github}).}
\begin{quote}
    \footnotesize
Measure the primary Lund jet-plane density in hadronic $Z$ decays using DELPHI data.

Recluster jets with the Cambridge/Aachen algorithm and follow the harder branch at each declustering step, recording the splitting variables ($k_T$, $\Delta R$) of the softer emission. Populate the Lund plane in the $(\ln(1/\Delta R), \ln k_T)$ representation.

This observable has been measured by ATLAS and ALICE at LHC energies but has never been measured at LEP. The clean $e^+e^-$ initial state (no underlying event, no pileup, known $\sqrt{s}$) makes this the definitive benchmark for the Lund plane: the perturbative region should match NLL DGLAP predictions exactly, while the nonperturbative boundary provides a clean measurement of the hadronization transition scale.

Data: [...]
\end{quote}

\paragraph{Energy-energy correlator, DELPHI (\href{https://github.com/jfc-mit/archive_v0_analysis_delphi_eec_two_point}{Github}).}
\begin{quote}
    \footnotesize
Measure the two-point energy-energy correlator (EEC) as a function of the angular separation $\chi$ between particle pairs in hadronic Z decays at DELPHI.

Cover the full angular range from the collinear limit ($\chi \to 0$) through the transition region to the back-to-back limit ($\chi \to \pi$). Correct for detector effects using the DELPHI simulation.

In the back-to-back region, extract $\alpha_s$ using the transverse-momentum-dependent (TMD) factorization framework, which provides excellent perturbative control.

In the collinear limit, probe the transition from perturbative to non-perturbative dynamics and test models of hadronization including recent analytic results connecting EECs to track functions.

Measure the three-point EEC as well, which probes the detailed radiation pattern in $e^+e^- \to 3$ jets. Compare with predictions from modern generators (Pythia 8, Sherpa, Herwig 7). This analysis has been done with ALEPH archived data but never with DELPHI.

Data: [...]
\end{quote}

\paragraph{Event shapes and $\alpha_s$ extraction, DELPHI (\href{https://github.com/jfc-mit/archive_v0_analysis_delphi_had_eventshape_alpha_s}{Github}).}
\begin{quote}
    \footnotesize
Perform a state-of-the-art determination of the strong coupling constant $\alpha_s(M_Z)$ using hadronic Z decay data from DELPHI at $\sqrt{s} = 91.2$~GeV. Measure the distributions of classic event shape variables --- thrust ($T$), heavy jet mass ($\rho_H$), wide and total jet broadenings ($B_W$, $B_T$), $C$-parameter, and the Durham $y_{23}$ jet resolution parameter --- using the full LEP1 dataset. Correct the data to particle level using the DELPHI simulation chain and unfold using modern techniques (e.g., iterative Bayesian or OmniFold).

Fit the unfolded distributions with NNLO QCD predictions matched to NNLL (or N$^3$LL where available) resummation, extracting $\alpha_s$ and non-perturbative power corrections simultaneously. Compare FOPT and CIPT prescriptions. Combine the individual extractions into a single DELPHI $\alpha_s$ result and compare with the current world average.

The theoretical landscape has improved dramatically since the original DELPHI publications: NNLO corrections are now available for all six event shapes, and resummation has been pushed to N$^3$LL for thrust. This reanalysis could yield one of the most precise $\alpha_s$ determinations from $e^+e^-$ data, directly relevant to the ongoing global $\alpha_s$ tension.

Data: [...]
\end{quote}
\fi

\end{document}